\documentclass[aps,pra,superscriptaddress,footnote, longbibliography,footinbib,twocolumn,notitlepage]{revtex4-2}

\usepackage{appendix}

\usepackage{tikz}  
\usetikzlibrary{arrows,shapes,positioning,shadows,backgrounds,fit}

\usepackage{subcaption}
\usepackage{relsize}
\usepackage{graphicx}
\usepackage{amsmath, bbm}
\usepackage{amssymb}
\usepackage{amsthm}
\usepackage{mathrsfs}
\usepackage{xcolor}
\usepackage{cancel}
\usepackage{titlesec}
\usepackage{enumitem}  
\usepackage{mathtools}
\usepackage{braket}

\usepackage[normalem]{ulem}

\titlespacing{\section}{0ex}{2ex}{0.4ex}
\titleformat{name=\section}{\bf}{\thesection.}{.3em}{}

\usepackage{subcaption}
\usepackage{soul}
\usepackage{dsfont}

\def\be{\begin{eqnarray}}
\def\ee{\end{eqnarray}}
\newcommand{\tr}[1]{\text{Tr}\left(#1\right)}

\newcommand{\ad}[1]{\textrm{ad}_{#1}}

\theoremstyle{plain}

\definecolor{myblue}{rgb}{0.2,0.2,0.8}
\definecolor{myblack}{rgb}{0,0,0}
\definecolor{myurl}{rgb}{0.1,0.1,0.4}

\usepackage[colorlinks=true,citecolor=myblue,linkcolor=myblack,urlcolor=myurl]{hyperref}

\begin{document}

\title{Restoring the second law to classical-quantum dynamics}

\author{Isaac Layton}
\affiliation{Department of Physics and Astronomy, University College London, Gower Street, London WC1E 6BT, UK}
\affiliation{Center for Quantum Computing, RIKEN, Wakoshi, Saitama, 351-0198, Japan}
\affiliation{Department of Applied Physics, The University of Tokyo,
7-3-1 Hongo, Bunkyo-ku, Tokyo 113-8656, Japan}

\author{Harry J.~D. Miller}
\affiliation{Department of Physics and Astronomy, The University of Manchester, Manchester M13 9PL, UK}

\begin{abstract}
All physical theories should obey the second law of thermodynamics. However, existing proposals to describe the dynamics of hybrid classical-quantum systems either 
violate the second law or lack a proof of its existence. Here we rectify this by studying classical-quantum dynamics that are (1) linear and completely-positive and (2) preserve the thermal state of the classical-quantum system. We first prove that such dynamics necessarily satisfy the second law.  We then show how these dynamics may be constructed, proposing dynamics that generalise the standard Langevin and Fokker-Planck equations for classical systems in thermal environments to include back-reaction from a quantum degree of freedom. Deriving necessary and sufficient conditions for completely-positive, linear and continuous classical-quantum dynamics to satisfy detailed balance, we find this property satisfied by our dynamics. To illustrate the formalism and its applications we introduce two models. The first is an analytically solvable model of an overdamped classical system coupled to a quantum two-level system, which we use to study the total entropy production in both quantum system and classical measurement apparatus during a quantum measurement. The second describes an underdamped classical-quantum oscillator system subject to friction, which we numerically demonstrate exhibits thermalisation in the adiabatic basis, showing the relevance of our dynamics for the mixed classical-quantum simulation of molecules. 
\end{abstract}

\maketitle

\section{Introduction}

A basic test of any theory is that it satisfies the second law of thermodynamics. By constructing for the first time consistent dynamics that preserve the classical-quantum thermal state and satisfy detailed balance, we here prove that theories of interacting classical and quantum degrees of freedom can also satisfy this basic requirement.

While many systems are well-described as entirely classical or entirely quantum, a number of regimes studied in physics and chemistry require the use of hybrid descriptions, where classical and quantum degrees of freedom coexist and interact. Common examples include semi-classical gravity \cite{wald1977back,Kuo:1993if,CalzettaHu},  molecular dynamics \cite{tully1990molecular,tully1998mixed,Kapral1999}, continuous measurement and feedback \cite{gagen1993continuous,jacobs2006straightforward,annby2022quantum} and systems at the classical-quantum boundary \cite{bertet2001complementarity,PhysRevLett.130.193603,PhysRevLett.89.018301}.

However, despite a huge range of theories having been suggested to describe the interactions between classical and quantum systems \cite{aleksandrov1981statistical,gerasimenko1982dynamical,boucher1988semiclassical,anderson1995quantum,diosi1995quantum,diosi1998coupling,diosi2000quantum,peres2001hybrid,hall2005interacting,salcedo2012statistical,barcelo2012hybrid,vachaspati2018classical,bondar2019koopman,barchielli2023hybrid,husain2023motivating,dammeier2023quantum,oppenheim2023postquantum,terno2023classical}, the question of whether any can be constructed to be compatible with the laws of thermodynamics has been largely left unanswered \footnote{A recent review \cite{terno2023classical} noted that most models do not demonstrate consistency with thermodynamics; those that were claimed to do so also fail the basic assumption of positivity, which means that the system cannot be interpreted as having an effectively classical subsystem \cite{layton2024classical}}. Indeed, this is in part because the common approaches used to study classical-quantum hybrids, namely mean-field methods \cite{boucher1988semiclassical,tully1998mixed} and reversible quantum-classical evolution laws \cite{aleksandrov1981statistical,gerasimenko1982dynamical,Kapral1999}, manifestly fail to satisfy both of the basic consistency requirements of a probability theory, namely linearity and positivity, which necessarily lead to violations of the second law of thermodynamics \cite{peres1989nonlinear,Argentieri_2014}. While a recent approach based on mean-field dynamics showed success for systems close to equilibrium using perturbation theory \cite{eglinton2024stochastic}, a form of classical-quantum dynamics that demonstrates consistency with thermodynamics far from equilibrium and for arbitrary coupling strengths has remained unknown.

To construct classical-quantum dynamics compatible with the second law of thermodynamics, we build on the theory of linear, completely-positive and continuous classical-quantum dynamics \cite{diosi1995quantum,diosi1998coupling,oppenheim2022two,oppenheim2023postquantum,weller2024classical}, which provide a consistent description of interacting classical and quantum systems at both the level of individual trajectories \cite{layton2022healthier} and at the level of the master equation \cite{oppenheim2022two}. The key property of such theories is that they are necessarily irreversible, with diffusion in the classical degrees of freedom and decoherence in the quantum ones \cite{oppenheim2023gravitationally}.  However, these additional sources of noise are typically understood to lead to unbounded energy production \cite{oppenheim2023objective,oppenheim2023postquantum,diosi2024classical}, making a consistent thermodynamic interpretation of such theories challenging.

To restore thermodynamic stability to classical-quantum theories, we study dynamics that preserve the thermal state of the combined classical-quantum system. A standard property in both classical and quantum non-equilibrium thermodynamics \cite{seifert2012stochastic,breuer2002theory}, this assumption naturally leads us to interpret the classical-quantum system as open, exchanging heat and entropy with a thermal environment. First suggested of the earliest works on this topic \cite{diosi1998coupling,halliwell1998effective}, the identification of a thermal environment provides a natural explanation of both the irreversibility of these theories and their emergence via environmental decoherence of a subsystem \cite{layton2024classical}. Most significantly, we show that for any completely-positive and linear dynamics, a positive entropy production may be defined \cite{landi2021irreversible}, proving the second law of thermodynamics for classical-quantum systems arbitrarily far from equilibrium.

The main technical contribution of our work is to provide tools that allow consistent thermal state preserving and detailed balance classical-quantum dynamics to be constructed. We first identify two classes of operators that allow the construction of dynamics that preserve arbitrary classical-quantum thermal states. We illustrate this by introducing two classes of dynamics that generalise the stochastic motion of overdamped and underdamped classical Fokker-Planck/Langevin equations \cite{risken1996fokker,itami2017universal} to the case in which the classical system experiences back-reaction from a quantum system.  By deriving necessary and sufficient conditions for detailed balance in classical-quantum systems, generalising the well-known classical \cite{graham1971generalized,risken1972solutions} and quantum \cite{fagnola2007generators} results, we show that this dynamics satisfies detailed balance. The framework not only applies at an ensemble level but also along individual stochastic trajectories. These results provide a step towards a general theory of non-equilibrium classical-quantum thermodynamics, combining the respective theories of classical \cite{seifert2012stochastic} and quantum \cite{sagawa2013second,vinjanampathy2016quantum} non-equilibrium thermodynamics  into one cohesive framework.

Our work has two main applications, which we illustrate by introducing and solving two models. The first core application is in the study of measurement and control of quantum systems. While the connection to classical-quantum dynamics has been known for some time  \cite{blanchard1993interaction,diosi1998coupling,milburn2012decoherence}, recent work has demonstrated how continuous measurement and feedback can be described using classical-quantum master equations \cite{annby2022quantum,desousa2025continuous} and proven the general equivalence of these two frameworks \cite{layton2022healthier,tilloy2024general}. By equipping classical-quantum theories with notions of detailed balance and the second law, we automatically extend these concepts to the study of continuous measurement and feedback. 
While a number of proposals exist to study detailed balance and the second law on the quantum system {alone}  \cite{manzano2022quantum,yada2022quantum,kumasaki2025thermodynamic,prech2025quantum}, {our contribution is to show how entire continuous measurement and feedback systems may in principle obey detailed balance and the second law{, and to provide a complete characterisation of their dynamics}.}   This is expected to be important as measurement and control systems are further developed into mesoscopic regimes \cite{clerk2003quantum,clerk2010introduction,oxtoby2006sensitivity}, where effectively classical degrees of freedom are unavoidably affected by thermal noise and back-reaction from the quantum degrees of freedom they manipulate. To illustrate this application, we introduce a simple toy  model of a quantum two level system coupled to a single classical degree of freedom, which we show may be interpreted either as an overdamped mesoscopic classical system, or as a macroscopic detector with finite bandwidth \cite{annby2022quantum}.  Providing to our knowledge the first analytical solution of a continuous classical-quantum master equation, we use this toy model to study the total entropy production of a quantum system plus measurement device during a quantum measurement, as well as the limits on quantum control when the control system is subject to fluctuations and quantum backreaction.

The second key application of our work is in the study of molecular dynamics. Here, the question of finding a detailed balance and thermal state preserving dynamics has been an open problem for many years \cite{mauri1993canonical,parandekar2005mixed,Kapral2006,schmidt2008mixed}, with even approximate solutions violating basic properties such as the linearity of dynamics, positivity of electronic populations, or the physicality of individual classical trajectories \cite{alonso2021computation, amati2023detailed,mannouch2023mapping}. The dynamics we provide not only solve these issues, but also reduce to consistent versions of two well-known methods in molecular dynamics, namely mean-field dynamics \cite{tully1998mixed} and the quantum-classical Liouville equation \cite{Kapral1999,Kapral2006}, in the high temperature limit. We illustrate this dynamics using a model of two coupled oscillators, one classical and one quantum, which we study by mapping the adiabatic basis to the standard harmonic oscillator basis. Using simulation methods from continuous measurement theory \cite{amini2011stability,rouchon2015efficient,ralph2016coupling}, we numerically solve the dynamics and show that it exhibits thermalisation in the adiabatic basis. The simulation is used to compute the statistics of heat dissipation arising from this equilibration process, which are confirmed to be consistent with the second law at the ensemble level. {Beyond this model, our conditions for detailed balance completely characterise the class of molecular dynamics that are linear, satisfy detailed-balance and guarantee well-defined classical and quantum trajectories. Our framework thus provides a natural setting for studying the non-equilibrium thermodynamics of molecules treated using hybrid classical-quantum methods.} 

Outside of the practical applications that we describe via these two models, our findings are also of relevance to recent foundational studies on the nature of gravity and decoherence. By proving that classical-quantum theories are compatible with the second law, our work puts recent attempts at constructing hybrid and stochastic alternatives to quantum gravity \cite{kafri2014classical,tilloy2016sourcing,oppenheim2023postquantum,layton2023weak,oppenheim2023path,layton2022healthier,diosi2024classical,angeli2025probing} on firmer ground, and provides a template for how friction may be added to these theories to ensure their stability \cite{carney2024classical}. Beyond gravitationally induced decoherence, the new forms of dynamics that we introduce demonstrate how including friction in the stochastic field driving collapse, rather than in the quantum system itself \cite{smirne2015dissipative,di2023linear,artini2025non}, may be used to construct collapse models provably compatible with the second law of thermodynamics.

\color{black}

The outline of the paper is as follows. In Section \ref{sec: framework} we review the formalism of classical-quantum dynamics, before describing in Section \ref{sec: entropyetc.} how the notions of entropy, energy and the second law can arise classical-quantum systems. In Section \ref{sec: construct} we introduce both overdamped and underdamped thermal state preserving dynamics, and demonstrate how our dynamics generalises the mean-field and quantum-classical Liouville approaches to dynamics that are completely-positive and linear. In Sections \ref{sec: analytic_model} and \ref{sec: oscillator_model} we introduce and solve two models to illustrate the general features of our dynamics and its applications. Finally, in Section \ref{sec: DB} we find general necessary and sufficient conditions for systems satisfying classical-quantum detailed balance, which we demonstrate are satisfied by both the overdamped and underdamped dynamics introduced in this paper.

To streamline the notation in the bulk of the paper, we will opt to denote operators on Hilbert space simply using capital letters (e.g. $H$, $L_z$, $M_{xy}$) or Greek letters (e.g. $\rho$, $\varrho$) that are otherwise undistinguished from other scalar quantities (e.g. $S$, $E$, $\Sigma$).  However, when studying the two models in Sections \ref{sec: analytic_model} and \ref{sec: oscillator_model}, we use hats to distinguish operator-valued quantities from real numbers for clarity and consistency with existing notation.

\section{Classical-quantum states and dynamics} \label{sec: framework}

\

We start by reviewing the framework for modelling combined classical-quantum systems \cite{aleksandrov1981statistical,gerasimenko1982dynamical,boucher1988semiclassical,layton2022healthier,weller2024classical}. The key feature we emphasise is that there exist two equivalent and interchangeable pictures: one in which the combined system is described by a pair of points in a classical state space and a quantum Hilbert space, the other where the total state of the system is described by the classical-quantum state, a hybrid object that generalises both the classical probability distribution and the quantum density operator. We introduce the general class of dynamics that allow both descriptions to be used consistently and interchangeably, taking the form of stochastic unravellings in the trajectory picture and classical-quantum master equations in the ensemble picture, and compare this to other common forms of classical-quantum dynamics, which fail to simultaneously guarantee consistent dynamics in both pictures.

\subsection{Classical-quantum kinematics}

We begin by recalling the formalism necessary to collectively describe the kinematics – i.e. states and observables – of a combined classical-quantum system. The classical degrees of freedom are characterised by points $z$ in a classical state space $\mathcal{M}\subseteq\mathbb{R}^n$ i.e. by $n$ real numbers $z=(z_1,\ldots,z_n)$. This may correspond to phase space, in the case of underdamped evolution, or configuration space when the evolution is overdamped. On the other hand, the quantum system is characterised by a separable Hilbert space $\mathcal{H}$, which may correspond to single qubit or bosonic quantum systems, or many interacting quantum degrees of freedom.

The most intuitive picture of classical-quantum systems is that on the level of individual trajectories \cite{boucher1988semiclassical,diosi1998coupling,jacobs2006straightforward,layton2022healthier}. Here, at any given time $t$, the classical system is described by a point in classical state space $z_t\in \mathcal{M}$, while the quantum system is described by a density operator $\rho_t$ i.e. a unit trace positive semi-definite operator acting on $\mathcal{H}$. Since the classical and quantum systems may be subject to noise, one must in general allow for $z_t$ and $\rho_t$ to be random variables. Considered as functions of time, the random variables $z_t$ and $\rho_t$ thus define stochastic processes, which we denote using a subscript $t$ and use $\mathbb{E}[\ \cdot\ ]$ to denote their expectation value as random variables. In general, each realisation of $z_t$ and $\rho_t$ generate distinct individual trajectories in the classical and quantum state spaces, providing an intuitive picture of classical-quantum dynamics using the standard classical and quantum frameworks.

An equivalent description at the ensemble level is provided by the classical-quantum state \cite{boucher1988semiclassical, diosi2014hybrid,oppenheim2023postquantum}. Here, the entire information about the classical-quantum system is contained in an operator-valued function of phase space $\varrho(z,t)$, known as the classical-quantum state, which must be (1) positive semi-definite at all points $z$, and (2) normalised to one after integrating over the classical state space and tracing over Hilbert space i.e. $\int dz \ \tr{\varrho(z,t)}=1$. This object is given physical meaning by identifying the classical probability distribution $P(z,t)$ with its trace 
\begin{equation}
    P(z,t)=\tr{\varrho(z,t)},
\end{equation} and identifying the quantum state conditioned on a given classical outcome $z$, which we denote $\rho(z,t)$, with its normalised value 
\begin{equation} \label{eq: rho_on_z_from_CQstate}
    \rho(z,t)=\frac{\varrho(z,t)}{\tr{\varrho(z,t)}},
\end{equation} each of which are guaranteed to satisfy the required positivity and normalisation properties by virtue of the two conditions on $\varrho(z,t)$. Using these, the classical-quantum state may also be written as
\begin{equation} \label{eq: CQstate_as_product}
    \varrho(z,t)=\rho(z,t)P(z,t),
\end{equation} which can be seen to be equivalent to the definition in terms of positivity and normalisation. From this, we see that the classical-quantum state provides a natural generalisation of the classical probability distribution or quantum density matrix to the combined classical-quantum case, and thus is important for characterising both the consistency and properties of classical-quantum dynamics.

A key feature of this framework is that the trajectory and ensemble descriptions can be directly related to each other \cite{layton2022healthier}. In particular, the two representations are related by the fundamental expression 
\begin{equation} \label{eq: CQ_state_from_ensembles}
    \varrho(z,t)=\mathbb{E}[\delta(z-z_t)\rho_t],
\end{equation} where here $\delta(z-z_t)$ denotes a delta function centred on the point $z_t$. To see how this relation arises, we first note that  the probability distribution may be written in terms of the random variable $z_t$ as
\begin{equation}
    P(z,t)=\mathbb{E}[\delta(z-z_t)],
\end{equation} which follows from the definition of the expectation value, while the quantum state conditioned on $(z,t)$ may be written as
\begin{equation} \label{eq: rho_on_z_from_trajectory}
    \rho(z,t)=\mathbb{E}[\rho_t|z_t=z]
\end{equation}where here $\mathbb{E}[\ \cdot\ |z_t=z]$ denotes the expectation conditioned on the outcome $z_t=z$. Substituting these into the definition of $\varrho(z,t)$ given in \eqref{eq: CQstate_as_product}, one sees that the two expectation values may be combined into one due to the presence of the delta function, thus recovering the expression \eqref{eq: CQ_state_from_ensembles}. 

A subtle but important conceptual feature of classical-quantum systems is that the state assigned to describe the quantum system depends on the degree of conditioning on the classical system \cite{layton2022healthier}. Taking first the extreme case, where no information about the classical system is available, the quantum system is described by the unconditioned state $\rho(t)$, which is found by integrating over the classical degrees of freedom in the classical-quantum state
\begin{equation}
    \rho(t)=\int dz \ \varrho(z,t). 
\end{equation} Using \eqref{eq: CQ_state_from_ensembles}, one can check that this may be written in the trajectory picture simply as 
$\rho(t)=\mathbb{E}[\rho_t]$. An intermediate case occurs when the only final state of the classical system $z$ is known, which gives the $z_t=z$ conditioned quantum state given in \eqref{eq: rho_on_z_from_CQstate} and \eqref{eq: rho_on_z_from_trajectory}. Finally, one may consider the other extreme case, where the entirety of the classical trajectory $\{z_s\}_{s\leq t}$ is known and conditioned upon. Since any remaining ambiguity in $\rho_t$ at this point would not be physical, we will always choose to represent dynamics such that $\rho_t$ corresponds to the state of the quantum system conditioned on the entire classical trajectory up to time $t$ i.e.
\begin{equation} \label{eq: CQ_rho_conditioned}
    \rho_t=\mathbb{E}[\rho_t|\{z_s\}_{s\leq t}].
\end{equation} This choice ensures that the state $\rho_t$ always has a physical meaning by virtue of the reality of the classical trajectories \cite{oppenheim2023objective,layton2022healthier}. In the special case that $\rho_t$ is pure, individual realisations of pure states are physically well-defined and unamibiguous, and we shall denote the quantum system here using the corresponding vector in Hilbert space $|\psi\rangle_t$. This may be understood as directly analogous to the case of perfectly efficient continuous quantum measurement, where the quantum system remains pure conditioned on the classical measurement signal.

Finally, turning to observables, we note that we may define expectation values on both the level of trajectories \cite{jacobs2006straightforward} and on the level of the ensemble \cite{aleksandrov1981statistical,gerasimenko1982dynamical,boucher1988semiclassical}. To start with, we define a classical-quantum observable to be a Hermitian operator-valued function of phase space, which we shall denote $A(z),B(z),\ldots$ etc. Since this defines a quantum observable, one may define a stochastic quantity simply by taking the standard quantum expectation value with respect to a given realisation of $z_t$ and $\rho_t$
\begin{equation} \label{eq: trajectory_expectation}
    \langle A(z) \rangle_t=\tr{A(z_t) \rho_t}.
\end{equation} Referring to this as the trajectory expectation value of $A(z)$, a given realisation of this random variable corresponds physically to averaging the outcomes of $A(z)$ measurements made on the quantum system when a specific classical trajectory occurs. On the other hand, we may also define the ensemble expectation value of a classical-quantum observable  
as
\begin{align} \label{eq: CQ_expectation_value}
    \langle \langle A(z) \rangle \rangle=\int dz \ \tr{A(z)\varrho(z)},
\end{align} where the double angled brackets indicate that here one must integrate over the classical state space and trace over the quantum Hilbert space. To relate the trajectory and ensemble expectation values, we may substitute \eqref{eq: CQ_state_from_ensembles} into the definition of the ensemble expectation value to see that 
\begin{equation} \label{eq: exp_traj_exp_ensemble}
    \langle \langle A(z) \rangle \rangle (t) = \mathbb{E}[\langle A(z) \rangle_t],
\end{equation} i.e. the ensemble expectation value of a classical-quantum observable at time $t$ is equal to the mean value of the corresponding trajectory expectation value.

\subsection{Classical-quantum dynamics}

Having defined the trajectory and ensemble descriptions, characterised either by $z_t$ and $\rho_t$ or by the classical-quantum state $\varrho(z,t)$, we now turn to studying dynamics in these two pictures, which correspond to stochastic unravellings and master equations respectively.

We start by discussing which properties of classical-quantum dynamics are necessary for the time evolution to be consistent with the trajectory and ensemble pictures of classical-quantum systems \cite{layton2022healthier}. Firstly, we note that if a trajectory level picture exists in terms of $z_t$ and $\rho_t$, then the corresponding ensemble level classical-quantum state $\varrho(z,t)$ is necessarily positive semi-definite everywhere in phase space. For the trajectory picture to remain valid over time, the dynamics must therefore preserve the positivity of the classical-quantum state. Moreover, for this to be valid when applied to just part of a quantum system, this dynamics must also be completely-positive \cite{layton2022healthier}. Finally, we note that since the classical-quantum state has a statistical interpretation, it is important to consider dynamics that is linear in $\varrho(z,t)$, just as one does when considering dynamics of either quantum density operators $\rho(t)$ or classical probability distributions $P(z,t)$. 

Alongside the necessary assumptions of complete-positivity and linearity, we will make two additional assumptions on the class of dynamics we work with. Firstly, we will focus our attention on dynamics that are Markovian in the classical-quantum state. Consistent with the vast majority of proposed classical-quantum dynamics \cite{terno2023classical}, the main advantage of assuming Markovianity is that it allows us to leverage a recent theorem known as the CQ Pawula theorem \cite{oppenheim2022two,oppenheim2023postquantum}, which characterised the general form of Markovian, completely-positive and linear classical-quantum dynamics. Secondly, to consistently describe classical degrees of freedom such as position and momentum, we additionally assume that the dynamics generate trajectories that are continuous in the classical degrees of freedom.

To write down the general form of classical-quantum dynamics in the ensemble picture, we use the formalism of classical-quantum master equations. First developed in \cite{aleksandrov1981statistical,boucher1988semiclassical,diosi1995quantum}, one may write their generic form as
\begin{equation} \label{eq: generic_master_equation}
     \frac{\partial\varrho}{\partial t}=\mathcal{L}\varrho,
\end{equation} where $\mathcal{L}$ is a classical-quantum superoperator that acts as the generator of dynamics. Under the assumptions made above, i.e. that the dynamics is completely-positive, linear, Markovian and continuous in phase space, the general form of this generator is characterised by a theorem known as the CQ Pawula theorem \cite{oppenheim2022two,oppenheim2023postquantum}. Denoting by $L_\alpha$ a set of $p$ operators acting on the Hilbert space, and assuming summation over repeated Roman letters $i,j=1,\dots,n$ or Greek letters $\alpha,\beta=1,\dots,p$, we may write this generator in the form
\begin{equation} \label{eq: L_general}
\begin{split}
\mathcal{L}\varrho=&  -\frac{\partial }{\partial z_i}(D_{1,i}^C  \varrho ) +\frac{1}{2} \frac{\partial^2 }{\partial z_i \partial z_j} ( D_{2, i j}  \varrho )\\
&-i[\bar{H},  \varrho ] + D_0^{\alpha \beta}\big( L_{\alpha}  \varrho L_{\beta}^{\dag} - \frac{1}{2}  \{ L_{\beta}^{\dag} L_{\alpha},  \varrho \}_+ \big) \\
&- \frac{\partial }{\partial z_{i}} \left( {D^{\alpha }_{1, i}}^* L_{\alpha}  \varrho +  \varrho D^{\alpha}_{1, i}  L_{\alpha}^{\dag} \right).
\end{split}
\end{equation}
The first line describes purely classical dynamics, with the classsical drift vector given by $D_1^C$, a real vector of length $n$, and the diffusion matrix denoted $D_2$, a real positive semi-definite $n\times n$ matrix. The purely quantum dynamics is determined by the Hermitian operator $\bar{H}$ describing the unitary evolution and the complex positive semi-definite $p\times p$ decoherence matrix $D_0$. Finally, the quantum back-reaction term appears on the final line, controlled by the $n\times p$ matrix $D_1$ with elements denoted $D_{1,i}^\alpha$. All of the $D$ matrices and operators $H$ and $L_\alpha$ may have dependence on $z$. 

For this dynamics to be completely-positive, two positivity conditions must be satisfied. The first is known as the decoherence-diffusion trade-off
\begin{align} \label{eq: dec-diff}
    D_0 \succeq D_1^{\dag} D_2^{-1} D_1,
\end{align} which states that when the back-reaction on the classical system is non-zero, there must be a minimum amount of decoherence and diffusion in the system. Here $D_2^{-1}$ denotes the pseudoinverse of the diffusion matrix $D_2$, and $A\succeq B$ is shorthand notation for the statement that $A-B\succeq 0$ i.e. the matrix $A-B$ is positive semi-definite. The second positivity condition controls the degrees of freedom in which diffusion is necessary, and is written as
\begin{align} \label{eq: pos_2}
   (\mathbb{I}- D_2 D_2^{-1})D_1 =0.
\end{align} where here $\mathbb{I}$ denotes the $n\times n$ identity matrix. When the operators $L_\alpha$ are orthogonal and traceless, positivity conditions \eqref{eq: dec-diff} and \eqref{eq: pos_2} provide sufficient and necessary conditions for positivity, but are sufficient to establish positivity for arbitrary $L_\alpha$ \cite{oppenheim2022two}.

This dynamics may also be written in the trajectories picture using the formalism of classical-quantum unravellings. Built using techniques from continuous quantum measurement theory \cite{belavkin1989new,carmichael1993open,gisin1989stochastic,wiseman2001complete,jacobs2006straightforward}, and appearing as a special case in \cite{diosi1998coupling}, the general form of classical-quantum unravellings was provided in \cite{layton2022healthier} (see also \cite{diosi2023hybrid,barchielli2023hybrid,tilloy2024general} for later discussion).  Defining $W^i_t$ to be a component of an $n$ dimensional Wiener process with corresponding increments $dW^i_{t}$ satisfying $dW^i_{t}dW^j_{t}=\delta_{ij}dt$, we may write the general form of classical-quantum unravelling as the following set of stochastic differential equations
\begin{equation}\label{eq: unravelingCQClass}
     \begin{split}
   dz^i_{t} =& D_{1,i}^{C} dt + \langle {D^{\alpha}_{1,i}}^*  L_{\alpha} + D^{\alpha }_{1,i} L_{\alpha}^{\dag} \rangle dt  + \sigma_{ij} d W^j_{t} 
 \end{split}
 \end{equation}
 and 
 \begin{equation}\label{eq: unravelingCQCQuantum}
 \begin{split}
  d \rho_t =& -i[\bar{H}, \rho_t]dt \\
  & + D_0^{\alpha \beta}(L_{\alpha} \rho L_{\beta}^{\dag} dt - \frac{1}{2} \{L_{\beta}^{\dag} L_{\alpha}, \rho_t \}_+ ) dt\\
   & + {D^{\alpha}_{1,j}}^* \sigma^{-1}_{ij}  (L_{\alpha} - \langle L_{\alpha} \rangle) \rho_t d W^i_t \\
 & +  D_{1,j}^{ \alpha} \sigma^{-1}_{ij}  \rho_t (L_{\alpha}^{\dag} - \langle L_{\alpha}^{\dag} \rangle)  d W^i_t
 \end{split}
 \end{equation} where here in the above $\langle A\rangle$ denotes $\tr{A\rho_t}$ i.e. the trajectory expectation value given in \eqref{eq: trajectory_expectation}. As before, the $D$ coefficients, $\bar{H}$ and $L$ may all depend on $z_t$, and this dynamics must satisfy both \eqref{eq: dec-diff} and \eqref{eq: pos_2} to be well-defined.

A special limit of the dynamics occurs when the decoherence in the system is minimal. In particular, when the decoherence-diffusion trade-off is saturated 
\begin{equation} \label{eq: saturating_dec_diff}
    D_0=D_1^\dag D_2^{-1} D_1,
\end{equation} one can show that the dynamics maintains the purity of initial pure states \cite{layton2022healthier}. In this case, one may rewrite the quantum part of dynamics entirely in terms of a pure quantum state $|\psi\rangle_t$. Moreover, it is straightforward to check using properties of the pseudoinverse $\sigma^{-1}$ that one may replace all of the appearances of $dW^i_t$ in Eq. \eqref{eq: unravelingCQCQuantum} with $dz^i_t$, verifying that $\rho_t=|\psi\rangle_t\langle\psi|_t$ satisfies Eq. \eqref{eq: CQ_rho_conditioned}. This guarantees that $|\psi\rangle_t$ may indeed be understood to be the quantum state conditioned on $\{z_s\}_{s\leq t}$ i.e. that the state $|\psi\rangle_t$ is uniquely and unambiguously determined from the observations of the classical trajectory up to time $t$.

Finally, we note the formal equivalence of the classical-quantum unravellings we consider with the diffusive unravellings of continuous measurement theory \cite{belavkin1989new,carmichael1993open,gisin1989stochastic,wiseman2001complete,jacobs2006straightforward}. In particular, one may show that the $z^i_t$ may always be interpreted as classical measurement signals, while the $\rho_t$ as the quantum state conditioned on these \cite{layton2022healthier,tilloy2024general}. Since the various $D$ matrices and $\bar{H}$ may depend on the measurement signals $z^i_t$, the dynamics in general include feedback onto the quantum system. The equivalence of the unravelling to the master equation thus provides an alternative representation for studying continuous measurement and feedback \cite{rosal2025deterministic}, which can be useful for solving the dynamics (c.f. Section \ref{sec: analytic_model}) or characterising its properties (c.f. Sec. \ref{sec: DB}).

\subsection{Other approaches to classical-quantum dynamics} \label{subsec: other_approaches}

It is worth contrasting this to other approaches that have been used to model classical-quantum dynamics that do not satisfy the basic consistency requirements of positivity and linearity.

An important example in the context of master equations is the quantum-classical Liouville equation, first suggested in  \cite{aleksandrov1981statistical}, with more recent use in the context of physical chemistry \cite{Kapral1999,Kapral2006}. Defined for when the classical system is described by a phase space i.e. $z=(q,p)$ and for a Hermitian operator-valued function of phase space known as the classical-quantum Hamiltonian $H(z)$, this approach is a straightforward generalisation of Hamiltonian dynamics and takes the form
\begin{equation}\label{eq: CQ_liouville}
   \frac{\partial \varrho}{\partial t}=-\frac{i}{\hbar}[H,\varrho]+\frac{1}{2}\big(\big\{H,\varrho\big\}-\big\{\varrho,H\big\}\big),
\end{equation} where here $\{\cdot,\cdot\}$ denotes the standard Poisson bracket. The first term describes the standard unitary evolution of quantum mechanics, while the second term, known as the Alexandrov bracket, provides a symmetrised version of the Poisson bracket describing both the pure classical evolution and the back-reaction of the quantum system on the classical one. While this contains both unitary and back-reaction terms, in contains neither the decoherence  nor the diffusion needed for the dynamics to be positive, and thus cannot be unravelled into individual classical and quantum trajectories. In fact, while such dynamics may be understood as an $\hbar\rightarrow 0$ limit of a bipartite quantum dynamics, the presence of entanglement means that the dynamics does not describe a genuinely classical subsystem \cite{layton2024classical}.

A second common approach to coupling classical and quantum systems in the trajectories picture is the mean-field approach, also sometimes referred to as the Ehrenfest or semi-classical approach \cite{boucher1988semiclassical,tully1998mixed}. Here, the back-reaction force on the momentum of a classical system is given by the expectation value, with a typical dynamics taking the form
\begin{align} \label{eq: mean-field_dynamics}
    d|\psi\rangle_t&=-\frac{i}{\hbar}H|\psi\rangle_t dt\\
    dq_t&=\frac{p_t}{m}dt\\
    dp_t&=-\langle\psi|\frac{\partial H}{\partial q}|\psi\rangle_t dt \label{eq: mean-field_dynamics_end}
\end{align} While this dynamics preserves the positivity of $\rho_t$ at the level of trajectories, and thus ensures that the corresponding dynamics for $\varrho(z,t)$ preserves positivity, this dynamics fails to be linear on the level of $\varrho(z,t)$ \cite{layton2022healthier}. The same holds for other stochastic modifications of this dynamics \cite{eglinton2024stochastic}, where a noise term is added to the classical degrees of freedom without including the necessary additional correlated stochastic terms in the quantum evolution. Aside from failing to respect a basic requirement of probability theories, which can lead to problems such as superluminal signalling \cite{gisin1989stochastic}, non-linear classical-quantum dynamics provide a more complex and difficult-to-solve alternative to the linear dynamics we have thus far presented.

\section{Entropy production and the second law for classical-quantum dynamics} \label{sec: entropyetc.}

\

In this section we introduce the main concepts relating classical-quantum dynamics to thermodynamics. After introducing notions of classical-quantum thermal states and entropy, we show how if a classical-quantum dynamics preserves the thermal state, and satisfies the basic consistency requirements of complete-positivity and linearity, it necessarily obeys the second law of thermodynamics.

\
\subsection{Classical-quantum entropy, energy and thermal states}
We begin by defining the entropy of a classical-quantum system. The entropy associated to a classical-quantum state $\varrho(z,t)$ can be defined as a hybrid of both the Shannon and von-Neumann entropies \cite{Alonso2020a}, namely
\begin{align}\label{eq:entropy}
   S(\varrho(z,t)):=-\int dz \ \tr{\varrho(z,t) \ \text{ln} \ \varrho(z,t)},
\end{align}
Just as with the standard Shannon or von-Neumann quantum entropy, one may use this to quantify the uncertainty of the classical-quantum state and similarly its information content. Since we deal with continuous variable classical systems, the entropy associated to the classical degrees of freedom is strictly speaking a differential entropy, and thus may be negative \cite{cover2012elements}. Its interpretation as an information measure can be restored through introduction of the classical-quantum \textit{relative entropy}, which is a positive divergence  between the actual state $\varrho(z,t)$ and a reference classical-quantum state $\sigma(z)$. We define relative entropy as
\begin{align}
    S[\varrho(z,t)||\sigma(z)]=\int dz  \tr{\varrho(z,t) \big[\text{ln} \ \varrho(z,t)-\text{ln} \ \sigma(z)\big]}.
\end{align}
This is a divergence measure that acts as a hybrid between the quantum relative entropy and classical Kullback-Leibler divergence. An important property of the classical-quantum relative entropy is that it is monotonic under the action of a completely-positive, trace non-increasing  linear map $\Lambda$ i.e. that
\begin{equation} \label{eq: CQ_DPI}
S[\Lambda(\varrho)|\Lambda(\pi)]\leq S[\varrho||\pi].
\end{equation} Known in quantum information theory as the data processing inequality \cite{sagawa2013second}, the fact that this holds for the classical-quantum states and maps that we have presented here follows from a straightforward embedding of the classical-quantum system into a fully quantum system – see Appendix \ref{app: CQmonotonicity} for further details. 

Alongside entropy, the other key ingredient of a theory of thermodynamics is that of energy. To define a notion of energy in the combined classical-quantum system, we assume the existence of a Hermitian operator valued function of phase-space that we refer to as the classical-quantum Hamiltonian, and denote by $H(z)$. This quantity determines the average energy $E$ in the classical-quantum system by the formula
\begin{equation}
    E=\int dz \tr{H(z)\varrho(z)}
\end{equation} which we write in the notation of \eqref{eq: CQ_expectation_value} as $E=\langle \langle H(z) \rangle \rangle$.
Rather than taking the classical-quantum Hamiltonian to directly determine the form of dynamics in the system, as with the other classical-quantum approaches described in \ref{subsec: other_approaches}, $H(z)$ here simply determines the energetics of the combined classical-quantum system. Assigning energy to both the individual classical and quantum systems, as well as to their interactions, $H(z)$ can be generically decomposed as
\begin{equation}
    H(z)=H^C(z)\mathds{1}+H^Q+H^I(z)
\end{equation} where the classical Hamiltonian $H^C(z)$ is a real-valued function, $\mathds{1}$ is the identity operator on the Hilbert space, the quantum Hamiltonian $H^Q$ is a Hermitian valued operator independent of phase space, and the interaction Hamiltonian $H^I(z)$ is traceless. The eigenbasis of the classical-quantum Hamiltonian as a function of phase space is commonly referred to as the adiabatic basis \cite{tully1990molecular} and will be denoted as
\begin{equation}
    H(z)|n(z)\rangle = \epsilon_n(z)|n(z)\rangle
\end{equation} where $\epsilon_n(z)$ are the corresponding eigenvalues, giving the energy for a given energy level and classical configuration.

The concepts of entropy and energy naturally lead to a notion of a classical-quantum thermal state. If a given classical-quantum system has a known average energy $E$, then applying the maximum entropy principle \cite{Alonso2020a,tronci2025entropy} with the constraint $\langle \langle H(z) \rangle \rangle = E$ one finds the operator valued distribution 
\begin{equation} \label{eq: thermal_state_def}
    \pi(z)=\frac{e^{-\beta H(z)}}{\mathcal{Z}},
\end{equation} which is known as the classical-quantum thermal state \cite{mauri1993canonical,parandekar2005mixed,schmidt2008mixed,amati2023detailed}. Here $\mathcal{Z}$ is given by 
\begin{equation}
    \mathcal{Z}=\int dz \tr{e^{-\beta H(z)}}.
\end{equation} which together with $H(z)=H(z)^\dag$ ensures that $\pi(z)$ defines a valid classical-quantum state, while $\beta$ defines the inverse temperature. It is straightforward to see that when $H(z)$ reduces to $H^C(z)\mathds{1}$ or $H^Q$, the above thermal state definition reduces to standard classical and quantum thermal states.

\subsection{Heat exchange and entropy production}

In thermodynamic systems, an important role is played by the environment, which allows the system to gain or dissipate heat and become more or less ordered. While we shall remain agnostic about the nature of the environment to the classical-quantum system, other than assuming that it can always be assigned a fixed temperature and that it is sufficiently large that the dynamics of the classical-quantum system are well-approximated as Markovian, we will need to define a number of quantities that implicitly rely on the ability of the system to exchange energy and information with its surroundings.

The first such quantity that we shall define is the heat exchanged with the environment. Taking the initial time to be $t_i$, we define the average heat exchanged between then and time $t$ on the level of the ensemble as
\begin{equation}
    \mathcal{Q}(t)= \langle \langle H(z) \rangle \rangle(t) - \langle \langle H(z) \rangle \rangle(t_i).
\end{equation} At the same time, on the level of trajectories, we define the stochastic heat exchanged via the difference in trajectory expectation values of the classical-quantum Hamiltonian
\begin{equation} \label{eq: stochastic_heat}
    Q_t=\langle H(z)\rangle_t - \langle H(z) \rangle_{t_i}.
\end{equation} When the quantum state is pure, this amounts to computing the change in $\langle\psi|H(z)|\psi\rangle$ along a trajectory. These ensemble and trajectory definitions of heat are related by taking the expectation value over trajectories 
\begin{equation} \label{eq: heat_ensemble_and_fluctuations}
    \mathcal{Q}(t)=\mathbb{E}[Q_t],
\end{equation} which follows from \eqref{eq: exp_traj_exp_ensemble}. We thus see that this framework allows one to study both the average transfer of heat with the environment, as well as study the stochastic fluctuations of this quantity.

The second quantity that we shall define is the entropy production in a classical-quantum system. A central quantity in non-equilibrium thermodynamics \cite{landi2021irreversible}, entropy production provides a measure of the irreversibility of a process, and is defined by balancing the rate of change of the system entropy with outgoing heat flux. Using the above definitions of entropy and heat for the classical-quantum system, we may define entropy production $\Sigma$ in the classical-quantum context as
\begin{equation} \label{eq: entropy-production}
    \Sigma(t)= \Delta S(t) - \beta \mathcal{Q}(t)
\end{equation} where here $\Delta S(t)$ denotes the change in the system entropy from time $t_i$ to time $t$.

\subsection{Thermal state preservation and the second law}

We first establish a basic requirement to make of any classical-quantum dynamics that is compatible with thermodynamics. If two systems at the same temperature are put into contact, one expects, on average, that no energy flows between the two. In our case, this means that the state of a classical-quantum system in contact with a thermal environment at inverse temperature $\beta$ should not change if the initial state of the classical-quantum system is a thermal state at the same inverse temperature. Written in terms of a generic classical-quantum generator $\mathcal{L}$ as introduced in \eqref{eq: generic_master_equation}, this leads us to the basic requirement that
\begin{equation} \label{eq: thermo_consistency}
    \mathcal{L}(\pi)=0,
\end{equation} i.e. that the thermal state is preserved in time by the dynamics. A typical assumption in both classical and quantum non-equilibrium thermodynamics \cite{seifert2012stochastic,breuer2002theory}, and sometimes discussed in the classical-quantum context \cite{Kapral2006,alonso2020entropy,alonso2021computation}, dynamics satisfying Eq. \eqref{eq: thermo_consistency} represent a subset of generic classical-quantum dynamics which have a well-defined fixed point.

The preservation of the thermal state has an important consequence for classical-quantum dynamics that also satisfy the basic properties of complete-positivity and linearity. We start by noting that it is well known from open quantum system theory any linear, completely positive dynamics with a fixed point is sufficient to guarantee a non-negative entropy production rate consistent with the second law of thermodynamics \cite{spohn1978entropy,lindblad2001non,alicki2007quantum,sagawa2013second}. To see that the same holds in the classical-quantum case, we first rewrite the entropy production rate in terms of the classical-quantum relative entropy as
\begin{align}\label{eq:entprod}
    \dot{\Sigma}(t)=-\frac{\partial}{\partial t}S[\varrho(z,t)||\pi(z)],
\end{align} which follows from the definitions of the classical-quantum thermal state $\pi$ and heat transfer $\mathcal{Q}$. 

Using the fact that the dynamics satisfies $\mathcal{L}(\pi)=0$, this may be rewritten as an infinitesimal change in relative entropy between the non-equilibrium state state $\varrho(z,t)$ and the thermal state
\begin{align}
    \dot{\Sigma}(t)&=\lim_{\delta t\to 0} \frac{S[\varrho(z,t)||\pi(z)]-S[e^{\delta t \mathcal{L}} \varrho(z,t)||e^{\delta t \mathcal{L}} \pi(z)]}{\delta t}.
\end{align}
Provided the generator $\mathcal{L}$ is completely-positive and linear, the right hand side is therefore necessarily positive by the data-processing inequality \eqref{eq: CQ_DPI}, and thus we see that the classical-quantum entropy production rate is necessarily non-negative 
\begin{equation}
    \dot{\Sigma}(t)\geq 0.
\end{equation} This provides a general formulation of the second law in a classical-quantum system, with the entropy production rate $\dot{\Sigma}(t)$ quantifying the irreversibility of the dynamics. Integrating this over time, and comparing to Eq. \eqref{eq: entropy-production}, we can rewrite this as a Clausius inequality 
\begin{equation} \label{eq: clausius_inequality_2nd_law}
    \Delta S(t)\geq\beta\mathcal{Q}(t),
\end{equation} recovering the standard  formulation of the second law as bounding the change in the system entropy by the heat transfers into an external environment.

We thus see that classical-quantum dynamics that are simultaneously linear, completely-positive and preserve the thermal state are necessarily compatible with the second law of thermodynamics – we shall refer to such dynamics as \textit{thermodynamically stable}. 
It is important to emphasise that the same argument for obeying the second law cannot be made for dynamics failing the basic requirements of complete-positivity or linearity. In the case of the quantum-classical Liouville equation \eqref{eq: CQ_liouville}, since initially positive states can evolve to negative states, the relative entropy will not increase monotonically under the evolution, and indeed will generically not be well-defined. The lack of positivity also precludes any statistical interpretation of entropy production and heat at the stochastic level, and one must resort to using quasi-probabilities in stochastic thermodynamics \cite{deffner2013quantum,santos2017wigner}. Similarly, for the mean-field dynamics of Eqs. \eqref{eq: mean-field_dynamics} to \eqref{eq: mean-field_dynamics_end}, the failure of the evolution to generate a linear map on the initial classical-quantum state means that it fails a basic assumption needed to apply the data processing inequality. Furthermore, the non-linearity at the level of the unravelling leads to non-linear evolution of the quantum state, known to be in violation of the second law of thermodynamics \cite{peres1989nonlinear}. We thus see that complete-positivity and linearity are natural assumptions to make on classical-quantum dynamics, purely on thermodynamic grounds.

\section{Thermal state preserving classical-quantum dynamics}
\label{sec: construct}

\

In this section we introduce two general classes of completely-positive and linear classical-quantum dynamics that preserve the classical-quantum thermal state. We show that these dynamics can be understood as extensions of the standard overdamped and underdamped dynamics, i.e. the Smoluchowski and Klein–Kramers equations for Brownian motion \cite{risken1996fokker}, that can now include additional quantum degrees of freedom. In the high temperature limit, the dynamics takes the form of a completely-positive completion of the standard forms of coupling between classical and quantum systems. 

\subsection{Overview of the problem}

In the previous section, we saw that if a completely-positive and linear classical-quantum dynamics preserves the thermal state in time, the system has a well-defined second law of thermodynamics. Since the general form of classical-quantum dynamics that is Markovian and continuous is known to take the form of Eq. \eqref{eq: L_general}, the problem amounts to finding matrices $D$ and operators $L_\alpha$ and $\bar{H}$ such that $\mathcal{L}(\pi)=0$. For such dynamics to be useful, it must be applicable to arbitrary $\pi(z)$ i.e. arbitrary classical-quantum Hamiltonians $H(z)$, rather than those satisfying special properties e.g. being simultaneously diagonaliseable everywhere in phase space in a fixed basis.

In contrast to the purely classical or quantum cases, there are a number of features that make even finding an example of such a dynamics extremely challenging. Firstly, the generator $\mathcal{L}$ includes both classical and quantum dissipative processes as well as coupling between the two systems via the unitary and back-reaction parts of the dynamics – each of these terms or combinations of them must vanish when applied to $\pi$ in order to preserve the thermal state. Secondly, the fact that the thermal state is an operator-valued function of phase space means that the diffusive and back-reaction parts of the dynamics involve derivatives of $\pi(z)$ that do not in general commute with $\pi(z)$ itself. This means that the operators $L_\alpha$ and $\bar{H}$ must necessarily be dependent on phase space and in general non-diagonaliseable in the same basis as $\pi(z)$. Finally, any dynamics must also simultaneously satisfy the two non-trivial positivity constraints \eqref{eq: dec-diff} and \eqref{eq: pos_2}.

Surprisingly, we demonstrate in this section that such a dynamics may indeed be found. Moreover, the dynamics satisfies a number of desirable properties, such as reducing to the correct classical limit, preserving the purity of quantum states conditioned on classical trajectories, and recovering a consistent form of standard approaches to classical-quantum coupling in the high temperature $\beta \rightarrow 0$ limit. The key feature that allows us to construct such dynamics is the identification of two kinds of operators defined in terms of the thermal state $\pi$, which ultimately will be seen in Sec. \ref{sec: DB} to be related to the general conditions for a classical-quantum dynamics to satisfy detailed balance.

\subsection{L and M operators}

To construct classical-quantum dynamics that preserve the classical-quantum thermal state, i.e. $\mathcal{L}(\pi)=0$, we first introduce two classes of operators. The first of these is a phase-space dependent operator, defined for each classical coordinate $z$ as
\begin{equation} \label{eq: L_op_def}
    L_{z}=-\frac{2}{\beta}\frac{\partial \pi^\frac{1}{2}}{\partial z}\pi^{-\frac{1}{2}}.
\end{equation} These operators determine both the back-reaction and decoherence in the dynamics we consider. Although these operators are not in general Hermitian, they each satisfy the important property
\begin{equation} \label{eq: L_op_property}
    L_{z}\pi^{\frac{1}{2}}=\pi^{\frac{1}{2}}L_{z}^{\dag},
\end{equation} which will be frequently used to verify thermodynamic properties of the resulting dynamics. 

The second important class of operators to introduce are phase-space dependent operators defined for each ordered pair of classical coordinates (x,y) as
\begin{equation} \label{eq: M_op_def1}
    M_{xy}=\frac{i\hbar}{2}\int_0^\infty  e^{-s\pi^\frac{1}{2}} [L_x^{\dag}L_y,\pi^\frac{1}{2}] e^{-s\pi^\frac{1}{2}} ds.
\end{equation} This operator is Hermitian when $x=y$, and controls part of the unitary dynamics of the quantum system. Since $M_{xy}$ takes the form of a solution to a Lyapunov equation, it is equivalently defined by the equation
\begin{equation} \label{eq: M_op_def2}
    M_{xy}\pi^\frac{1}{2} + \pi^\frac{1}{2} M_{xy} = \frac{i\hbar}{2}[L_x^{\dag}L_y,\pi^\frac{1}{2}].
\end{equation} As with \eqref{eq: L_op_property}, this implicit definition of $M_{xy}$ will useful to prove properties of these dynamics. 

While both $L_z$ and $M_{xy}$ are defined in terms of the square root of $\pi$, there are two useful relations that relate these operators to $\pi$ itself. The first of these is 
\begin{equation} \label{eq: L_pi_identity}
    L_z \pi + \pi L_z^\dag=-\frac{2}{\beta}\frac{\partial \pi}{\partial z},
\end{equation} follows directly from the definition \eqref{eq: L_op_def}  while the second 
\begin{equation} \label{eq: M_pi_identity}
    -\frac{i}{\hbar}[M_{xy},\pi]=-L_x \pi L_y^\dag +\frac{1}{2}\{L_x^\dag L_y,\pi\}_+,
\end{equation} follows from \eqref{eq: L_op_property} and \eqref{eq: M_op_def2}. These two relations are important for proving that the dynamics that we construct satisfies $\mathcal{L}(\pi)=0$.

Finally, it is important to recognise a particular limiting form of these operators. To see how these arise, we first note that we may use the definition of the derivative of the exponential map \cite{hall2013lie} to rewrite $L_z$ as
\begin{equation} \label{eq: L_series}
    L_z=\frac{e^{\ad{\frac{-\beta H}{2}}}-1}{\ad{\frac{-\beta H}{2}}} \bigg(\frac{\partial H}{\partial z}\bigg)
\end{equation} where $\ad{A}B = [A,B]$ and the above fraction is interpreted as the series
\begin{equation} \label{eq: frac_to_series}
    \frac{e^{\ad{\frac{-\beta H}{2}}}-1}{\ad{\frac{-\beta H}{2}}}=\sum_{n=0}^\infty \frac{1}{(n+1)!}\ad{\frac{-\beta H}{2}}^n.
\end{equation} For sufficiently simple commutation relations, this series may be computed explicitly even when the series does not truncate. However, it is useful to note that in two cases, the series truncates to zeroth order. The first case occurs for a special class of Hamiltonians that satsify the property that $H(z)$ and $H(z^\prime)$ commute for all $z$, $z^\prime \in\mathcal{M}$, which we refer to as \textit{self-commuting}. The second case occurs in the high temperature $\beta\rightarrow 0$ limit of the dynamics for arbitrary $H(z)$. In both cases, all of the terms in the series with $n>0$ vanish, with $L_z$ reducing to $\partial_z H$. For our other class of operators,  $M_{xy}$, we note that \eqref{eq: L_op_property} implies that the right-hand side of \eqref{eq: M_op_def2} vanishes when $L_x$ and $L_y$ are Hermitian. Since the above form of $L_z$ is Hermitian for any $z$, it must also be the case that $M_{xy}=0$. In summary, we thus arrive at 
\begin{align} \label{eq: high_temp_limits}
L_z=\frac{\partial H}{\partial z}\quad \quad \quad \quad \quad \quad M_{xy}= 0 \\[5pt]
\nonumber\text{if}\ H\ \text{is}\   \text{self-commuting}\ 
\text{or}\   \beta\rightarrow 0.
\end{align}These limiting forms of the $L_z$ and $M_{xy}$ operators are useful for studying dynamics of simple models, such as that given in Section \ref{sec: analytic_model}, as well as studying the classical and high-temperature limits of the dynamics we will present.

\subsection{Overdamped dynamics}

The first class of dynamics we introduce is an overdamped dynamics. Taking a single one-dimensional classical degree of freedom $x$, with mobility given by $\mu$, we describe its interactions with a quantum system either via the following master equation
\begin{equation} \label{eq: overdamped}
\begin{split}
\frac{\partial \varrho}{\partial t}=&-\frac{i}{\hbar} [H + \frac{\mu \beta}{8}M_{xx},\varrho ]+\frac{\mu}{2}\frac{\partial}{\partial x}(L_x \varrho + \varrho L_x^\dag) \\
&+ \frac{\mu}{\beta} \frac{\partial^2 \varrho }{\partial x^2} 
+\frac{\mu \beta }{8}(L_x \varrho L_x^\dag -\frac{1}{2}\{L_x^\dag L_x, \varrho  \}_+),
\end{split}
\end{equation}
or via a stochastic unravelling as
\begin{equation} \label{eq: overdamped_class}
    dx_t=-\frac{\mu}{2} \langle L_x + L_x^{\dag}\rangle dt + \sqrt{\frac{2\mu}{\beta}} dW_t
\end{equation}
\begin{equation}\label{eq: overdamped_quantum}
    \begin{split}
        d\rho_t=&-\frac{i}{\hbar}[H+\frac{\mu\beta }{8}M_{xx},\rho_t]dt\\
        &+ \frac{\mu \beta }{8} (L_x \rho_t L_x^{\dag} - \frac{1}{2}\{ L_x^{\dag}L_x,\rho_t\}_+)dt\\
        &-\sqrt{\frac{ \mu \beta}{8}}(L_x \rho_t + \rho_t L_x^{\dag} - \langle L_x +L_x^{\dag} \rangle \rho_t)dW_t,
    \end{split}
\end{equation}  where here $dW_t$ defines the increment of a one dimensional Wiener process. This dynamics describes how an overdamped classical system subject to thermal noise is affected by the back-reaction from a quantum system, as well as how this interaction leads to noise in the otherwise unitary evolution of the quantum system. It provides a \textit{minimal} model of such an interaction – all of the noise on the classical system is assumed thermal, while all of the noise on the quantum system arises from the interaction with the classical system, rather than any due to direct coupling to the thermal environment.

While the above dynamics is ultimately postulated, it is straightforward to check that it satisfies a number of desirable properties. Firstly, the dynamics is completely-positive and linear at the level of the master equation. Secondly, the dynamics is thermodynamically stable i.e. the thermal state is preserved in time. Additionally, the dynamics saturates the decoherence-diffusion trade-off, meaning that the quantum state of the system remains pure conditioned on the classical trajectory. Finally, the model correctly reproduces the standard overdamped classical dynamics in the classical limit.

To see how these properties arise, we first compare the form of \eqref{eq: overdamped} to the general form of completely-positive generator given in \eqref{eq: L_general}. Doing so, we see that it is of the same form, guaranteeing that the dynamics is norm-preserving and linear, with parameters given
\begin{equation} \label{eq: overdamped_params}
\begin{split}
    \bar{H}=\frac{H}{\hbar}+\frac{\mu\beta}{8\hbar}M_{xx} \quad \quad \quad \quad L=L_x\\
    D_0=\frac{\mu\beta}{8}\quad\quad D_1=-\frac{\mu}{2} \quad\quad D_2=\frac{2\mu}{\beta}.
\end{split}
\end{equation} In order for the dynamics to be completely-positive, the two positivity constraints \eqref{eq: dec-diff} and \eqref{eq: pos_2} must also be satisfied. The second of these trivially holds since $D_2$ has an inverse, and multiplying the scalar $D$ coefficients we see that \eqref{eq: dec-diff} also holds. Since here $D_0=D_1^2/D_2$, we see that the dynamics saturates the decoherence-diffusion trade-off, meaning that the dynamics has minimal decoherence and keeps intially pure quantum states pure \cite{layton2022healthier}.

It is also straightforward to see that this dynamics preserves the thermal state i.e. satisfies $\mathcal{L}(\pi)=0$. To do so, one must evaluate the right hand side of \eqref{eq: overdamped} with $\varrho=\pi$
and check that the result is zero. Doing so, one sees using \eqref{eq: L_pi_identity} that the back-reaction and diffusion terms cancel, and using \eqref{eq: M_pi_identity} that the $M_{xx}$ unitary term cancels with the decoherence term. Since the rest of the unitary term vanishes, due to $H$ commuting with $\pi$, we see that indeed $\mathcal{L}(\pi)=0$ for this dynamics.

To see that this dynamics reduces to the standard dynamics in the classical limit, we consider the case where the classical-quantum Hamiltonian is proportional to the identity operator, $H(x)=H^C(x)\mathds{1}$. Since $H(z)$ here is self-commuting, we may use the simplified forms of $L_x$ and $M_{xx}$ given in \eqref{eq: high_temp_limits}. Substituting these into the above dynamics, and using the fact that the operator $H$ is proportional to the identity operator, we find that the above dynamics reduces to
\begin{equation} 
\begin{split}
\frac{\partial \varrho}{\partial t}=\mu\frac{\partial}{\partial x}(\frac{\partial H}{\partial x }\varrho) + \frac{\mu}{\beta} \frac{\partial^2 \varrho }{\partial x^2} 
\end{split}
\end{equation} in the master equation picture or
\begin{equation}
    dx_t=-\mu\frac{\partial H}{\partial x} dt +  \sqrt{\frac{2\mu}{\beta}} dW_t
\end{equation} in the unravelling picture. We thus see that our dynamics reduces to that of a single overdamped classical system in a potential, with a diffusion coefficient that satisfies the Einstein relation  i.e. that described by the standard Langevin/Smoluchowski equations \cite{risken1996fokker}.

Having established the basic properties of this dynamics, we emphasise that it provides a minimal model that can be easily adapted to more general settings. Firstly, 
in certain settings  it is natural for the quantum system to be directly coupled to the thermal environment, such that it experiences direct thermal dissipation \cite{annby2022quantum,eglinton2024stochastic,prech2025quantum}. To incorporate these effects within a Markovian approximation, one may include a dissipation term $\mathcal{D}(\varrho)$ in the master equation \eqref{eq: overdamped} and $\mathcal{D}(\rho_t)$ in the quantum part of the unravelling \eqref{eq: overdamped_quantum}, where $\mathcal{D}(\cdot)=\tilde{L}_\alpha \cdot \tilde{L}_\alpha^\dag - \frac{1}{2}\{\tilde{L}_\alpha^\dag \tilde{L}_\alpha,\cdot\}_+$ is a superoperator that captures the system-bath coupling via Lindblad operators $\tilde{L}_\alpha$. In Sec. \ref{sec: DB} we provide constraints that must be satisfied by this additional dissipation in order for the thermal state of the combined system to be preserved by the dynamics and for the combined system to satisfy detailed balance.  Secondly, it is also easy to adapt this dynamics to describe multiple overdamped classical degrees of freedom. This is shown in Appendix \ref{app: generalised_dynamics}, where we use the $L_z$ and $M_{xy}$ operators to construct a dynamics that saturates the decoherence-diffusion trade-off for $n$ overdamped particles, as well as allowing for $x$-dependent correlations in the noise. 

As a final remark, we note a basic inequality implied by our set-up. In particular, by applying the decoherence-diffusion trade-off \eqref{eq: dec-diff} with the parameters of the dynamics given in \eqref{eq: overdamped_params}, we find that the decoherence $D_0$ in the $L_x$ basis is constrained by
\begin{equation} \label{eq: overdamped_D0_bound}
    D_0\geq\frac{\mu\beta}{8}.
\end{equation} This provides a \textit{lower bound} on the amount of decoherence in a quantum system interacting with an overdamped effectively classical system, that arises by assuming that (1) the Einstein relation holds and that (2) the dynamics are completely-positive. The first condition only holds in regimes in which the classical system is subject to purely thermal noise, and thus will fail to hold for certain mesoscopic systems at low temperatures when shot noise dominates \cite{blanter2000shot,kobayashi2021shot,magrini2021real}. The second condition only holds while the degree of freedom $x$ behaves effectively classically, which breaks down when the decoherence on this subsystem is not sufficiently strong \cite{layton2024classical}. This inequality thus provides a natural regime of validity of the current theory. The only exception to this is when $\beta$ may be interpreted as an effective temperature, as discussed in Sec. \ref{sec: analytic_model}.

\subsection{Underdamped dynamics}

The second class of dynamics we will introduce is an underdamped dynamics. Taking the position of the classical degree of freedom to be $q$ and its conjugate momentum $p$, we make the standard assumption that the only dependence of the Hamiltonian on the classical momentum $p$ is a classical kinetic term $(p^2/2m) \mathds{1}$. Choosing $\gamma$ to denote the friction coefficient, the dynamics takes the form
\begin{equation} \label{eq: underdamped}
\begin{split}
\frac{\partial \varrho}{\partial t}=&-\frac{i}{\hbar} [H + \frac{ \beta}{8 \gamma}M_{qq},\varrho ]\\
&+\frac{1}{2}\frac{\partial}{\partial p}(L_q \varrho + \varrho L_q^\dag) - \frac{p}{m}\frac{\partial \varrho}{\partial q} \\
&+\gamma\frac{\partial}{\partial p}(\frac{ p}{m}\varrho)+ \frac{\gamma}{\beta} \frac{\partial^2 \varrho }{\partial p^2} \\
&+\frac{\beta }{8\gamma}(L_q \varrho L_q^\dag -\frac{1}{2}\{L_q^\dag L_q, \varrho  \}_+),
\end{split}
\end{equation} which for initially pure states may also be unravelled as
\begin{equation} \label{eq: underdamped_unravelling_q}
    dq_t=\frac{p_t}{m}dt 
\end{equation}
\begin{equation} \label{eq: underdamped_unravelling_p}
    dp_t=-\frac{1}{2}\langle L_q + L_q^\dag \rangle dt - \frac{\gamma}{m}p dt + \sqrt{\frac{2 \gamma}{\beta}} dW_t
\end{equation}
\begin{equation} \label{eq: underdamped_unravelling_psi}
\begin{split}
    d|\psi\rangle_t=&-\frac{i}{\hbar} (H+\frac{\beta}{8\gamma}M_{qq})|\psi\rangle dt\\
    &-\frac{\beta}{16\gamma} (L_q^\dag L_q - 2 \langle L_q^\dag \rangle L_q + \langle L_q^\dag \rangle \langle L_q\rangle ) |\psi \rangle dt \\
    & -\sqrt{\frac{\beta}{8\gamma}}(L_q - \langle L_q \rangle)|\psi\rangle dW_t.
\end{split}
\end{equation} The above describe in the ensemble and trajectory pictures how a classical particle subject to thermal noise and friction responds to a quantum potential, as well as how the quantum system sourcing this potential is affected by the decoherence that arises from this interaction.

As in the overdamped case, this underdamped dynamics satisfies a natural set of properties: (1) complete-positivity and linearity; (2) preserves the classical-quantum thermal state; (3) preserves pure quantum states when conditioned on the classical trajectory; and (4) recovers the correct classical limit. 

Looking first at the properties of complete-positivity and pure-state preservation, we compare the master equation dynamics to \eqref{eq: L_general} to find that the dynamics is characterised by 
\begin{equation}
\begin{split}
    \bar{H}=\frac{H}{\hbar} + \frac{\beta}{8\gamma} M_{xx}\quad \quad L=L_q\quad \quad
    D_1^C=\frac{p}{m}\begin{pmatrix}
        1\\
        -\gamma 
        \end{pmatrix}\\[4pt]
         D_0=\frac{\beta}{8\gamma}\quad \quad
    D_1=\frac{1}{2}\begin{pmatrix}
        0 \\
        -1
    \end{pmatrix} \quad \quad D_2=\begin{pmatrix}
        0 & 0\\
        0 &  2\gamma/\beta
        \end{pmatrix}.
\end{split}
\end{equation} Computing the pseudoinverse of $D_2$ and multiplying the $D$ matrices, we see that the dynamics satisfies  \eqref{eq: dec-diff}, \eqref{eq: pos_2} and \eqref{eq: saturating_dec_diff}, ensuring that the dynamics both preserves the positvity of the classical-quantum state, and the purity of any initial state $\rho_t$ that starts off in a pure state. This latter property ensures that the unravelling given in Eqs. \eqref{eq: underdamped_unravelling_q} to \eqref{eq: underdamped_unravelling_psi} indeed is equivalent to the master equation \eqref{eq: overdamped}.

To see that the dynamics preserves the thermal state, we again replace $\varrho$ with $\pi$ on the right hand side of the master equation and check that all the terms cancel. In particular, we see here that the combination of the first and fourth lines of \eqref{eq: underdamped} vanish due to \eqref{eq: M_pi_identity} and $[H,\pi]=0$, while the second and third lines each vanish independently due to the inclusion of $(p^2/2m) \mathds{1}$ in $H$ and the identity \eqref{eq: L_pi_identity}, ensuring that the dynamics preserves arbitrary thermal states $\pi$. 

Finally, we check that the dynamics correctly reduces in the classical limit to the standard underdamped dynamics. Taking again the simplified forms appearing in \eqref{eq: high_temp_limits} and taking $H$ proportional to the identity, we find the dynamics reduce in the master equation and unravelling pictures to
\begin{equation}
    \frac{\partial \varrho}{\partial t}=\{H,\varrho\}+\gamma\frac{\partial}{\partial p}(\frac{ p}{m}\varrho)+ \frac{\gamma}{\beta} \frac{\partial^2 \varrho }{\partial p^2}
\end{equation} and 
\begin{equation} 
    dq_t=\frac{p_t}{m}dt 
\end{equation}
\begin{equation} 
    dp_t=- \frac{\partial H}{\partial q}  dt - \frac{\gamma}{m}p_t dt + \sqrt{\frac{2 \gamma}{\beta}} dW_t
\end{equation} as expected, describing the standard underdamped dynamics of a diffusing particle satisfying the Einstein relation, i.e. the Klein-Kramers equation \cite{risken1996fokker}.

As in the case of the overdamped dynamics, this dynamics may be generalised to include a range of additional phenomena not captured in the above model. Firstly, it straightforward to generalise the above dynamics to multiple particles and dimensions, by including an $L_q$ operator for each degree of freedom and direction and replacing each $q$ and $p$ with a sum over $q_i$ and $p_i$. Secondly, in the above, we assume that $H(q,p)$ contains no coupling between the momentum $p$ and the quantum degrees of freedom, which guarantees that the only necessary noise in the system is in $p$ directly. However, one may also write down models with noise in $q$ and $p$ such that arbitrary Hamiltonians $H(q,p)$ may be considered, and we provide such an example in Appendix \ref{app: generalised_dynamics}.  Finally, we note that as in the overdamped case, one may include excess decoherence in the above model. Aside from using the general formalism given in Section \ref{sec: DB}, it is simple to see that one may include excess decoherence in the $L_q$ basis by replacing the $\beta/(8\gamma)$ coefficient of $M_{qq}$ and the quantum disspator with a generic $D_0$. In this case, we see that the decoherence rate in the $L_q$ basis must necessarily obey
\begin{equation} \label{eq: underdamped_D0_bound}
    D_0\geq \frac{\beta}{8\gamma},
\end{equation} which as in the overdamped case, provides a lower bound on the decoherence of a quantum system interacting with a classical system that is subject to purely thermal noise and obeys the Einstein relation.

\subsection{The high temperature limit: recovering consistent mean-field and quantum-classical Liouville equations }\label{subsec: back_reaction_limit}

Our dynamics can also be understood as providing a consistent version of the standard approaches to coupling classical and quantum systems described in Section \ref{subsec: other_approaches} that is (1) completely-positive and linear and (2) preserves the classical-quantum thermal state even at low temperatures.

To see this, we first consider the high temperature, $\beta \rightarrow 0$, limit of the underdamped dynamics in the master equation representation. One may do so using the limiting forms of the operators $L_z$ and $M_{xy}$ provided in \eqref{eq: high_temp_limits}, which we can substitute into our dynamics to find our dynamics to lowest order in $\beta$ as
\begin{equation} \label{eq: underdampedMEhightemp}
\begin{split}
\frac{\partial \varrho}{\partial t}=&-\frac{i}{\hbar}[H,\varrho]+\frac{1}{2}\big(\big\{H,\varrho\big\}-\big\{\varrho,H\big\}\big) \\
&+\gamma\frac{\partial}{\partial p}\big(\frac{ p}{m}\varrho\big)+ \frac{\gamma}{\beta} \frac{\partial^2 \varrho }{\partial p^2} \\
&+\frac{\beta }{8\gamma}\bigg(\frac{\partial H}{\partial q} \varrho \frac{\partial H}{\partial q} -\frac{1}{2}\{\frac{\partial H}{\partial q}^2 , \varrho  \}_+\bigg),
\end{split}
\end{equation} where we have rewritten two of the drift terms using the Poisson bracket. It is straightforward to see that the top line exactly coincides with the quantum-classical Liouville equation, given previously in \eqref{eq: CQ_liouville}. Since the completely-positivity of the dynamics is unchanged by the limit of the operators, we can understand the additional diffusion and decoherence terms as providing the minimal additional decoherence and diffusion required to supplement the quantum-classical Liouville equation to be completely-positive, as first noted was possible in \cite{diosi1995quantum}. However, while the dynamics of \eqref{eq: underdampedMEhightemp} satisfies complete-positivty and linearity, the thermal state $\pi$ will only be preserved approximately at high temperatures. The full underdamped dynamics given in \eqref{eq: underdamped} may therefore be understood as a completely-positive generalisation of the classical-quantum Liouville equation, that \textit{additionally} satisfies the important requirement of preserving the classical-quantum thermal state $\pi$ for arbitrary inverse temperature $\beta$.

Moving now to the trajectory representation, we find that a similar conclusion may be found in the $\beta \rightarrow 0$ limit of the stochastic unravelling of the underdamped dynamics. Using again the limiting forms of operators given in \eqref{eq: high_temp_limits}, we find that the unravelling equations \eqref{eq: underdamped_unravelling_q} to \eqref{eq: underdamped_unravelling_psi} take the form
\begin{equation} \label{eq: underdamped_unravelling_qhightemp}
    dq=\frac{p}{m}dt 
\end{equation}
\begin{equation} \label{eq: underdamped_unravelling_phightemp}
    dp=-\langle \frac{\partial H}{\partial q} \rangle dt - \frac{\gamma}{m}p dt + \sqrt{\frac{2 \gamma}{\beta}} dW_t
\end{equation}
\begin{equation} \label{eq: underdamped_unravelling_psihightemp}
\begin{split}
    d|\psi\rangle_t=&-\frac{i}{\hbar} H|\psi\rangle dt\\
    &-\frac{\beta}{16\gamma} \bigg(\frac{\partial H}{\partial q}- \langle\frac{\partial H}{\partial q} \rangle \bigg)^2 |\psi \rangle dt \\
    & -\sqrt{\frac{\beta}{8\gamma}}(\frac{\partial H}{\partial q} - \langle \frac{\partial H}{\partial q} \rangle)|\psi\rangle dW_t.
\end{split}
\end{equation} Comparing to \eqref{eq: mean-field_dynamics}, we see that this dynamics takes the form of the mean field dynamics, with additional temperature and friction dependent terms in the classical momentum and quantum state evolution. Previously written down as a ``healed version" of the mean field equations \cite{diosi1998coupling,layton2022healthier}, we note again that since the limit occurs at the level of the operators, this dynamics is necessarily still linear at the level of the classical-quantum state. However, as an unravelling of \eqref{eq: underdampedMEhightemp}, it will only preserve the thermal state approximately at sufficiently high temperatures. We thus can understand the full unravelling dynamics \eqref{eq: underdamped_unravelling_q} to \eqref{eq: underdamped_unravelling_psi} as a generalisation of the mean-field equations that both satisfies linearity  and preserves the thermal state of the combined classical-quantum system.

\section{Model I} \label{sec: analytic_model}

\

In this section we introduce an analytically solveable toy model consisting of an overdamped classical degree of freedom that interacts with a quantum two-level system. To distinguish operators from real functions we introduce hats $H(x)\mapsto \hat{H}(x)$ for clarity of notation in this section (and also Section~\ref{sec: oscillator_model}). 

\subsection{Set-up}

In what follows, we consider a one-dimensional classical degree of freedom $x$, coupled to a quantum two-level system. Here $x$ may be interpreted as a mechanical or circuit degree of freedom, such as a position or voltage, that behaves effectively classically, but in a sufficiently mesoscopic regime to experience both thermal noise and quantum back-reaction. The thermal noise is controlled by the inverse temperature $\beta$, while response of the classical system to external forces is characterised by the mobility $\mu$.

To couple this effective classical degree of freedom to a quantum two-level system, we assume the energetics of the joint system are described by a classical-quantum Hamiltonian of the form
\begin{equation} \label{eq: H_overdamped}
    \hat{H}(x)=\lambda (x\hat{\mathds{1}}-l\hat{\sigma}_z)^2,
\end{equation} where here $\hat{\sigma}_z$ is the standard Pauli spin-$Z$ operator. We may understand this Hamiltonian as a classical quadratic potential that depends on whether the quantum system is in the $|0\rangle$ or $|1\rangle$ state. The strength of the potential is controlled by the parameter $\lambda$, while the minimum of the potential is either $l$ or $-l$ depending on whether the quantum state is $|0\rangle$ or $|1\rangle$ respectively. The dynamics governing this system are then assumed to be given by the overdamped evolution described by either the master equation \eqref{eq: overdamped} or the equivalent stochastic unravelling of \eqref{eq: overdamped_class} and \eqref{eq: overdamped_quantum}. 

This toy model is naturally interpreted as a direct interaction of a mesoscopic classical system with a quantum degree of freedom, that may model measurement or control in specific parameter regimes. However, it can also be understood as describing a specific macroscopic continuous measurement and feedback procedure. In Appendix \ref{app: measfeedbackmodel1} we show explicitly how the master equation for this model coincides with a special case of the recently proposed ``quantum Fokker-Planck master equation" for a quantum system measured by a detector with measurement strength $k$ and finite bandwidth $g$ \cite{annby2022quantum}. In this case, despite having no intrinisic source of thermal noise or dissipation, the macroscopic measurement and feedback system can be thought of having an effective temperature given by $\beta=8 k/g s$, where $s$ is the feedback strength, which allows for a consistent thermodynamic treatment of the combined system.

The thermal state of this model is given by $\hat{\pi}(x)=\mathcal{Z}^{-1} e^{-\beta \hat{H}(x)}$, where the partition function is
\begin{equation}\mathcal{Z}=\int^\infty_{-\infty} dx \ \tr{e^{-\beta \hat{H}(x)}}=2\sqrt{\frac{\pi}{\beta \lambda}},
\end{equation} which ensures that $\hat{\pi}(x)$ is normalised. Although this state does not explicitly appear in the following analysis, its existence as a fixed point of the dynamics guarantees that the dynamics of the joint classical-quantum system obey the second law.

\color{black}

\subsection{Analytic solution}
Since the Hamiltonian we study in this case takes a simple form, it turns out that the dynamics may be analytically solved in the master equation representation of Eq. \eqref{eq: overdamped}. Our first step is to compute the $\hat{L}_x$ and $\hat{M}_{xx}$ operators for this model. Since the classical-quantum Hamiltonian $\hat{H}(x)$ is self-commuting, i.e. $[\hat{H}(x),\hat{H}(x')]=0$ for all $x,x'$, we may use the limiting forms provided in \eqref{eq: high_temp_limits}. Here, the $\hat{L}_x$  operator may be computed by simply taking the derivative of $\hat{H}(x)$ with respect to $x$, while the $\hat{M}_{xx}$ vanishes. The dynamics of this model is thus summarised by 
\begin{equation}
    \hat{L}_x=2\lambda( x\hat{\mathds{1}}- l \hat{\sigma}_z), \quad \quad \hat{M}_{xx}=0.
\end{equation} Plugging these operator definitions into \eqref{eq: overdamped} and expanding the classical-quantum state in the eigenbasis of $\hat{\sigma}_z$ we obtain three independent equations, 
\begin{align}
    \nonumber\frac{\partial \varrho_{00}}{\partial t}&=\mu\frac{\partial}{\partial x}(2\lambda (x-l) \varrho_{00})+\frac{\mu }{\beta}\frac{\partial^2 \varrho_{00}}{\partial x^2}\\
    \nonumber\frac{\partial \varrho_{01}}{\partial t}&=\mu\frac{\partial}{\partial x}(2\lambda x \varrho_{01})+\frac{\mu }{\beta}\frac{\partial^2 \varrho_{01}}{\partial x^2} + (\frac{i}{\hbar}4\lambda x l  - \mu  \lambda^2 l^2 \beta)\varrho_{01} \\
     \frac{\partial \varrho_{11}}{\partial t}&=\mu\frac{\partial}{\partial x}(2\lambda (x+l) \varrho_{11})+\frac{\mu }{\beta}\frac{\partial^2 \varrho_{11}}{\partial x^2} 
\end{align} where we leave out the dynamics of $\varrho_{10}$ since the solution is given by $\varrho_{10}(x,t)=\varrho_{01}^*(x,t)$. From the above dynamics, we can see explicitly that each component of the classical-quantum state evolves under different dynamics. The $\varrho_{00}$ component, corresponding to the positive eigenvalue of $\hat{\sigma}_z$, experiences diffusion with a restoring force to the point $x=l$, while the $\varrho_{11}$ component experiences diffusion instead with a restoring force to the point $x=-l$. The component corresponding to coherence, given by $\varrho_{01}$, experiences diffusion with a restoring force given by the average value of the two i.e. to the point $x=0$. At the same time, the coherence simultaneously picks up a complex phase and is damped by a term corresponding to the decoherence in the system, with larger damping as any of $\mu$, $\beta$, $\lambda$ or $l$ increase. 

Assuming that initial state of the quantum system is known to be in the state $\hat{\rho}_0$ with components $\rho_{00}, \rho_{01}, \rho_{10}$ and $\rho_{11}$, and that the classical system starts at the point $x_0$, the combined classical-quantum state at $t=0$ is given $\hat{\varrho}(x,0)=\hat{\rho}_0 \delta(x-x_0)$. It is straightforward to check that with this initial condition, the above set of equations have an analytic solution of the form
\begin{widetext}
\begin{equation} \label{eq: analytic_sol_varrho00}
    \varrho_{00}(x,t)= \rho_{00}\sqrt{\frac{\beta \lambda}{\pi(1-e^{-4 \mu \lambda t})}}\exp{ \big[-\frac{\beta \lambda(x- l(1- e^{-2 \mu \lambda  t})-x_0 e^{-2 \mu \lambda  t})^2}{1-e^{-4 \mu \lambda t}}\big]}
\end{equation}
\begin{equation}
    \varrho_{01}(x,t)= \rho_{01}\sqrt{\frac{\beta \lambda}{\pi(1-e^{-4 \mu \lambda t})}}\exp{\big[-\frac{\beta \lambda (x-x_0 e^{-2 \mu \lambda  t})^2}{1-e^{-4 \mu \lambda t}}+\frac{i}{\hbar}\frac{ 2 l (x+x_0)}{\mu} \tanh{\mu \lambda t}- \mu\lambda^2 l^2 \beta  t -\frac{4 l^2(\mu \lambda t- \tanh{\mu \lambda t})}{ \mu^2 \lambda \beta \hbar^2 }\big]}
\end{equation}
\begin{equation}\label{eq: analytic_sol_varrho11}
    \varrho_{11}(x,t)= \rho_{11}\sqrt{\frac{\beta \lambda}{\pi(1-e^{-4 \mu \lambda t})}}\exp{ \big[-\frac{\beta \lambda(x + l(1- e^{-2 \mu \lambda t})-x_0 e^{-2 \mu \lambda  t})^2}{1-e^{-4 \mu \lambda t}}\big]}
\end{equation}
\end{widetext} Since the initial quantum state $\hat{\rho}_0$ can be taken to depend on $x_0$, this solution also provides a Green's function for the dynamics with arbitrary initial conditions.

The $\rho_{00}$ and $\rho_{11}$ components describe classical probability distributions relaxing to a probability distribution peaked around $z=\pm l$. The first term of the off-diagonal components shows relaxation around the origin, while the second term gives the average phase accumulated by the unitary part of the quantum dynamics when the classical system reaches $x$ at time $t$. However, there is also simultaneously decoherence of this part of the classical-quantum state, given by the third and fourth terms. The first of these is primary decoherence, due to the action of the operators $\hat{L}_x$, while the other is secondary decoherence, which arises from destructive interference of the different phases picked up by the various classical paths that end up at $(x,t)$.

\subsection{Measurement and entropy production}\label{subsec: toy_model_measurement}

We first consider a regime in which the classical degree of freedom acts as to measure the quantum system in the $z$-basis. In doing so, we can study the total entropy production of the measuring apparatus and quantum system during a quantum measurement.

We start by reviewing the concept of a quantum measurement in a classical-quantum formalism. Here, a quantum system in an initial state $\hat{\rho}_0$ is allowed to interact with a classical system that acts as a measurement device. The interaction with the classical system causes the quantum system to decohere in a particular basis, while the quantum back-reaction on the classical system causes the final classical configuration to be correlated with the quantum state. Conditioning on the final classical state, the observer deduces information about the final state of the quantum system, which may correspond to a projective measurement if the set of final states are orthogonal pure states. 

To see that this arises in this model, we note that at long times, the above analytic solution tends to a stationary distribution $\hat{\varrho}^{st}(x)$, which we may normalise locally in phase space $\hat{\varrho}^{st}(x)/\tr{\hat{\varrho}^{st}(x)}$ to find the quantum state conditioned on the final classical position $x$, denoted $\hat{\rho}^{st}(x)$. This takes the form
\begin{equation}
\hat{\rho}^{st}(x)=\begin{pmatrix}
    \rho_{00}{(\rho_{00}+\rho_{11}e^{ - 4 \beta \lambda x l})^{-1}} \quad \quad \quad 0 \quad \quad \ \ \   \\
    \quad \quad \quad 0 \quad \quad \ \  \rho_{11}{(\rho_{11}+\rho_{00}e^{4 \beta \lambda x l})^{-1}}
\end{pmatrix}.
\end{equation} In the limit that $l\rightarrow0$, the final quantum state contains no dependence on the final classical position, and simply corresponds to the intial quantum state $\hat{\rho}_0$ decohered in the $\hat{\sigma}_z$ eigenbasis. However, in general, the dependence on $x$ indicates that observing the classical configuration provides information about the quantum state. In the limit of large $l$, the conditioned quantum state reduces to being either the pure state $|0\rangle$ or $|1\rangle$, depending on whether $x>0$ or $x<0$. This provides the projective measurement regime of this model, in which the observations $x>0$ or $x<0$ of the final classical position $x$ correspond to the measurement outcomes $\pm 1$ of the operator $\hat{\sigma}_z$ in the standard quantum observables formalism. In general, the parameter $l$ here provides a measure of how successful a given measurement is at distinguishing the $|0\rangle$ and $|1\rangle$ states.

\begin{figure}
    \centering
    \includegraphics[width=1\linewidth]{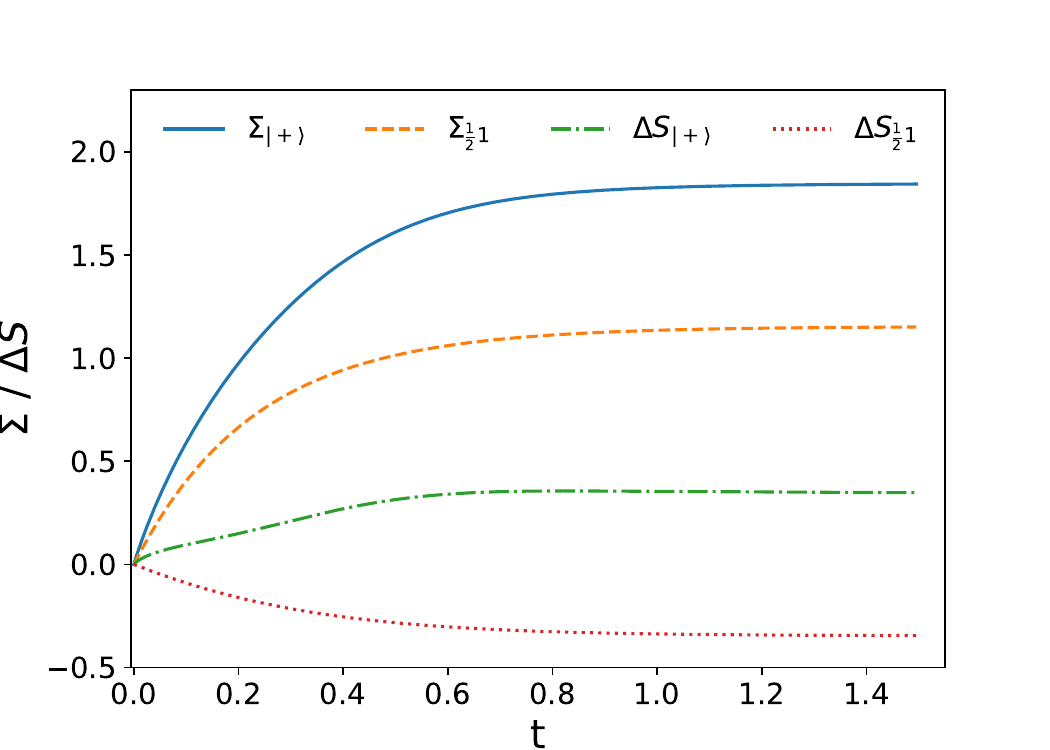}
    \caption{} The entropy production $\Sigma$ and change in entropy $\Delta S$ over time for a classical-quantum system with the quantum state either in a pure state $|+\rangle$ or a maximally mixed state $\frac{1}{2}\hat{\mathds{1}}$, over the course of a $\hat{\sigma}_z$ measurement. Here the classical system starts in a Gaussian state centered on the origin with variance $\sigma$. We here plot this with free parameters $\beta,\lambda,l,\mu,\hbar$ and $\sigma$ all equal to 1.
    \label{fig: entropy_prod_for_analtic_model}
\end{figure}

Having established that the dynamics of this model resemble that of a (possibly imperfect) measurement of the operator $\hat{\sigma}_z$, we now may turn to understand the entropy production in such a model. To do so, we initialise the classical state in a Gaussian probability distribution around the origin with variance $\sigma^2$, and take the initial quantum state to be $\hat{\rho}_0$. Using the analytic solution as a Green's function, we may integrate over the $x_0$ variable for this initial condition to find the subsequent evolution of the classical-quantum state. Numerically evaluating the integrals needed to compute the change in entropy and heat at each time $t$,  we may plot the entropy production over time to find $\Sigma(t)$=$\Delta S(t) - \beta \mathcal{Q}(t)$, for different initial conditions and parameters of the model. In Figure \ref{fig: entropy_prod_for_analtic_model} we plot the entropy production and change in entropy over time for two different initial states, $\hat{\rho}_0=|+\rangle\langle+|$ and $\hat{\rho}_0=\frac{1}{2} \hat{\mathds{1}
}$. We see here that the maximally mixed state configuration has decreasing entropy, due to the classical configuration becoming more ordered as the system relaxes. By contrast, the initially pure state configuration has an overall gain of entropy, with the entropy gain due to a loss of coherence outweighing the entropy loss due to the changes in the classical degrees of freedom. In both cases, the combined classical-quantum system experiences the same loss of heat into its surroundings over time, leading to an overall positive entropy production, that tends to a steady value as the system relaxes.  

The above model demonstrates explicitly that measurements of quantum states with coherence lead to greater entropy production than those without. While intuitively reasonable, given the clear difference in entropy changes in the two cases, the current framework provides a real time description of this process. Understanding the consequences of this in general settings, or in more physically motivated models, is likely to be an interesting area of future study.

\subsection{Coherent control via relaxation}

The previous example showed how this toy model may be used to study the non-equilibrium thermodynamics of quantum measurements. We now turn to illustrate how one may also use this model to understand quantum control in a setting in which fluctuations, dissipation and quantum back-reaction affect a mesoscopic classical control system. 

As with the previous example, we first review the concept of quantum control in the classical-quantum setting. Here, a quantum system evolves with a Hamiltonian that depends on the state of a classical system. Since the state of the classical system can change with time, it may act as a controller of the quantum dynamics, allowing the quantum system's Hamiltonian to be switched on or off such that a specific unitary operation is performed after a desired time. In addition to this unitary part of the dynamics, the quantum system will experience some additional decoherence, either due to its environment or due to noise in the classical control system. To correctly describe this latter type of decoherence, one must average over all possible realisations of the noise in the classical controller, i.e. study the unconditioned quantum state. 

To see how the overdamped classical degree of freedom controls the implementation of a unitary on the two-level system in our toy model, we first revisit the analytic solution.  
Integrating over the classical degree of freedom in the solution for the classical-quantum state $\hat{\varrho}(x,t)$ given in equations \eqref{eq: analytic_sol_varrho00} to \eqref{eq: analytic_sol_varrho11}, we find the unconditioned quantum state $\hat{\rho}(t)$ takes the form
\begin{equation}
    \hat{\rho}(t)=\begin{pmatrix}
        \rho_{00} & \rho_{01} \exp{[i\theta(t)+\Gamma(t)]}\\
        \rho_{10}\exp{[-i\theta(t)+\Gamma(t)]} & \rho_{11}, 
    \end{pmatrix}
\end{equation} where the phase $\theta$ and damping factor $\Gamma$ are given as the following functions of time
\begin{equation}
    \theta(t)=\frac{2l x_0}{\mu \hbar}(1-e^{-2\mu\lambda t}),
\end{equation}
\begin{equation}
    \Gamma(t)=-l^2\bigg(\mu \lambda^2 \beta +\frac{4}{\mu \beta \hbar^2}\bigg)t + \frac{l^2(3+e^{-4 \mu \lambda t} - 4e^{-2 \mu \lambda t})}{\mu^2 \lambda \beta \hbar^2}.
\end{equation} Considering first the phase $\theta(t)$, we see that the quantum state undergoes a unitary transformation corresponding to a rotation about the $z$ axis of the Bloch sphere. Moreover, since at long times $\theta(t)$ tends to fixed value, the evolution of the phase corresponds to applying the following unitary on the initial state
\begin{equation}
\hat{U}=e^{i\phi\hat{\sigma}_z}\quad \quad \phi=\frac{x_0 l }{\mu \hbar},
\end{equation} provided  $t\gg (\mu \lambda)^{-1}$. The ``switching off" of the Hamiltonian evolution that enables this specific unitary to be implemented at long times is due to the time evolution of the classical system, which turns off the interaction as it relaxes towards the origin. Since the total unitary applied is dependent on the $x_0$, we see that the toy model provides a model of quantum control where choosing an initial non-equilibrium classical state, and allowing the system to relax towards a fixed point, allows for a specific unitary gate to be performed on a quantum system.

The success of performing this unitary transformation is determined by the loss of coherence. In this toy model the decoherence is in the $\hat{\sigma}_z$ eigenbasis and is entirely determined by $\Gamma(t)$, which must remain small over a timescale $t\gg (\mu \lambda)^{-1}$ in order for the unitary to be performed with low noise. Immediately, we see that a key parameter to achieve this is the spacing $l$ between the two potential sites, which must remain small in order for the unitary to be performed accurately. Since the spacing of the potentials does not affect the timescale after which the unitary has been implemented, we see that provided $l$ can be made arbitrarily small, and that $x_0$ may be made sufficiently large to compensate this to achieve a given choice of unitary, the model describes a unitary control operation with arbitrary accuracy. 

Although this toy model provides a proof-of-principle of how quantum control may be implemented via an relaxation process, it fails a number of requirements needed for a realistic description of an experimental platform. For starters, the Hamiltonian of the model we consider is decomposeable in terms of $\hat{\mathds{1}}$ and $\hat{\sigma}_z$, meaning that it is limited to performing rotations around the $z$-axis. Moreover, since $l$ must be small, a high degree of control of the parameters $l$ and $x_0$ would be needed here to achieve a high precision unitary. However, the above method, where the dynamics are solved to find the unconditioned quantum state $\hat{\rho}(t)$, which may have its phase and decoherence compared to find optimal parameters of performance, provides a blueprint for future studies in more complex realistic models.

\section{Model II} \label{sec: oscillator_model}

In this section we numerically study a model of a classical oscillator interacting with a quantum oscillator via a linear coupling. As we shall see, this model features a classical-quantum Hamiltonian that is not self-commuting in phase space, and illustrates a number of interesting features, including thermalisation.

\subsection{Set-up}

We consider here a one-dimensional model of an underdamped classical oscillator coupled to a quantum oscillator. The position and momentum of the classical system are denoted $q,p$, while the corresponding operators for the quantum system will be denoted $\hat{q},\hat{p}$ and satisfy the canonical commutation relation. The classical-quantum Hamiltonian that governs the dynamics of this system is given

\begin{equation} \label{eq: H_oscillator}
    \hat{H}(q,p)= \frac{\hat{p}^2}{2m_q}+\frac{1}{2}m_q\omega^2(q\hat{\mathds{1}}-\hat{q})^2 + \big(\frac{p^2}{2 m_c}+\frac{1}{2}m_c\Omega^2 q^2 \big)\hat{\mathds{1}}.
\end{equation} Here the first two terms describe the quantum oscillator and its coupling to the classical system, while the last two terms proportional to the identity describe the classical harmonic oscillator. As such, $m_q$ and $m_c$ are the quantum and classical particle masses, while $\omega$ represents the angular frequency of the coupling between the classical and quantum systems in terms of the quantum mass, and $\Omega$ denotes the classical oscillator's angular frequency. In this system, the classical oscillator will experience friction, with a corresponding friction coefficient $\gamma$. The whole classical-quantum system is also assumed to be in an environment with inverse temperature $\beta$. The dynamics are then taken to be governed by the coupled stochastic equations \eqref{eq: underdamped_unravelling_q} to \eqref{eq: underdamped_unravelling_psi}, which are equivalent to the master equation \eqref{eq: underdamped}.

{This model provides a basic version of the problem faced in molecular dynamics, where electronic degrees of freedom are modelled using quantum theory, and the nuclei are treated classically \cite{tully1990molecular,tully1998mixed,Kapral1999}. Here, transitions between adiabatic energy levels are relevant to processes such as photochemical reactions that are not well-described under the Born-Oppenheimer approximation, with finding the correct dynamics to describe this an ongoing area of active research \cite{curchod2018ab,alonso2021computation,amati2023detailed,mannouch2023mapping}. }

To find the adiabatic basis for this model, we must solve the eigenvalue problem for the operator $\hat{H}(q,p)$. Since the non-trivial part of this Hamiltonian corresponds to a quantum harmonic oscillator Hamiltonian displaced from the origin by $q$, it is intuitive that the adiabatic basis is given by displaced eigenstates of the quantum harmonic oscillator Hamiltonian. Letting $|n\rangle$ denote the number states, the adiabatic basis takes the form
\begin{equation} \label{eq: adiabatic_basis_oscillator}
    |n(q,p)\rangle=e^{-\frac{i}{\hbar}q\hat{p}}|n\rangle
\end{equation} i.e. the standard number eigenstates of the quantum harmonic oscillator $|n\rangle$ displaced by distance $q$. The corresponding eigenvalues of $\hat{H}(q,p)$ are 
\begin{equation} \label{eq: oscillator_H_eigenvalues}
    \epsilon_n(q,p)=\hbar \omega (n+\frac{1}{2})+\frac{p^2}{2m_c} + \frac{1}{2}m_c\Omega^2 q^2.
\end{equation} which can easily be seen to be the sum of the quantum oscillator energy for a given energy eigenstate $|n\rangle$ and the classical oscillator energy for a given $q$, $p$.

The thermal state corresponding to this system is given 
\begin{equation} \label{eq: thermal_state_oscillators}
\hat{\pi}(q,p)=\frac{1}{\mathcal{Z}}e^{-\beta \hat{H}(q,p)}
\end{equation} where here
\begin{equation} \label{eq: Z_oscillator}
    \mathcal{Z}= \mathcal{Z}_C \mathcal{Z}_Q
\end{equation} for the classical and quantum thermal partition functions
\begin{equation}
    \mathcal{Z}_C=\frac{2 \pi}{\beta \Omega}, \quad\quad \mathcal{Z}_Q=\frac{1}{2\sinh{\frac{\beta \omega \hbar}{2}}}.
\end{equation} These may be found by representing the thermal state in the adiabatic basis, and taking the trace and integrating over phase space to ensure normalisation.

\subsection{Computing $\hat{L}_q$ and $\hat{M}_{qq}$} \label{subsec: explicit_comp_of_L_oscillator}
Before studying properties of this dynamics, the first step is to explicitly compute the operators $\hat{L}_q$ and $\hat{M}_{qq}$ for the classical-quantum Hamiltonian $\hat{H}$ of this model. To do so, we shall need to exploit a number of relations describing power series of  $\ad{\frac{-\beta \hat{H}}{2}}=-(\beta/2)[\hat{H},\cdot]$ .

To begin, we first note that the commutator of $\hat{H}$ with operators linear in $q\hat{\mathds{1}}-\hat{q}$ and $\hat{p}$ takes a particularly simple form. In particular,  the commutator of $\hat{H}$ with $q\hat{\mathds{1}}-\hat{q}$ gives $-i \omega^2 \hat{p}/m_q$, while when applied to $\hat{p}$ gives $i m_q \omega^2 \hbar (q\hat{\mathds{1}}-\hat{q})$. We thus see that representing the operators as vectors
\begin{equation} \label{eq: op_to_vector}
q\mathds{1}-\hat{q} \ \mapsto \begin{pmatrix} 1\\ 0 \end{pmatrix}\quad \quad\hat{p}\ \mapsto \begin{pmatrix} 0\\ 1 \end{pmatrix},
\end{equation}
we can represent the adjoint $\ad{\frac{-\beta \hat{H}}{2}}$ as the following $2\times 2$ matrix
\begin{equation} \label{eq: oscillator_H_adjoint_to_matrix}
\ad{\frac{-\beta H}{2}}\ \mapsto
    \begin{pmatrix}
         0 & \frac{i m_q  \omega^2\beta \hbar}{2}\\
         -\frac{i \beta \hbar}{2 m_q} & 0
    \end{pmatrix}.
\end{equation} Moreover, since the adjoint action acting on an operator linear in  $q\hat{\mathds{1}}-\hat{q}$ and $\hat{p}$ produces another operator linear in these operators, $\ad{\frac{-\beta \hat{H}}{2}}$ closes on these operators, and thus the action of arbitrary series of  $\ad{\frac{-\beta \hat{H}}{2}}$ on linear combinations of $\hat{q}$ and $\hat{p}$ may be computed by finding the corresponding series for the $2\times 2$ matrix \eqref{eq: oscillator_H_adjoint_to_matrix}.

Having established this, we first consider $\hat{L}_q$. Using the series form of $\hat{L}_z$ given in \eqref{eq: L_series}, we may compute $\hat{L}_q$ as a series of $\ad{\frac{-\beta \hat{H}}{2}}$ acting on the derivative of the classical-quantum Hamiltonian
\begin{equation} \label{eq: H_oscillator_q_derivative}
    \frac{\partial \hat{H}}{\partial q}=m_c\Omega^2 q \hat{\mathds{1}}+m_q\omega^2 (q\hat{\mathds{1}}-\hat{q}).
\end{equation} While the first term commutes with $\hat{H}$, and thus appears unmodified in $\hat{L}_q$, the rest of $\hat{L}_q$ must be computed by acting with the series of $\ad{\frac{-\beta \hat{H}}{2}}$ given in Eq. \eqref{eq: frac_to_series} on the second term in Eq. \eqref{eq: H_oscillator_q_derivative}.  To find a closed form expression for this, we simply compute the corresponding series for the $2\times 2$ matrix \eqref{eq: oscillator_H_adjoint_to_matrix}, which gives
\begin{equation}
\begin{split}
    \frac{e^{\ad{\frac{-\beta \hat{H}}{2}}}-1}{\ad{\frac{-\beta \hat{H}}{2}}}\ \mapsto 
    \begin{pmatrix}
         \frac{2 \sinh{\frac{\hbar\omega\beta}{2}}}{\beta \omega \hbar} & \frac{-2 i m_q (1-\cosh{\frac{\hbar\omega\beta}{2}})}{\beta \hbar}\\
         \frac{2 i  (1-\cosh{\frac{\hbar\omega\beta}{2}})}{m_q \beta \hbar \omega^2} & \frac{2\sinh{\frac{\hbar\omega\beta}{2}}}{\beta \omega \hbar} 
    \end{pmatrix}.
\end{split}
\end{equation} Applying this matrix to $(m_q \omega^2\  0)^T$, the vector representing the second term of Eq. \eqref{eq: H_oscillator_q_derivative}, one finally arrives at the form of $\hat{L}_q$ as
\begin{equation} \label{eq: L_oscillator}
\begin{split}
    \hat{L}_q=& \frac{2 m_q \omega}{\hbar \beta } \sinh{\frac{\hbar \omega \beta}{2}} (q\hat{\mathds{1}}-\hat{q}) + \frac{2i}{\hbar \beta }(1 -\cosh{\frac{\hbar \omega \beta}{2}})\hat{p}\\
    &+ m_c\Omega^2 q \hat{\mathds{1}}.
\end{split}
\end{equation} Taking the $\beta\rightarrow 0$ limit of this operator, one finds that $\hat{L}_q$ indeed reduces to the expression in Eq. \eqref{eq: H_oscillator_q_derivative}, in agreement with the general result of Eqs. \eqref{eq: high_temp_limits}.

To compute $\hat{M}_{qq}$, we will exploit two identities that allow one to swap the position of $\hat{q}$ and $\hat{p}$ with the square root of the thermal state $\hat{\pi}^\frac{1}{2}$. These are given
\begin{equation} \label{eq: sqrtpi_q_identity}
    \hat{\pi}^\frac{1}{2}(q\hat{\mathds{1}}-\hat{q}) = \big[\cosh{\frac{\hbar\omega \beta}{2}} (q \hat{\mathds{1}}-\hat{q}) -\frac{i}{m_q \omega} \sinh{\frac{\hbar\omega \beta}{2}}\hat{p}\big] \hat{\pi}^\frac{1}{2},
\end{equation} and
\begin{equation}  \label{eq: sqrtpi_p_identity}
    \hat{\pi}^\frac{1}{2}\hat{p} = \big[i m_q \omega \sinh{\frac{\hbar\omega \beta}{2}}(q \hat{\mathds{1}}-\hat{q})+\cosh{\frac{\hbar\omega \beta}{2}}  \hat{p}\big] \hat{\pi}^\frac{1}{2}.
\end{equation} To prove these, we note that
\begin{equation} \label{eq: exp_of_adjoint}
    e^{-\frac{\beta}{2}\hat{H}}\hat{A} e^{\frac{\beta}{2}\hat{H}}=e^{\ad{\frac{-\beta \hat{H}}{2}}} \hat{A}
\end{equation} for any operator $\hat{A}$, and that for operators linear in $q\hat{\mathds{1}}-\hat{q}$ and $\hat{p}$ we may represent the exponential of the adjoint as
\begin{equation}
    e^{\ad{\frac{-\beta \hat{H}}{2}}} \mapsto 
    \begin{pmatrix}
         \cosh{\frac{\hbar\omega\beta}{2}} &  i m_q\omega \sinh{\frac{\hbar\omega\beta}{2}}\\
         -\frac{i }{m_q \omega} \sinh{\frac{\hbar\omega\beta}{2}} & \cosh{\frac{\hbar\omega\beta}{2}}
    \end{pmatrix}.
\end{equation} Applying this matrix to the two vectors in \eqref{eq: op_to_vector} to compute the right hand side of \eqref{eq: exp_of_adjoint}, and then acting both sides on $\hat{\pi}^{\frac{1}{2}}$, we recover the two identities \eqref{eq: sqrtpi_q_identity} and \eqref{eq: sqrtpi_p_identity}.

To compute $\hat{M}_{qq}$, we assume that $\hat{M}_{qq}$ takes the form of a Hermitian operator that is at most quadratic in $q\hat{\mathds{1}}-\hat{q}$ and $\hat{p}$. Plugging this form of $\hat{M}_{qq}$ into the left hand side of \eqref{eq: M_op_def2}, and the previously computed $\hat{L}_q$ into the right hand side, we may use the commutation relations \eqref{eq: sqrtpi_q_identity} to rearrange both left and right hand sides of \eqref{eq: M_op_def2} to have $\hat{\pi}^\frac{1}{2}$ on the right hand side of the expression. Removing this by acting with $\hat{\pi}^{-\frac{1}{2}}$ on both sides, and comparing terms, we find the solution as
\begin{equation}
\begin{split}
    \hat{M}_{qq}=&\frac{2 m_q \omega}{\hbar \beta^2}(\sinh{\frac{\hbar \omega \beta}{2}}-\tanh{\frac{\hbar \omega \beta}{2}})\{ \hat{q}-q\hat{\mathds{1}},\hat{p}\}_+ \\
    &+\frac{2 m_c \Omega^2}{\beta}(1-\cosh{\frac{\hbar \omega \beta}{2}})\hat{p}.
\end{split}
\end{equation} 
Since the solutions to Eq. \eqref{eq: M_op_def2} are unique, we see that the original assumption that $\hat{M}_{qq}$ was at most quadratic in $q\hat{\mathds{1}}-\hat{q}$ and $\hat{p}$ was correct. In Appendix ~\ref{app:Mop} we also provide a constructive derivation without relying on this assumption, by combining the identities~\eqref{eq: sqrtpi_q_identity} and~\eqref{eq: sqrtpi_p_identity} with a series expansion \cite{liu2016quantum} of the exponential operators appearing in~\eqref{eq: M_op_def1}. It is straightforward to check that in the high temperature limit, $\beta \rightarrow 0$,  $\hat{M}_{qq}$ correctly reduces to zero, consistent with the previously derived expression in Eqs. \eqref{eq: high_temp_limits}. 

\subsection{Relative position representation}

Having found the operators $L_q$ and $M_{qq}$, one may in principle directly study the dynamics of Eqs. \eqref{eq: underdamped_unravelling_q} to \eqref{eq: underdamped_unravelling_psi} to understand properties of the system.  However, we will first introduce an alternate representation for describing the quantum system, which makes the dynamics simpler to both solve and interpret.

In the unravelling picture of classical-quantum dynamics, $|\psi\rangle$ represents the state of the quantum system in absolute space. However, given that the interactions between the two systems depend on their relative positions, it is convenient to describe the quantum system using a state vector that represents the quantum system relative to the classical position $q$. We refer to this as the \textit{relative position representation}, and denote the quantum state in this representation $|\psi^\mathfrak{r}\rangle$.  The absolute and relative position descriptions are related by the standard unitary transformation
\begin{equation} \label{eq: rel_pos_states}
    |\psi^{\mathfrak{r}}\rangle=e^{\frac{i}{\hbar} q \hat{p}} |\psi\rangle.
\end{equation} In order for expectation values to be correctly computed in this representation, observables must also be transformed by a unitary transformation 
\begin{equation} \label{eq: rel_pos_operators}
\hat{A}^{\mathfrak{r}}(q,p)=e^{\frac{i}{\hbar} q \hat{p}} \hat{A}(q,p) e^{-\frac{i}{\hbar} q \hat{p}},
\end{equation} which ensures that $\langle \psi^{\mathfrak{r}}|\hat{A}^{\mathfrak{r}}|\psi^{\mathfrak{r}}\rangle=\langle \psi|\hat{A}|\psi\rangle$. To compute this in practice, we will make use of a version of the identity \eqref{eq: exp_of_adjoint}, namely that
\begin{equation}
\label{eq: exp_of_adjoint}
    e^{\frac{i}{\hbar}q \hat{p}}\hat{A} e^{-\frac{i}{\hbar}q \hat{p}}=e^{\ad{\frac{i}{\hbar}q \hat{p}}} \hat{A}.
\end{equation} It is straightforward to check using this that the map $\hat{A} \mapsto \hat{A}^\mathfrak{r}$ amounts to replacing every appearance of $\hat{q}-q\hat{\mathds{1}}$ with $\hat{q}$.

The utility of this representation is apparent when we consider the relative position representation of the states and operators specific to our current model. Taking first the adiabatic basis, we see that
\begin{equation}
    |n^\mathfrak{r}(q,p)\rangle=|n\rangle
\end{equation}
i.e. that the adiabatic basis in the relative position representation is simply the number basis of the quantum harmonic oscillator. Representing dynamics in terms of $|\psi^\mathfrak{r}\rangle$ thus allows us decompose the dynamics in a fixed basis, independent of the current state of the classical system, that nevertheless describes the adiabatic basis of the system.

This representation also simplifies the form of operators. Making the substitution $\hat{q}-q\hat{\mathds{1}}\mapsto \hat{q}$ it is straightforward to see that the classical-quantum Hamiltonian in this representation takes the form
\begin{equation}
    \hat{H}^{\mathfrak{r}}(q,p)=H_{C}^{\mathfrak{r}}(q,p) \hat{\mathds{1}} + \hat{H}_{Q}^{\mathfrak{r}}
\end{equation} where
\begin{equation}
H_{C}^\mathfrak{r}(q,p)=\frac{p^2}{2 m_c}+\frac{1}{2}m_c\Omega^2 q^2 
\end{equation} is the Hamiltonian of a classical harmonic oscillator and 
\begin{equation}
\hat{H}_{Q}^\mathfrak{r}=\frac{\hat{p}^2}{2m_q}+\frac{1}{2}m_q\omega^2 \hat{q}^2 
\end{equation} is the Hamiltonian of a quantum harmonic oscillator. Solving the eigenvalue problem in this representation, it is therefore straightforward to see both why the adiabatic basis is given by \eqref{eq: adiabatic_basis_oscillator}, and why the energy eigenvalues $\epsilon_n(q,p)$ are those written in \eqref{eq: oscillator_H_eigenvalues}.

Finally, we note an intuitive relation that connects unravellings to classical-quantum states in the relative position representation. If one takes $\hat{\varrho}(q,p)$, and maps this using \eqref{eq: rel_pos_operators}, we find the classical-quantum state in the relative position representation $\hat{\varrho}^\mathfrak{r}(q,p)$. Using the fact that $\mathbb{E}[f(z_t)\delta(z-z_t)]=f(z)\mathbb{E}[\delta(z-z_t)]$, it is simple to see that this may equivalently be found using
\begin{equation} \label{eq: CQ_state_rel_position}
    \hat{\varrho}^{\mathfrak{r}}(q,p,t)=\mathbb{E}[|\psi^\mathfrak{r}\rangle_t\langle \psi^{\mathfrak{r}}|_t \delta(q-q_t)\delta(p-p_t)],
\end{equation} i.e. using the relation \eqref{eq: CQ_state_from_ensembles} and replacing $|\psi\rangle_t$ with $|\psi^{\mathfrak{r}}\rangle_t$. This identity is useful since it means that we may compute the classical-quantum state in the relative position representation directly from the distribution of trajectories in terms of $|\psi^\mathfrak{r}\rangle_t$, $q_t$ and $p_t$.

\subsection{Numerical solution}
To study the dynamics of this model, we shall use the unravelling approach introduced in Sec. \ref{sec: framework}, studying solutions to equations \eqref{eq: underdamped_unravelling_q} to \eqref{eq: underdamped_unravelling_psi}. To study these numerically in an efficient manner, we will use a simplified set of dynamics, employing both the relative position representation introduced in the previous section, and a particular invariance property of the dynamics under changes to the classical part of the Hamiltonian.

We begin by finding the equations of motion in the relative state representation. Since the expectation values remain the same provided both the states and operators are transformed as in \eqref{eq: rel_pos_states} and \eqref{eq: rel_pos_operators}, the classical equations for $q_t$ and $p_t$ remain unchanged. To find the equation of motion for $|\psi^{\mathfrak{r}}\rangle_t$, we take the derivative of \eqref{eq: rel_pos_states} with respect to time. Doing so is straightforward to see two changes in the dynamics of the relative quantum state $|\psi^{\mathfrak{r}}\rangle_t$ versus the absolute quantum state $|\psi^{\mathfrak{r}}\rangle_t$. The first is that the time dependence of $q_t$ leads to an additional unitary term in the dynamics, with Hamiltonian $-p_t \hat{p}/m_c$. The second is that, in order to put write the dynamics in terms of $|\psi^{\mathfrak{r}}\rangle_t$, the operators $\hat{L}_q$, $\hat{M}_{qq}$ and $\hat{H}$ must all be replaced with their transformed versions $\hat{L}_q^{\mathfrak{r}}$, $\hat{M}_{qq}^{\mathfrak{r}}$ and $\hat{H}^{\mathfrak{r}}$.

To further simplify the form of the dynamics, we first use a particular invariance property of the general dynamics presented in \ref{sec: construct}. Namely, as we show in Appendix \ref{app: invariance}, one may always remove the part of $\hat{L}_z$ proportional to the identity and a corresponding term in $\hat{M}_{zz}$, and instead include it as an additional drift term in the dynamics. Guaranteeing that the form of decoherence and back-reaction is independent of any additional purely classical dynamics, in this case it allows us to remove the dependence of $\hat{L}_q^{\mathfrak{r}}$ and $\hat{M}_{qq}^{\mathfrak{r}}$ on the classical angular frequency $\Omega$ and replace it with an additional classical drift term $-m_c \Omega^2 q dt$ in equation \eqref{eq: underdamped_unravelling_p}. Finally, we drop from $\hat{H}^{\mathfrak{r}}(q,p)$ the term proportional to the identity, $H^\mathfrak{r}_C(q,p)\hat{\mathds{1}}$.

Taken together, the two steps lead to the following form of equations for $q_t,p_t$ and $|\psi^\mathfrak{r}\rangle_t$
\begin{equation} \label{eq: SDE_for_sim_q}
    dq_t=\frac{p_t}{m_C}dt 
\end{equation}
\begin{equation}
    dp_t=-\frac{1}{2}\langle \hat{L}_q^{\mathfrak{r}} + \hat{L}_q^{\mathfrak{r}\dag} \rangle dt -m_c \Omega^2 q_t dt -  \frac{\gamma p_t}{m_c} dt + \sqrt{\frac{2 \gamma}{\beta}} dW
\end{equation}
\begin{equation}
\begin{split} \label{eq: SDE_for_sim_psi}
    d|\psi^{\mathfrak{r}}\rangle_t=&-\frac{i}{\hbar} (\hat{H}_Q^\mathfrak{r}+\frac{\beta}{8\gamma}\hat{M}_{qq}^{\mathfrak{r}}-\frac{p_t}{m_c}\hat{p})|\psi^{\mathfrak{r}}\rangle_t dt\\
    &-\frac{1}{2}\frac{\beta}{8\gamma} (\hat{L}_q^{\mathfrak{r} \dag} \hat{L}_q^{\mathfrak{r}} - 2 \langle \hat{L}_q^{\mathfrak{r} \dag} \rangle \hat{L}_q^{\mathfrak{r}} + \langle \hat{L}_q^{\mathfrak{r} \dag} \rangle \langle \hat{L}_q^{\mathfrak{r}}\rangle ) |\psi^{\mathfrak{r}} \rangle_t dt \\
    & -\sqrt{\frac{\beta}{8\gamma}}(\hat{L}_q^\mathfrak{r} - \langle \hat{L}_q^{\mathfrak{r}} \rangle)|\psi^{\mathfrak{r}}\rangle_t dW_t
\end{split}
\end{equation} where here the expectation values are all taken with respect to $|\psi^{\mathfrak{r}}\rangle_t$, and the operators $\hat{L}_q^{\mathfrak{r}}$ and $\hat{M}_{qq}^{\mathfrak{r}}$ are defined as
\begin{equation} \label{eq: L_rel_pos_oscillator}
    \hat{L}_q^{\mathfrak{r}}= -\frac{2 m_q \omega}{\hbar \beta } \sinh{\frac{\hbar \omega \beta}{2}} \hat{q} + \frac{2i}{\hbar \beta }(1 -\cosh{\frac{\hbar \omega \beta}{2}})\hat{p},
\end{equation}
\begin{equation}
    \hat{M}_{qq}^{\mathfrak{r}}=\frac{2 m_q \omega}{\hbar\beta^2}(\sinh{\frac{\hbar \omega \beta}{2}}-\tanh{\frac{\hbar \omega \beta}{2}})\{ \hat{q},\hat{p}\}_+.
\end{equation} To simulate these equations efficiently, we employ a numerical method introduced in the context of continuous measurement theory, known as Rouchon's method \cite{amini2011stability,rouchon2015efficient,ralph2016coupling}. In this case, it allows one to simulate the coupled set of stochastic differential equations \eqref{eq: SDE_for_sim_q} to \eqref{eq: SDE_for_sim_psi} in discrete time. Taking $t_{i}=0$ and $t_{f}=N_{steps}\Delta t$, where $N_{steps}$ is the number of timesteps and $\Delta t$ is the time increment, we define $\Delta W^{(n)}$ as the $n$th sample of a Gaussian random variable with mean zero and variance $\Delta t$. We then compute the values of $q$, $p$ and $|\psi^\mathfrak{r}\rangle$ at time $t=(n+1)\Delta t$ by iterating the following set of equations
\begin{equation}
    q^{(n+1)}=q^{(n)} + \frac{p^{(n)}}{m_c}\Delta t
\end{equation}
\begin{equation}
\begin{split}
    p^{(n+1)}=&-\frac{1}{2}\langle \psi^{\mathfrak{r}}_{(n)}| \hat{L}_q^\mathfrak{r} + \hat{L}_q^{\mathfrak{r}\dag}|\psi^{\mathfrak{r}}_{(n)} \rangle \Delta t -m_c \Omega^2 q^{(n)} \Delta t \\
    &- \frac{\gamma p^{(n)}}{m_c} \Delta t + \sqrt{\frac{2 \gamma}{\beta}} \Delta W^{(n)}
\end{split}
\end{equation}
\begin{equation}
\begin{split}
    R_{(n)}=&\mathds{1}-\frac{i}{\hbar}(\hat{H}^\mathfrak{r}_Q+\frac{\beta}{8\gamma}\hat{M}_{qq}^\mathfrak{r}-\frac{p^{(n)}}{m_c}\hat{p})\Delta t\\
    & -\frac{\beta}{16 \gamma}\hat{L}_q^{\mathfrak{r}\dag}\hat{L}_q^{\mathfrak{r}}\Delta t\\ 
    & +\frac{\beta}{8\gamma} \langle \psi^{\mathfrak{r}}_{(n)}| \hat{L}_q^\mathfrak{r} + \hat{L}_q^{\mathfrak{r}\dag}|\psi_{(n)}^{\mathfrak{r}} \rangle \hat{L}_q^\mathfrak{r}  \Delta t\\
    & -\sqrt{\frac{\beta}{8\gamma}} \hat{L}_q^\mathfrak{r} \Delta W^{(n)}\\
    &+\frac{\beta}{16\gamma} (\hat{L}_q^\mathfrak{r})^2  ((\Delta W^{(n)})^2-\Delta t)
\end{split}
\end{equation}
\begin{equation}
|\psi^\mathfrak{r}_{(n+1)}\rangle=\frac{\hat{R}_{(n)}|\psi^{\mathfrak{r}}_{(n)}\rangle}{\sqrt{\langle\psi^\mathfrak{r}_{(n)}| \hat{R}_{(n)}^\dag \hat{R}_{(n)}|\psi^\mathfrak{r}_{(n)}\rangle}}.
\end{equation} To represent the quantum state and operators for numerical simulation, we use the adiabatic basis in this representation i.e. the standard harmonic oscillator number states $|n\rangle$. Truncating these at a finite maximum energy level $N_{max}$, we may thus represent $|\psi^\mathfrak{r}_{(n)}\rangle$ as a complex vector of length $N_{max}$, and the operators $ \hat{q}$ and $\hat{p}$ as $N_{max}\times N_{max}$ restrictions of their standard infinite dimensional number state representations \cite{shankar2012principles}.

\begin{figure*}[!tbp]
  
  \centering
  \begin{subfigure}[b]{0.3215625\textwidth}
  \includegraphics[width=\textwidth]{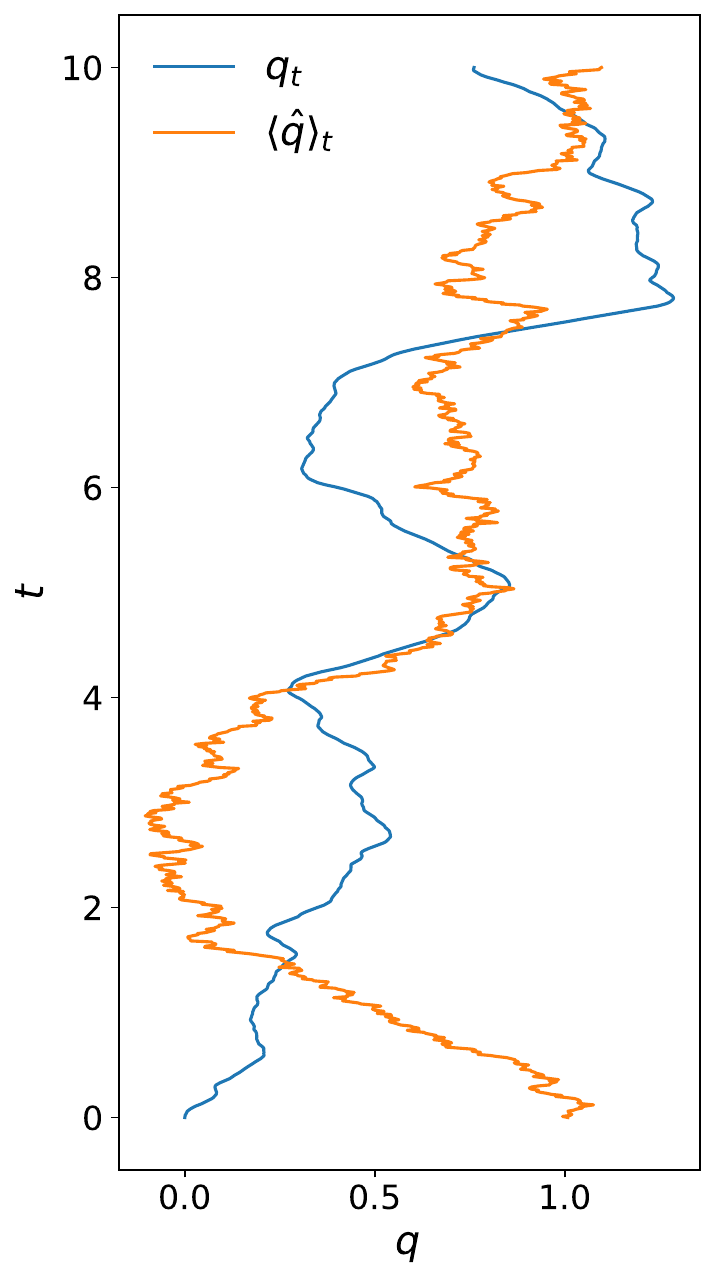}
  \end{subfigure}
  \begin{subfigure}[b]{0.3\textwidth}
  {\includegraphics[width=\textwidth]{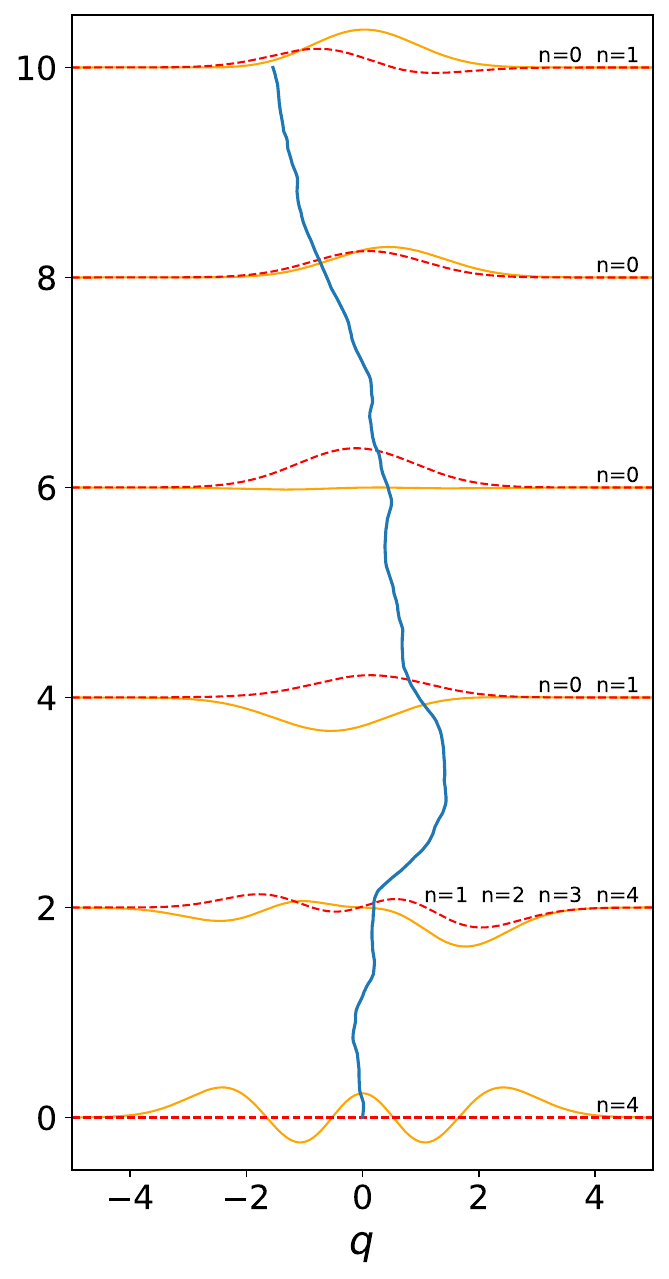}}
  \end{subfigure}
  \begin{subfigure}[b]{0.3\textwidth}
  {\includegraphics[width=\textwidth]{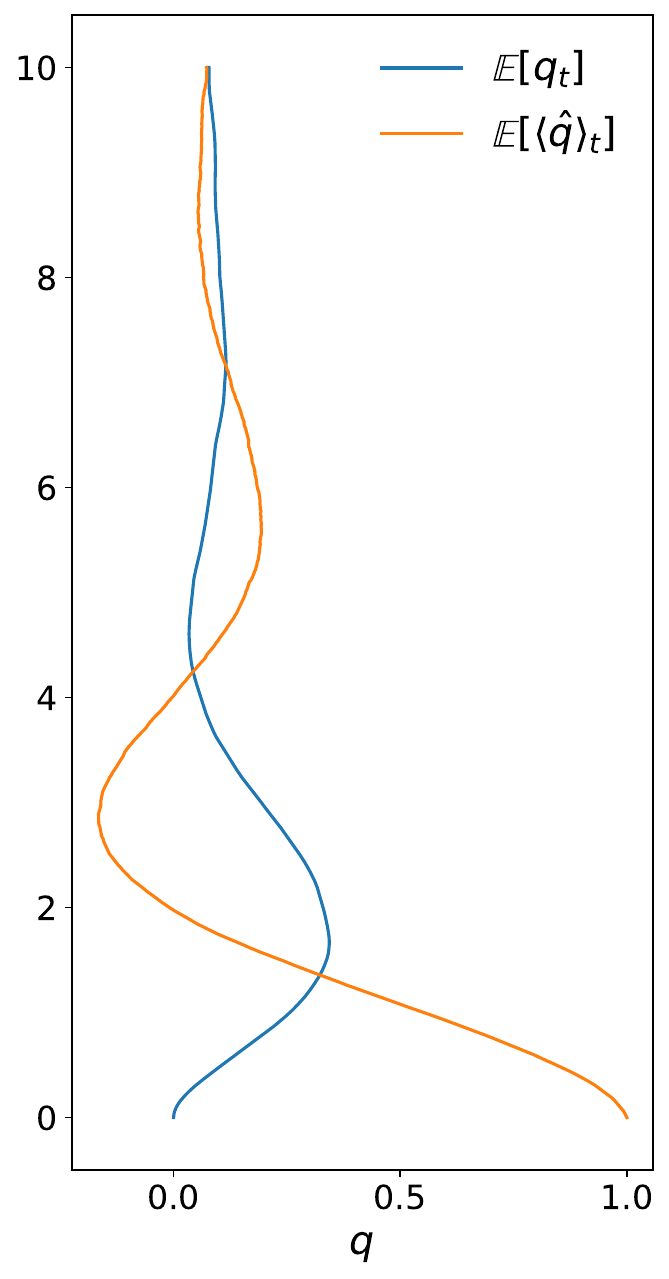}}
  \end{subfigure}
  \caption{\small{Simulations of trajectories for two coupled oscillators, one classical and one quantum, for parameters $\omega=m_C=m_Q=\hbar=1$, $\beta=\gamma=3$ and $\Omega=10^{-5}$ between $t=0$ and $t=10$. \textit{(left)} The classical position $q_t$ and expected position of the quantum state $\langle{q}\rangle_t$ is simulated and plotted for a single realisation, with $N_{max}=20$ and $N_{steps}=2\times 10^{4}$. Here the classical system starts at the origin of phase space, while the quantum system starts in a coherent state with $\langle \hat{q}\rangle_0=1$ and $\langle \hat{p}\rangle_0=0$. \textit{(middle)} The classical position $q_t$ is plotted with 6 distinct snapshots of the real (solid line) and imaginary (dashed line) parts of the wavefunction for a pair of trajectories starting at the origin in phase space and the $3$rd excited adiabatic state, with $N_{max}=50$ and $N_{steps}=10^5$. On the right of each wavefunction, we denote the main components of the wavefunction in the adiabatic basis.  \textit{(right)} The expected values of the classical position $q_t$ and the trajectory expectation value $\langle \hat{q}\rangle_t$ are plotted, performed by averaging over $2000$ individual trajectories with the same initial conditions, $N_{max}$ and $N_{step}$ as in the left hand panel.}}
  \label{fig: trajectories}
\end{figure*}

\subsection{Individual and average trajectories}

Having established a numerical method for studying the dynamics of this system, we begin to gain some insights into these dynamics by considering both individual and average trajectories of the joint classical-quantum system. 

We begin by considering a typical trajectory, for which we plot both $q_t$ and $\langle  \hat{q} \rangle_t$ in the left panel of Figure \ref{fig: trajectories}. Here we start the classical system at the origin, and the quantum system in a coherent state with $\langle \hat{p} \rangle=0$ and $\langle \hat{q}\rangle=1$. The mean position of the quantum system is noticeably less continuous than that of the classical system, which is due to the fact that the quantum state is conditioned on the classical momentum, which itself is experiencing fluctuations due to the environment.

It is also insightful to see how the wavefunction evolves along an individual trajectory, and how the quantum state is decomposed in the adiabatic basis. In the middle panel of Figure \ref{fig: trajectories} we plot $q_t$, and additionally plot the real and complex parts of the wavefunction $\psi(x,t)=\langle q|\psi\rangle_t$ at equal instances in time, for an initially excited state beginning in $|\psi^{\mathfrak{r}}\rangle_0=|4\rangle$ i.e. the third excited adiabatic state. The number of components for which the probability of occupation $|\langle n|\psi^\mathfrak{r}\rangle|^2$ is higher than $0.1$ is indicated on the right hand side. We see here that the dynamics damps the excited components of the wavefunction, with the state after $t=4$ being predominantly made up of the adiabatic ground state $|\psi^\mathfrak{r}\rangle=|0\rangle$.

Finally, while each individual classical trajectory will differ due to fluctuations, it is possible to see the effect of quantum backreaction on the classical system by averaging the trajectories over many realisations. In the right hand panel of Figure \ref{fig: trajectories} we plot both $\mathbb{E}[q_t]$ and $\mathbb{E}[\langle\psi| \hat{q} |\psi \rangle_t]$ i.e. the ensemble averages over trajectories. Choosing the parameter $\Omega$ to be small as in the previous simulations such that the purely classical potential is negligible, we see from this plot that the quantum backreaction has a non-trivial effect on the mean classical evolution, with the two systems appearing to accelerate back and forth towards each other. The damping of the oscillations occurs due to the friction in the classical system, which dissipates the original energy stored in the interaction with the quantum oscillator.

\subsection{Thermalisation} \label{subsec: thermalisation}

The dynamics describes a quantum system that remains pure conditioned on the classical trajectory. It is perhaps surprising therefore, that the joint classical-quantum system simultaneously appears to demonstrate thermalisation at the level of the classical-quantum state.

To begin, we first compute the form of the thermal state in the relative position representation. In particular, one may check that it takes the particularly simple form
\begin{equation} \label{eq: thermal_state_oscillators_rel_pos}
     \hat{\pi}^{\mathfrak{r}}(q,p)=\frac{1}{\mathcal{Z}}e^{-\beta \hat{H}_Q}e^{-\beta H_C(q,p)},
\end{equation} where here $\mathcal{Z}$ is the same as in Eq. \eqref{eq: Z_oscillator}. Since this expression factorises, we see by comparison to \eqref{eq: CQ_state_rel_position} that we may check that the classical-quantum state $\hat{\varrho}(q,p)$ approaches the thermal state $\hat{\pi}(q,p)$ by checking in the relative position representation that 
\begin{equation} \label{eq: thermalisation_condition_1}
    \lim_{t\rightarrow\infty}\mathbb{E}[|\psi^\mathfrak{r}\rangle_t\langle \psi^{\mathfrak{r}}|_t]=\frac{1}{\mathcal{Z}_Q} e^{-\beta \hat{H}^\mathfrak{r}_Q}
\end{equation}
\begin{equation} \label{eq: thermalisation_condition_2}
    \lim_{t\rightarrow\infty}\mathbb{E}[\delta(q-q_t)\delta(p-p_t)]=\frac{e^{-\beta H_C(q,p)}}{\mathcal{Z}_C}
\end{equation}
\begin{equation}\label{eq: thermalisation_condition_3}
\lim_{t\rightarrow\infty}\hat{\varrho}^{\mathfrak{r}}(q,p)=\mathbb{E}[|\psi^\mathfrak{r}\rangle_t\langle \psi^{\mathfrak{r}}|_t]\ \mathbb{E}[\delta(q-q_t)\delta(p-p_t)]
\end{equation} i.e. that the reduced quantum and classical distributions approach the standard thermal states of the corresponding classical and quantum oscillators, and that the correlations between the classical and quantum degrees of freedom vanish. 

\begin{figure}[!tbp]
    \centering
    \begin{subfigure}[t]{0.45\textwidth}
    \includegraphics[width=\textwidth]{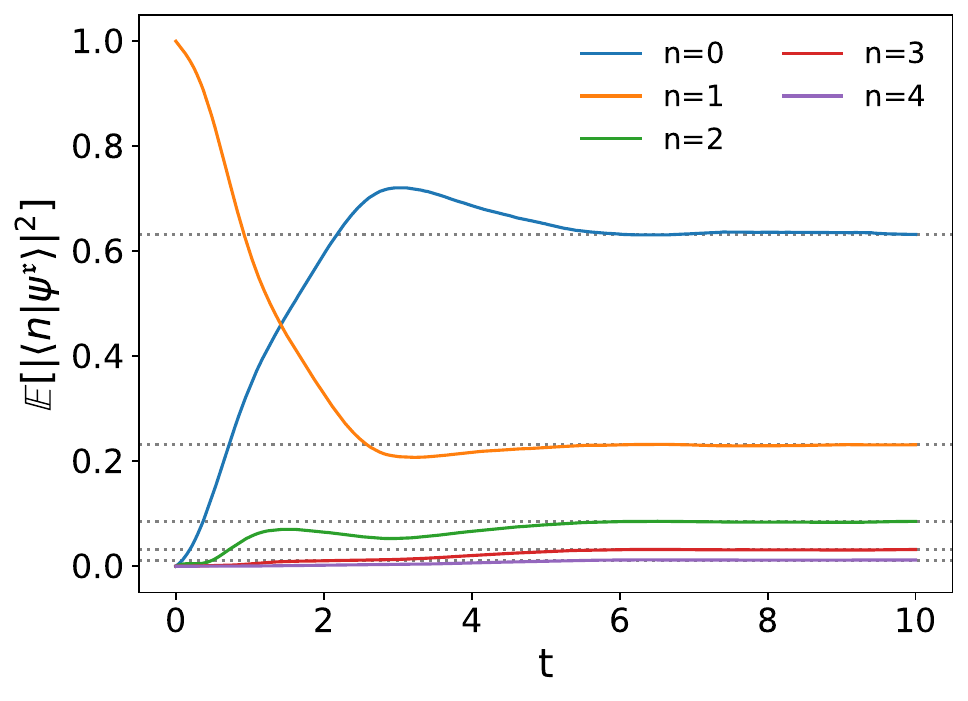}
    \end{subfigure}
    \begin{subfigure}[t]{0.45\textwidth}
    \includegraphics[width=\textwidth]{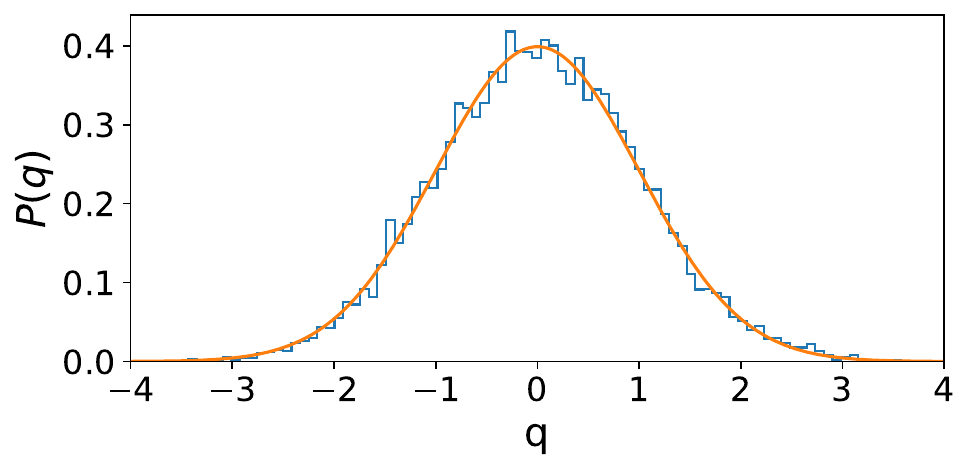}
    \end{subfigure}
     \begin{subfigure}[t]{0.45\textwidth}
    \includegraphics[width=\textwidth]{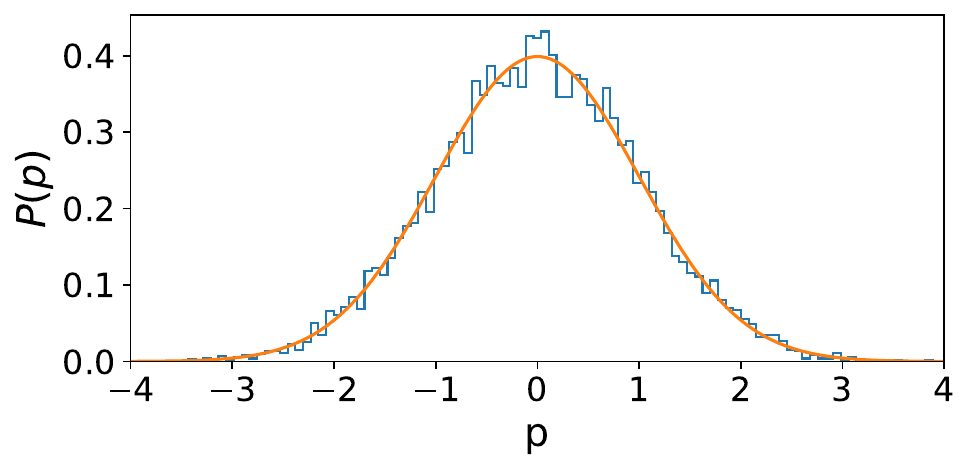}
    \end{subfigure}
    \caption{\small{Three plots demonstrating the thermalisation of the classical-quantum state over $10^4$ numerically simulated trajectories. Here the initial quantum state is in the first excited state $|\psi^{\mathfrak{r}}\rangle_0=|1\rangle$, while the classical system is in a Gaussian probability distribution centered at the origin with standard deviations $\sigma_q=\sigma_p=10^{-3}$. The model parameters are given $\omega=m_C=m_Q=\hbar=\beta=\gamma=\Omega=1$, while the numerical simulation has $N_{max}=10$ and $N_{steps}=5000$, and is run from from $t=0$ to $t=10$. \textit{(top)} The average populations in the adiabatic basis of the density operator in the relative position representation are shown up to the 4th excited state, which appear to converge to the corresponding values predicted from Eq. \eqref{eq: thermalisation_condition_1}. \textit{(middle and bottom)} The marginal distributions (solid orange line) of the classical thermal state appearing in Eq. \eqref{eq: thermalisation_condition_2}  are plotted against a histogram of the numerically simulated final position and momentum of the classical system.}}
    \label{fig: thermalisation}
\end{figure}

Considering first the quantum dynamics, we first plot in the top of Figure \ref{fig: thermalisation} the numerically simulated ensemble average of the populations in the adiabatic basis i.e. $\mathbb{E}[|\langle n|\psi^\mathfrak{r}\rangle_t|^2]$. Starting in the state $|\psi^\mathfrak{r}\rangle_0=|1\rangle$, we find that these approach the corresponding populations of the thermal state of the quantum harmonic oscillator. Despite the quantum state along any trajectory remaining pure, the diagonal part of the density matrix $\hat{\rho}^\mathfrak{r}(t)=\mathbb{E}[|\psi^\mathfrak{r}\rangle_t\langle \psi^\mathfrak{r}|_t]$ thus appears to thermalise in the adiabatic basis. Turning to the classical dynamics, we verify that the marginals of the classical thermal state are reached by plotting histograms of the position and momentum against the theoretical prediction, as shown in the middle and bottom of Figure \ref{fig: thermalisation}.  

While the above provides a strong indication of thermalisation, it is necessary to also check the remaining properties: (i) that the coherences of the quantum state in the adiabatic basis are also damped in time (ii) that the correlations between position and momentum vanish (iii) that correlations between the classical and quantum degrees of freedom vanish. We explore these in Appendix \ref{app: thermalisation_additional}, and find clear evidence that the dynamics also causes coherences in the adiabatic basis to relax to zero, as well as the correlations (in the relative position representation) between the various degrees of freedom.

While the above numerical evidence does not rule out the existence of particular initial states which do not thermalise, it does suggest strongly that typical initial configurations  relax to the thermal state given in Eq. \eqref{eq: thermal_state_oscillators}. Proving this rigorously, such as using techniques from the theory of stochastic differential equations to demonstrate that Eqs. \eqref{eq: thermalisation_condition_1} to  \eqref{eq: thermalisation_condition_3} indeed hold, is an interesting question that we leave to future work.

\subsection{Heat fluctuations and the second law}

Having developed both a numerical framework for simulating trajectories, and establishing that the model we describe appears to demonstrate thermalisation, we finally turn to combine these to demonstrate how the model illustrates fluctuations in the heat it exchanges with the environment, and how the average value of these fluctuations are bounded by the second law of thermodynamics.

We begin by computing the entropy of a particular initial classical-quantum configuration.  We first note that since the classical-quantum entropy in \eqref{eq:entropy} is invariant under phase-space dependent unitary transformations, we may compute the entropy directly in the relative position basis.  Assuming the quantum system is in a pure quantum state in the relative position representation $|\psi^\mathfrak{r}\rangle_0$ and the classical probability distribution is normally distributed around the origin in phase space with variances $\sigma_q^2$ and $\sigma_p^2$, the combined system is uncorrelated between the classical and quantum degrees of freedom, and thus the entropy is simply a sum of the entropies of the classical and quantum degrees of considered separately. Since the quantum state is initially pure, the quantum entropy is zero. Computing the classical part of the entropy is here well-defined since the classical probability distribution is Gaussian. This gives the initial entropy of the classical-quantum system as
\begin{equation}
S(\hat{\varrho}_0)=1+\ln{2\pi \sigma_q \sigma_p}.
\end{equation} For an arbitrary final state, one may compute the entropy to find $\Delta S$. Here we shall assume we are considering timescales long enough that the system has relaxed to the thermal state given in Eq. \eqref{eq: thermal_state_oscillators}. Again utilising the fact that the entropy $S$ is invariant under unitary transformations, we may compute the entropy of $ \hat{\pi}^\mathfrak{r}$, the thermal state in the relative position representation, which takes the simple uncorrelated form given in Eq. \eqref{eq: thermal_state_oscillators_rel_pos}. As before, it is straightforward to compute the entropy here as the sum of the classical and quantum contributions
\begin{equation}
S( \hat{\pi})=1+\ln{\frac{4\pi}{\beta \Omega}}+\frac{\hbar \omega \beta}{2}\coth{\frac{\hbar \omega \beta}{2}}-\ln{\big[2\sinh{\frac{\hbar \omega \beta}{2}}\big]},
\end{equation} which gives the total change in entropy as 
\begin{equation}
    \Delta S= \ln{2}-\ln{\beta \Omega \sigma_q \sigma_p} +\frac{ \hbar  \omega \beta}{2}\coth{\frac{\hbar \omega \beta}{2}}-\ln{\big[2\sinh{\frac{\hbar \omega \beta}{2}}\big]}.
\end{equation} From Eq. \eqref{eq: clausius_inequality_2nd_law} and Eq. \eqref{eq: heat_ensemble_and_fluctuations}, we see that this provides a bound on the ensemble average of the heat exchanged into the system on along trajectory, given these initial conditions.

\begin{figure}
    
    \centering
    \begin{subfigure}[t]{0.45\textwidth}
    \includegraphics[width=1\linewidth]{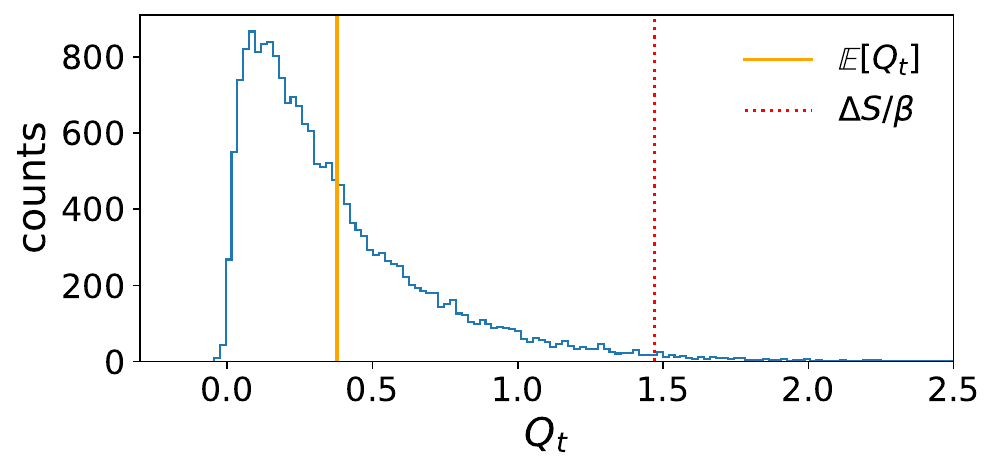}
    \end{subfigure}
    \begin{subfigure}[t]{0.45\textwidth}
        \includegraphics[width=1\linewidth]{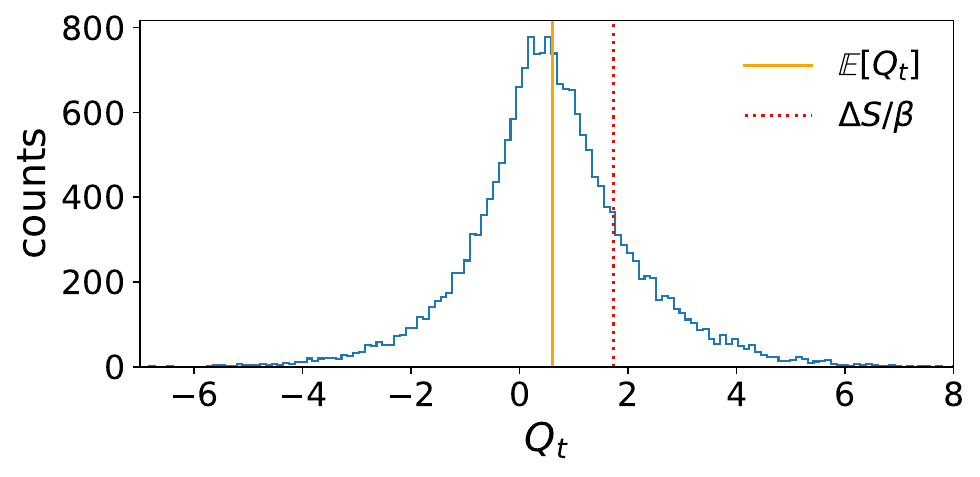}
    \end{subfigure}
    \caption{\small{Distributions of the heat exchanged into the classical-quantum system as the system thermalises, with model parameters $\omega=m_C=m_Q=\hbar=\beta/\gamma=\Omega=1$. \textit{(top)} Heat distribution with $\beta=\gamma=3$,  $\sigma_q=\sigma_p=0.1$, $N_{max}=10$, $N_{steps}=5000$ over $2\times 10^4$ trajectories simulated from $t=0$ to $t=10$. \textit{(bottom)} Heat distribution with $\beta=\gamma=1$, $\sigma_q=\sigma_p=1$, $N_{max}=10$, $N_{steps}=10^4$ over $2\times 10^4$ trajectories simulated from $t=0$ to $t=20$. }}
    \label{fig: heat_distribution}
\end{figure}

We may now compare the bound on heat exchange derived from the second law to the statistics of the heat exchange along individual trajectories. In the top of Figure \ref{fig: heat_distribution} we plot the distribution of heat $Q_t$, as defined in Eq. \eqref{eq: stochastic_heat}, for an initial state in which $|\psi^\mathfrak{r}\rangle_0=|0\rangle$ i.e. in which the quantum system starts in the adiabatic ground state, and where the classical system is close to its ground state i.e. the origin in phase space. As expected, the vast majority of realisations lead to the system gaining energy, with the mean energy increase given by the vertical solid line. This can be seen to be lower than the maximum upper limit of $\mathbb{E}[Q_t]$ allowed by the second law, which is marked by the dotted line. While the average satisfies the second law, individual realisations may gain more heat than their mean value, as evidenced by the small but non-zero set of counts to the right of the dashed line, which appear to fall off exponentially. In the bottom half of Figure \ref{fig: heat_distribution} we plot the heat distribution at a higher temperature and lower friction, in the same quantum state but with a classical state with greater variance in phase space. Here we see more pronounced fluctuations of energy increase above the entropy bound, though a greater number of realisations dissipating heat into the environment ensure than the average satisfies the second law.

\section{Detailed balance} \label{sec: DB}

While an equilibrium state is a minimal requirement for thermodynamic stability, a wide range of open systems in both classical and quantum regimes satisfy a stronger notion of \textit{detailed balance}.  In this section we extend the concept of detailed balance to hybrid classical-quantum systems, deriving the general sufficient and necessary conditions under which a classical-quantum system satisfies detailed balance. In doing so, we completely characterise both measurement-based feedback, and hybrid models of molecular dynamics that have linear master equations with well-defined trajectories, that satisfy detailed balance. Using these conditions, we show that the two classes of dynamics introduced in Section \ref{sec: construct} both satisfy classical-quantum detailed balance.

\subsection{Alternative representation of the classical-quantum generator and the classical-quantum adjoint}

Before starting our discussion of detailed balance in classical-quantum systems, we first introduce two important forms of classical-quantum generators that will be needed in the following sections.  

The first is an alternative representation of the generator of completely-positive classical quantum dynamics.  Letting $K_i$ denote phase-space dependent operators associated to each classical degree of freedom $z_i$, $\tilde{L}_\alpha$ denote traceless phase-space Lindblad operators, and $G$ denoting a generic phase-space dependent operator, we may write the general classical-quantum generator in the form
\begin{equation} \label{eq: L_alternate}
\begin{split}
\mathcal{L}(\varrho)=& -\frac{i}{\hbar}[G,  \varrho ]- \frac{\partial }{\partial z_{i}} \left( K_i  \varrho +  \varrho K_i^\dag \right) \\
&+\frac{1}{2} \frac{\partial^2 }{\partial z_i \partial z_j} ( D_{2, i j}  \varrho )\\
&+ {D_{2,ij}^{-1}} \big( K_i  \varrho K_j^{\dag} - \frac{1}{2}  \{ K_j^{\dag} K_i,  \varrho \}_+ \big) \\
& + \tilde{L}_\alpha\varrho \tilde{L}_\alpha^\dag - \frac{1}{2}\{\tilde{L}_\alpha^\dag \tilde{L}_\alpha,\varrho \}_+
\end{split}
\end{equation} which is completely positive \textit{if and only if} 
\begin{equation} \label{eq: alt_pos_conditions_1}
     \quad D_2\succeq 0, \quad G=G^\dag
\end{equation} and there exists a complex-valued phase space dependent vector $v$ of length $n$ such that if $K$ denotes the operator valued vector $K=(K_1,\ldots, K_n)^T$ then
\begin{equation} \label{eq: alt_pos_conditions_2}
    (\mathbb{I}-D_2^{-1}D_2)K=v\mathds{1}.
\end{equation} We show the equivalence of this generator and its positivity conditions to Eqs. \eqref{eq: L_general} to  \eqref{eq: pos_2} in Appendix \ref{app: L_equivalent}. 

In this representation, the Hermitian operator $G$ determines the purely unitary quantum evolution. To avoid confusion with the classical-quantum Hamiltonian, which $G$ does not necessarily coincide with, we refer to this operator as the generator of unitary dynamics, or unitary generator for short. The back-reaction on each classical degree of freedom $z_i$ is determined by phase-space dependent operators $K_i$, which we shall refer to as the ``backreaction operators". Rather than assuming these operators are traceless, and describing purely classical evolution by a separate drift term, the component of $K_i$ proportional to the identity operator describes the purely classical drift on the $i$th classical degree of freedom. The necessary decoherence corresponding to this backreaction is encoded in the decoherence term involving $K_i$ and the pseudoinverse of the diffusion matrix $D_2$. Finally, the Lindblad operators $\tilde{L}_\alpha$ determine any additional decoherence on top of the minimum required by the coupling to the classical system. These are thus necessarily zero if the decoherence-diffusion trade-off is saturated i.e. when the unravelling preserves the purity of initially pure quantum states. 

The advantage of this representation is that sufficient and necessary conditions for complete-positivity may be checked without relying on a decomposition of the back-reaction operators in an orthonormal basis of operators. It turns out that this is extremely useful for checking detailed balance, as it allows operators such as the $L_z$ to be studied without needing to compute their explicit forms (such as was done for the classical-quantum oscillators model in Sec. \ref{subsec: explicit_comp_of_L_oscillator}). Aside from this, this form of dynamics provides some conceptual simplicity, such as explicitly separating out the decoherence necessary for positivity from additional kinds, and making it explicit which operator back-reacts on each classical degree of freedom.

The second generator we shall define is the adjoint generator of classical-quantum dynamics. As with the adjoint of the Fokker-Planck equation \cite{risken1972solutions} or the GKLS equation \cite{alicki1976detailed}, one may define the adjoint in the classical-quantum setting by studying the evolution of classical-quantum observables. Considering the definition of the expectation value given in \eqref{eq: CQ_expectation_value}, we define the adjoint of the generator $\mathcal{L}$ as the classical-quantum superoperator $\mathcal{L}^\dag$ that satisfies
\begin{equation}
    \int dz \tr{\mathcal{L}(\varrho)A}=\int dz \tr{\varrho \mathcal{L}^\dag(A)},
\end{equation} i.e. that governs the evolution of classical-quantum observables in the Heisenberg representation. Assuming that the classical-quantum state $\varrho$ satisfies vanishing boundary conditions at infinity, it is straightforward using integration by parts and properties of the trace to show that this can be written in the general form
\begin{equation} \label{eq: L_adjoint}
\begin{split}
\mathcal{L}^\dag(A)=& \ \frac{i}{\hbar}[G,  A ]+ K_i^\dag \frac{\partial A}{\partial z_{i}} + \frac{\partial A}{\partial z_{i}} K_i     \\
&+\frac{1}{2}D_{2, i j}  \frac{\partial^2 A}{\partial z_i \partial z_j}\\
&+ {D_{2,ij}^{-1}} \big( K_i^\dag  A K_j - \frac{1}{2}  \{ K_j^{\dag} K_i,  A \}_+ \big) \\
& + \tilde{L}_\alpha^\dag A \tilde{L}_\alpha - \frac{1}{2}\{\tilde{L}_\alpha^\dag \tilde{L}_\alpha, A \}_+.
\end{split}
\end{equation} As should be expected, this dynamics takes the form of a generalisation of the standard classical and quantum adjoint dynamics, and as such satisfies many of the same properties. A particular property that is useful to note is that this map is unital i.e. that $\mathcal{L}^\dag(\mathds{1})=0$, since in this case the derivative, dissipator and commutator terms all vanish.

\subsection{Defining detailed balance in classical-quantum systems}

In this section we begin our discussion of classical-quantum detailed balance, providing a general definition of detailed balance in hybrid systems. This definition arises as a natural generalisation of the definitions of detailed balance of classical and quantum systems when characterised in terms of their generators, and we refer the reader to \cite{graham1971generalized,risken1972solutions,fagnola2007generators} for further details on this topic.

To define detailed balance in the classical-quantum setting, we must first introduce the concept of time-reversal. Taking a video of a particle in flight, and playing it in reverse, one sees that while the position $q$ of the particle remains unchanged, the momentum $p$ reverses sign. In a similar manner, for any set of classical variables $z=(z_1,\ldots,z_n)$, a subset known as the even variables will remain constant under time-reversal i.e. $z_i\mapsto z_i$, while the odd variables will each be multiplied by minus one i.e. $z_i\mapsto -z_i$. To capture this, we follow \cite{risken1972solutions} and use the notation
\begin{equation}
   \epsilon z=(\epsilon_1 z_1,\ldots \epsilon_n z_n)
\end{equation} where here $\epsilon_i=\pm1$ depending on whether the $i$th classical variable is even or odd. Using this notation, a function $f(z)$ of the classical variables is thus transformed under time-reversal to $f(\epsilon z)$. To denote the time-reversal of differential operators, such as the generator of dynamics $\mathcal{L}$, we use the shorthand notation $\mathcal{L}_\epsilon$, which implies changing all occurrences of $z_i$ to $\epsilon_i z_i$, including in derivatives.

An important feature of detailed balance is that it is always defined with respect to a given state. Although up to now we have focused on classical-quantum fixed points $\pi$ that are of the standard thermal form $\pi\propto\exp(-\beta H)$, for the remainder of our discussion of detailed balance we allow $\pi$ to be a generic fixed point. Aside from generalising the discussion, this is also a natural choice given the earlier form of our dynamics, where the key operators $L_z$ and $M_{xy}$ depended explicitly on $\pi$ rather than the $H.$ For this section, we therefore allow $\pi$ to be a generic fixed point, that may be written generally in the form
\begin{equation} \label{eq: pi_general_def}
    \pi(z)=\frac{1}{\mathcal{Z}}e^{-\Phi(z)},
\end{equation} where $\Phi(z)$ is a Hermitian operator that we refer to as the potential. In order to define detailed balance with respect to this state, we shall need to make three assumptions that are standard in the classical and quantum cases \cite{graham1971generalized,risken1972solutions, fagnola2007generators}. Firstly, we will assume that $\pi$ has boundary conditions such that it vanishes at infinity in the classical configuration/phase space; this ensures that the state can be normalised, and that the representation of the adjoint generator introduced in \eqref{eq: L_adjoint} may be used. Secondly, we assume that the state $\pi$ is invertible at every point in phase space i.e. the quantum degrees of freedom must always have a non-zero probability of being in an arbitrarily excited state. Finally will assume that $\pi$ is invariant under time-reversal i.e. $\pi(\epsilon z)=\pi(z)$, as is the case for the thermal state for time-reversal invariant Hamiltonians.

Having defined the notion of time-reversal and the requirements of the fixed point $\pi$, we may define detailed balance in the hybrid setting. A classical-quantum dynamics described by the generator $\mathcal{L}$ will be said to satisfy detailed balance with respect to $\pi$ if and only if there exists a Hermitian, time-reversal invariant operator $X$ that commutes with $\pi$ such that
\begin{equation} \label{eq: classical-quantum_DB}
\pi^{-1/2}\mathcal{L}(\pi^{1/2} A \pi^{1/2}) \pi^{-1/2}=\mathcal{L}_{\epsilon }^\dag (A) - 2i [X,A]
\end{equation} for all operator-valued functions of phase space $A$. 

To motivate this definition as the correct definition of detailed balance in the classical-quantum setting, we consider two limiting cases. Firstly, taking $\pi$ and $A$ to be proportional to the identity operator, one may check that this definition reduces to the requirement that
\begin{equation} \label{eq: classical_DB}
    \pi^{-1}(z)\mathcal{L}(\pi(z) f(z))=\mathcal{L}_\epsilon^\dag(f(z))
\end{equation} holds for all functions $f(z)$, where here $\mathcal{L}$ denotes a general generator of Fokker-Planck dynamics. This equation, alongside the assumption that $\pi$ is invariant under time-reversal, exactly coincides with the definition of detailed balance in the classical setting, when using the characterisation of detailed balance in terms of generators developed by Risken \cite{risken1972solutions}. Secondly, taking $\pi$ and $A$ to have no dependence on $z$, we find that the detailed balance condition reduces to \begin{equation} \label{eq: quantum_DB}
    \pi^{-\frac{1}{2}}\mathcal{L}(\pi^{\frac{1}{2}}A\pi^{\frac{1}{2}})\pi^{-\frac{1}{2}}=\mathcal{L}^\dag(A)-2i[X,A]
\end{equation} for all operators $A$, where here $X$ is any Hermitian operator such that $[X,\pi]=0$. This provides a definition of detailed balance in the quantum setting \cite{fagnola2007generators}. Although other definitions of detailed balance are available in the quantum setting due to alternative operator orderings of $ \pi$, this ``symmetric" definition is the weakest possible, and thus encompasses the broadest class of dynamics. For open quantum systems the symmetric detailed balance condition emerges naturally from the requirement of thermodynamic consistency \cite{soret2022thermodynamic}, and from rigorous treatments of weakly-coupled many-body systems  \cite{scandi2025thermalization}. 

From the above, it is straightforward to see that the definition of classical-quantum detailed balance we provide is a straightforward generalisation of these two definitions, generated by combining the uniquely classical and quantum structures associated to each. Namely, by using the left-hand and right-hand sides of definition \eqref{eq: quantum_DB} which respect operator ordering, while including the time-reversal operation arising in definition \eqref{eq: classical_DB} we arrive at the definition of detailed balance provided in \eqref{eq: classical-quantum_DB}. The only ambiguity that arises is whether $X$ is chosen to be invariant under time-reversal or not, since $X$ and time-reversal only simultaneously appear in the full classical-quantum case. In our definition we opt for the former, which ultimately is supported by the identification of $X$ with the classical-quantum Hamiltonian $H$ when we come to study detailed balance for the main dynamics introduced in Sec. \ref{sec: construct}.

Finally, we note an important feature of classical-quantum detailed balance that connects this concept to the earlier sections of this paper. Taking $A$ to be the identity operator, and independent of $z$, we see that the left hand side of \eqref{eq: classical-quantum_DB} reduces to $\pi^{-\frac{1}{2}}\mathcal{L}(\pi)\pi^{-\frac{1}{2}}$. Doing the same for the right hand side, we see that the commutator with $X$ vanishes, leaving simply $\mathcal{L}^\dag_\epsilon(\mathds{1})$. Since the adjoint map under time-reversal remains unital, we see that the entirety of the right hand side vanishes. Acting with $\pi^{1/2}$ on either side of the lefthand side, we thus see that that $\mathcal{L}(\pi)=0$. It therefore follows that classical-quantum detailed balance implies thermal state preservation i.e.
\begin{equation}
    \begin{matrix}
     \mathcal{L} \ \textrm{satisfies DB}  \\
     \textrm{w.r.t. }\pi
 \end{matrix} \implies \mathcal{L}(\pi)=0.
\end{equation} Understanding how in practice to relate this condition to the thermal state preserving dynamics we have already provided in terms of $L_z$ and $M_{xy}$ will be the main aim of the rest of this section.

\subsection{Conditions for purity-preserving dynamics} \label{subsec: pure_DB_conditions}

In this section we derive conditions on the basic components of the classical-quantum master equation, such as the diffusion matrix and back-reaction operators, that are satisfied if and only if a dynamics satisfies classical-quantum detailed balance. To derive this general set of conditions, we first restrict ourselves to the subset of dynamics that saturate the decoherence-diffusion trade-off – this is later generalised in  Section \ref{subsec: general_DB_conditions} to cases where the quantum state does not remain pure conditioned on the classical trajectory.

In order to derive a set of sufficient and necessary conditions, we first define a set of operators that are characterised by how they transform under time-reversal. In particular, we may define the symmetric and anti-symmetric parts of the Hamiltonian as
\begin{align} \label{eq: symmetric_G_def}
    G^{S}(z)&=\frac{G(z)+G(\epsilon z)}{2}\\
    G^{A}(z)&=\frac{G(z)-G(\epsilon z)}{2}
\end{align}
and the reversible and irreversible parts of the classical drift and backreaction operators $K_i$ as
\begin{align}
K_i^{rev}(z)&=\frac{K_i(z)-\epsilon_i K_i(\epsilon z)}{2}\\
K_i^{irr}(z)&=\frac{K_i(z)+\epsilon_i K_i(\epsilon z)}{2}. \label{eq: irreversible_K_def}
\end{align} These operators satisfy the following relations
\begin{align} \label{eq: time_reversal_relations_beginning}
    G^{S}(\epsilon z)&=G^{S}(z)\\
    G^{A}(\epsilon z)&=-G^{A}(z)\\
    {K_i^{rev}}(\epsilon z)&=-\epsilon_i {K_i^{rev}}(z)\\
    {K_i^{irr}}(\epsilon z)&=\epsilon_i {K_i^{irr}}(z),\label{eq: time_reversal_relations_end}
\end{align} and hence allow the dynamics under time-reversal to be written in terms of quantities dependent on $z$ rather than $\epsilon z$.

Having defined this alternate set of operators, we are now able to derive necessary and sufficient conditions on these in order for classical-quantum detailed balance to hold. Since we are initially considering dynamics that saturates the
trade-off i.e. that preserves the purity of quantum states
in the unravelling, we first assume that all of the Lindblad
operators responsible for additional decoherence vanish i.e.
$\tilde{L}_\alpha=0$ for all $\alpha$. Rewriting this simplified form of the generator $\mathcal{L}$ given in \eqref{eq: L_alternate} in terms of $K_i^{rev}$,$K_i^{irr}$,$G^S$ and $G^A$, it is straightforward to compute the form of $\mathcal{L}^\dag_\epsilon$ by performing the time-reversal operation on the $\mathcal{L}^\dag$ and using the relations  given in \eqref{eq: time_reversal_relations_beginning} to \eqref{eq: time_reversal_relations_end}.
Substituting in the forms of $\mathcal{L}$ and $\mathcal{L}_\epsilon$ into the definition of detailed balance \eqref{eq: classical-quantum_DB}, we find a sum of terms containing $A$, $\partial_i A$ and $\partial_i \partial_j A$ that must equal zero. Since these must hold for all $A$, the Hermitian and anti-Hermitian parts of each expression containing different derivatives of $A$ lead to independent conditions on the dynamics for detailed balance. Using the positivity conditions \eqref{eq: alt_pos_conditions_1} and \eqref{eq: alt_pos_conditions_2}, in Appendix \ref{app: DB_derivation} we show that this leads to the following sufficient and necessary conditions for classical-quantum dynamics that saturate the decoherence-diffusion trade-off to also satisfy detailed balance 
\begin{equation} \label{eq: diffusion_constraint}
{D_{2,ij}}(\epsilon z)=\epsilon_i \epsilon_j D_{2,ij} (z)
\end{equation}
\begin{equation} \label{eq: K_rev_constraint}
    K_i^{rev} \pi^{1/2}   - \pi^{1/2} {K_i^{rev}}^\dag =  i a_i \pi^{1/2},
\end{equation} 
\begin{equation} \label{eq: K_irr_constraint}
    K_i^{irr} \pi^{1/2} + \pi^{1/2} {K_i^{irr}}^\dag=\frac{1}{2}\frac{\partial{D_{2,ij}}}{\partial z_j}\pi^{1/2}+ D_{2,ij}\frac{\partial \pi^{1/2}}{\partial z_j}
\end{equation}
 
\begin{equation} \label{eq: symmetric_effective_H}
    -\frac{i}{\hbar}\{G^{S},\pi^{1/2}\}_+=O^{S}-i\{X,\pi^{1/2}\}_+ 
\end{equation}
\begin{equation} \label{eq: antisymmetric_effective_H}
    -\frac{i}{\hbar}[G^{A},\pi^{1/2}]=O^{A},
\end{equation} where here $a_i$ are the elements of a real vector $a(z)$ that satisfy $a_i(\epsilon z)=-\epsilon_i a_i(z)$, $X$ is a Hermitian operator that is symmetric under time-reversal and commutes with $\pi$, and the operators $O^{S}$ and $O^{A}$ are defined as
\begin{equation}
\begin{split} \label{eq: OS_operator_def}
    O^{S}=&\ \frac{1}{2}\frac{\partial}{\partial z_i}(K_i^{irr}\pi^{1/2}-\pi^{1/2}{K_i^{irr}}^\dag)\\
    &-\frac{1}{2}(\mathbb{I}-D_2 D_2^{-1})_{ij}(K_i^{irr}\frac{\partial \pi^{\frac{1}{2}}}{\partial z_j}-\frac{\partial \pi^{\frac{1}{2}}}{\partial z_j}{K_i^{irr}}^\dag) \\
    &-\frac{1}{4}D_{2,ij}^{-1}\frac{\partial D_{2,ik}}{\partial z_k}(K_j^{irr} \pi^{1/2}- \pi^{1/2}{K_j^{irr}}^\dag) \\
    &+\frac{i}{2}D_{2,ij}^{-1}a_i (K_i^{rev} \pi^{1/2} + \pi^{1/2} {K_i^{rev}}^\dag)\\
    &+\frac{1}{2}D_{2,ij}^{-1}[{K_i^{rev}}^\dag K_j^{rev}+{K_i^{irr}}^\dag K_j^{irr},\pi^{1/2}],\\
\end{split}
\end{equation}
and 
\begin{equation}
\begin{split}\label{eq: OA_operator_def}
    O^{A}=&\ \frac{1}{2}\frac{\partial}{\partial z_i}(K_i^{rev}\pi^{1/2}+\pi^{1/2}{K_i^{rev}}^\dag)\\
    &+\frac{1}{2}(\mathbb{I}-D_2 D_2^{-1})_{ij}(K_i^{rev}\frac{\partial \pi^{\frac{1}{2}}}{\partial z_j}+\frac{\partial \pi^{\frac{1}{2}}}{\partial z_j}{K_i^{rev}}^\dag) \\
    &-\frac{1}{4}D_{2,ij}^{-1}\frac{\partial D_{2,ik}}{\partial z_k}(K_j^{rev} \pi^{1/2}+ \pi^{1/2}{K_j^{rev}}^\dag) \\
    &+\frac{i}{2}D_{2,ij}^{-1}a_i (K_i^{irr} \pi^{1/2} - \pi^{1/2} {K_i^{irr}}^\dag)\\
    &+\frac{1}{2}D_{2,ij}^{-1}\{{K_i^{rev}}^\dag K_j^{irr}+{K_i^{irr}}^\dag K_j^{rev},\pi^{1/2}\}_+.
\end{split}
\end{equation}
We refer to these five conditions as the diffusion constraint, reversible back-reaction constraint, irreversible back-reaction constraint, symmetric unitary generator constraint and anti-symmetric unitary generator constraint respectively. These are extended to the case where classical-quantum dynamics does not necessarily saturate the decoherence-diffusion trade-off in the next section. 

Before moving on, we observe some basic consequences of these conditions. Firstly, we see that the diffusion constraint is identical to the known classical condition on the diffusion matrix for detailed balance \cite{graham1971generalized,risken1972solutions} – we shall further investigate how these conditions reduce to the classical case in \ref{subsec: C_and_Q_DBlimit}. Secondly, we see that the reversible and irreversible back-reaction constraints, which state that the back-reaction operators must be related to both $\pi^{1/2}$ and its first derivatives, are reminiscent of those for $L_z$ given in \eqref{eq: L_op_property} and \eqref{eq: L_op_def} respectively. Similarly, we see that as with the definition of the Hermitian $M_{xy}$ operators of \eqref{eq: M_op_def2}, the equation for $G^{S}$ takes the form of a Lyapunov equation, and thus can be directly solved to give
\begin{equation} \label{eq: symmetric_generator_solution}
    G^{S}=i\hbar \int_0^\infty e^{-s\pi^{1/2}} O^{S} e^{-s\pi^{1/2}} ds + X .
\end{equation} Turning now to the antisymmetric unitary generator constraint, it is important to note that while the constraint for $G^S$ purely constraints the quantum unitary evolution, the equation for $G^{A}$ includes a constraint on the form of $K^{irr}$, $K^{rev}$ and $D_2$ independent of $G^A$, namely that
\begin{equation} \label{eq: antisymmetric_generator_constraint_part}
    \langle n(z) | O^A | n(z) \rangle = 0 \ \  \forall n
\end{equation} where $|n(z)\rangle$ is the adiabatic basis defined for the generic fixed point $\pi$ i.e. the eigenstates of the potential operator $\Phi(z)$. Although the solution for $G^{A}$ is not unique, since any time-reversal anti-symmetric Hermitian operator that commutes with $\pi^{1/2}$ may be added to it, the part of the operator that does not commute with $\pi^{1/2}$ may be uniquely solved in terms of the adiabatic basis, giving the off-diagonal elements of $G^A$ in this basis as
\begin{equation} \label{eq: antisymmetric_generator_solution}
    G^A_{nm}=i\hbar \frac{\langle n(z) | O^A | m(z)\rangle }{\sqrt{p_m}-\sqrt{p_n}} \ \ n\neq m
\end{equation} where here $p_n$ is the eigenvalue of $\pi$ corresponding to the adiabatic state $|n(z)\rangle$. Finally, we note that the time-reversal property of the vector $a$ ensures that each of the detailed balance conditions is consistent under time-reversal.

\subsection{Conditions for general dynamics} \label{subsec: general_DB_conditions}

To extend the derivation from the case where the decoherence diffusion trade-off is saturated, to the case where it is not, we exploit a recent result that states that a generic classical-quantum dynamics may always be embedded in a larger classical phase space where the trade-off is saturated \cite{layton2022healthier}. This so called ``temple of the larger phase space" is useful, because it allows us to leverage the expressions previously derived in the comparatively simpler setting where the trade-off is saturated.

To begin with, we take $\mathcal{L}$ to be an arbitrary completely-positive continuous CQ generator i.e. of the form of \eqref{eq: L_alternate} with $\tilde{L}_\alpha$ non-zero, and fix a given fixed point $\pi$ to consider detailed balance with respect to. We then introduce an auxiliary classical degree of freedom $y_\alpha$ for each independent traceless Lindblad operator $\tilde{L}_\alpha$. These classical degrees of freedom may be taken to be even under time reversal symmetry without loss of generality. In this enlarged phase space, we then define a new classical-quantum generator $\tilde{\mathcal{L}}$, which takes the form
\begin{equation}
\begin{split} \label{eq: L_purification_def_1}
    \tilde{\mathcal{L}}(A)=\mathcal{L}(A)+ \frac{1}{2}\frac{\partial^2 A}{\partial y_\alpha\partial y_\alpha}-\frac{\partial}{\partial y_\alpha}(\tilde{K}_\alpha A + A \tilde{K}_\alpha^\dag),
\end{split}
\end{equation} where the traceless auxiliary back-reaction operators $\tilde{K}_\alpha$ have no dependence on the auxiliary degrees of freedom $y_\alpha$, and satisfy
\begin{equation} \label{eq: L_purification_def_2}
    \mathcal{D}(\tilde{K}_\alpha) =\mathcal{D}(\tilde{L}_\alpha),
\end{equation}
and
\begin{equation} \label{eq: L_purification_def_3}
    \tilde{K}_\alpha(z)\pi^{\frac{1}{2}}+\pi^{\frac{1}{2}}\tilde{K}_\alpha^\dag(\epsilon z) = 0.
\end{equation} 
 The first of these conditions, \eqref{eq: L_purification_def_1} guarantees that the generator $\mathcal{\tilde{L}}$ recovers $\mathcal{L}$ upon tracing over the auxiliary degrees of freedom. The second condition \eqref{eq: L_purification_def_2} ensures that the $\tilde{\mathcal{L}}$ is both completely-positive and saturates the decoherence-diffusion trade-off, by requiring the GKLS dissipator $\mathcal{D}$ corresponding to $\tilde{L}_\alpha$ to be the same as that constructed from $\tilde{K}_\alpha$. Finally, \eqref{eq: L_purification_def_3} amounts to stating that the purification we have chosen is not arbitrary, but is adapted to the choice of fixed state $\pi$ that we wish to prove the detailed balance of $\mathcal{L}$ with respect to.

Turning now to detailed balance, it is straightforward to prove using \eqref{eq: L_purification_def_1} and \eqref{eq: L_purification_def_3} that 
\begin{equation}
    \pi^{-\frac{1}{2}}\tilde{\mathcal{L}}(\pi^{\frac{1}{2}}A\pi^{\frac{1}{2}})\pi^{-\frac{1}{2}}-{\tilde{\mathcal{L}}_\epsilon}^\dag(A)=\pi^{-\frac{1}{2}}\mathcal{L}(\pi^{\frac{1}{2}}A\pi^{\frac{1}{2}})\pi^{-\frac{1}{2}}-{\mathcal{L}}_\epsilon^\dag(A).
\end{equation} An immediate consequence of this is that
\begin{equation}
 \begin{matrix}
     \mathcal{L} \ \textrm{satisfies DB}  \\
     \textrm{w.r.t. }\pi
 \end{matrix} \iff  
 \begin{matrix}
     \tilde{\mathcal{L}} \ \textrm{satisfies DB}  \\
     \textrm{w.r.t. }\pi
 \end{matrix}
\end{equation} i.e. that checking detailed balance for $\tilde{\mathcal{L}}$ with respect to $\pi$ is sufficient and necessary to conclude the same for $\mathcal{L}$. 

Since $\tilde{\mathcal{L}}$ saturates the decoherence-diffusion trade-off, we can apply each of the detailed balance conditions already proven in equations \eqref{eq: diffusion_constraint} to \eqref{eq: antisymmetric_effective_H}, this time including the additional diffusion, decoherence and back-reaction associated to the auxiliary classical degrees of freedom $y_\alpha$, to find iff constraints for $\mathcal{L}$ to satisfy detailed balance. In the enlarged space, one can check from \eqref{eq: L_purification_def_1} that condition \eqref{eq: diffusion_constraint} is trivially satisfied on the auxiliary degrees of freedom, and thus provides no additional constraint. Similarly, computing the reversible and irreversible parts of the auxiliary back-reaction operators $\tilde{K}_\alpha$ and substituting into \eqref{eq: K_rev_constraint} and \eqref{eq: K_irr_constraint}, one finds that since $\pi$ is independent of $y_\alpha$ and the $\tilde{K}_\alpha$ are traceless, the constraints on the auxiliary backreaction operators $\tilde{K}_\alpha$ are equivalent to that of \eqref{eq: L_purification_def_3} and that the vector of real numbers $a_\alpha$ is zero. The only additional constraints come from considering the symmetric and anti-symmetric effective Hamiltonians, which in this case take the form
\begin{equation} \label{eq: aux_back_op_constraint_symmetric}
\begin{split}
      -\frac{i}{\hbar}\{G^S,\pi^{1/2}\}_+ =\  & O^S-i\{X,\pi^{1/2}\}_+ \\
    &+\frac{1}{4}[\tilde{K}_\beta^{\dag} \tilde{K}_\beta + \tilde{K}_\beta^{\dag}(\epsilon z)\tilde{K}_\beta(\epsilon z),\pi^{\frac{1}{2}}]
\end{split}
\end{equation}
\begin{equation} \label{eq: aux_back_op_constraint_antisymmetric}
    -\frac{i}{\hbar}[G^A,\pi^{1/2}]=O^A + \frac{1}{4}\{\tilde{K}_\alpha^\dag \tilde{K}_\alpha - \tilde{K}_{\alpha}^\dag(\epsilon
    z)\tilde{K}_{\alpha}(\epsilon z),\pi^{\frac{1}{2}}\}_+,
\end{equation} i.e. are modified by an additional term in terms of $\tilde{K}_\alpha$.

To relate the conditions \eqref{eq: L_purification_def_3}, \eqref{eq: aux_back_op_constraint_symmetric} and \eqref{eq: aux_back_op_constraint_antisymmetric} on $\tilde{K}_\alpha$ to conditions on $\tilde{L}_\alpha$, we utilise \eqref{eq: L_purification_def_2}. In particular, since the two GKLS disspators will be equal iff the Lindblad operators are related via unitary matrix \cite{gisin1992quantum,wiseman2001complete,fagnola2007generators}, we can always rewrite the auxiliary back-reaction operators $\tilde{K}_\alpha$ as
\begin{equation}
    \tilde{K}_\alpha = u_{\alpha \beta } \tilde{L}_\beta
\end{equation} where $u_{\alpha \beta}$ are the coefficients of some unitary matrix $u$ that may depend on phase space. Plugging this relation into the modified constraints \eqref{eq: L_purification_def_3}, \eqref{eq: aux_back_op_constraint_symmetric} and \eqref{eq: aux_back_op_constraint_antisymmetric}, we finally arrive at the final set of necessary and sufficient conditions for detailed balance with respect to $\pi$ as
\begin{equation} \label{eq: diffusion_constraint_full}
    {D_{2,ij}}(\epsilon z)=\epsilon_i \epsilon_j D_{2,ij}(z)
\end{equation}
\begin{equation} 
    K_i^{rev} \pi^{1/2}   - \pi^{1/2} {K_i^{rev}}^\dag = i a_i \pi^{1/2},
\end{equation}  
\begin{equation} 
    K_i^{irr} \pi^{1/2} + \pi^{1/2} {K_i^{irr}}^\dag=\frac{1}{2}\frac{\partial{D_{2,ij}}}{\partial z_j}\pi^{1/2}+ D_{2,ij}\frac{\partial \pi^{1/2}}{\partial z_j}
\end{equation}
\begin{equation} 
      -\frac{i}{\hbar}\{G^S,\pi^{1/2}\}_+=O^S-i\{X,\pi^{1/2}\}_+ +\frac{1}{2}[\tilde{N}^S,\pi^{\frac{1}{2}}]
\end{equation}
\begin{equation}
    -\frac{i}{\hbar}[G^A,\pi^{1/2}]=O^A + \frac{1}{2}\{\tilde{N}^A,\pi^{\frac{1}{2}}\}_+
\end{equation}
\begin{equation} \label{eq: decoherence_constraint}
    \tilde{L}_\alpha(z) \pi^{1/2}   = -(u^\dag(z) u^*(\epsilon z))_{\alpha \beta} \pi^{1/2} \tilde{L}_{\beta}^\dag (\epsilon z),
\end{equation} where here we have defined $\tilde{N}^{S}$ and $\tilde{N}^A$ as the symmetric and anti-symmetric parts under time-reversal of the operator 
\begin{equation}
    \tilde{N}=\tilde{L}_\alpha^{\dag} \tilde{L}_\alpha.
\end{equation}We thus see that when the decoherence-diffusion trade-off is not saturated, the symmetric and anti-symmetric unitary generator constraints found previously are modified to depend on the excess decoherence, while a sixth constraint \eqref{eq: decoherence_constraint}, that we refer to as the decoherence constraint, provides conditions on the Lindblad operators of the excess decoherence that must be satisfied in order for the dynamics to satisfy detailed balance. This new set of conditions satisfies similar properties to before, in particular with analogous relations to \eqref{eq: symmetric_generator_solution}, \eqref{eq: antisymmetric_generator_constraint_part} and \eqref{eq: antisymmetric_generator_solution} that are straightforward to find using the same arguments.

\subsection{Classical and quantum detailed balance} \label{subsec: C_and_Q_DBlimit}

Having found the general form of conditions for a classical-quantum dynamics to satsify detailed balance, we now turn to check that these indeed correctly reproduce the known conditions in the purely classical and quantum cases.

We first turn to the purely classical limit. To do so, we first set $\tilde{L}_\alpha=0$, allowing us to use the characterisation of detailed balance for dynamics that saturates the trade-off given in Eqs. \eqref{eq: diffusion_constraint} to \eqref{eq: antisymmetric_effective_H}. Setting $G=0$, $K_i=\frac{1}{2}D^C_i\mathds{1}$, where $D^C$ is a real-valued vector representing the classical drift, and letting $\pi$ be proportional to the identity, the constraints simplify considerably. We first note that in this case the reversible back-reaction constraint \eqref{eq: K_rev_constraint} is trivially satisfied and implies that $a_i$ is zero for all $z$. Since this means that $O^S$ also vanishes, we see that the symmetric unitary generator constraint is also necessarily satisfied. Turning now to the anti-symmetric unitary constraint, we note while $G^A$ vanishes, the additional condition \eqref{eq: antisymmetric_generator_constraint_part} nevertheless provides a non-trivial condition on the classical dynamics. In particular, we see that since $\pi$ is proportional to the identity, every vector is a valid adiabatic state, and thus that the entire expression for $O^A$ must vanish for detailed balance to hold. Writing this final constraint alongside the diffusion and irreversible back-reaction constraints, and making the appropriate simplifications given the form of $K_i$ and properties of the pseudoinverse, we find the following three constraints in the classical limit
\begin{equation}
    {D_{2,ij}}(\epsilon z)=\epsilon_i \epsilon_j D_{2,ij}(z)
\end{equation}
\begin{equation}
    D^{C,irr}_{1,i} =\frac{1}{2}\frac{\partial{D_{2,ij}}}{\partial z_j}+ D_{2,ij}\frac{\partial \pi^{1/2}}{\partial z_j}\pi^{-1/2}
\end{equation}
\begin{equation}
\begin{split}
    \frac{\partial}{\partial z_i}(D^{C,rev}_{1,i}\pi)=0.
\end{split}
\end{equation} Additionally assuming that $D_2$ is full rank, such that its inverse is well-defined, and using the explicit form of $\pi$ provided in \eqref{eq: pi_general_def}, it is straightforward to check that these conditions coincide with the necessary and sufficient conditions for classical detailed balance found in \cite{graham1971generalized,risken1972solutions}. 

To study the purely quantum dynamics, we take $K=0$ and $D_2=0$, and assume that $\tilde{L}$, $S$ and $\pi$ have no dependence on phase space. In this case the only non-trivial constraints are
\begin{equation} \label{eq: Q_DB_L_constraint}
    \tilde{L}_\alpha \pi^{1/2}   = -(u^\dag u^*)_{\alpha \beta} \pi^{1/2} \tilde{L}_{\beta}^\dag ,
\end{equation} and

\begin{equation} 
\begin{split}
      -\frac{i}{\hbar}\{G,\pi^{1/2}\}_+=&-\frac{i}{\hbar}\{X,\pi^{1/2}\}_+ +\frac{1}{2}[ \tilde{L}_\beta^{\dag}\tilde{L}_\beta,\pi^{\frac{1}{2}}].
\end{split}
\end{equation} Defining the additional unitary matrix $v=u^\dag u^*$, this exactly reproduces the known conditions of \cite{fagnola2007generators}. In fact, the form that we derive makes it explicit that the unitary matrix $v$ must also be symmetric, which follows by taking the complex conjugate of Eq. \eqref{eq: Q_DB_L_constraint}. 

\subsection{Proving detailed balance for $L_z$, $M_{xy}$ dynamics}

Having derived the form of sufficient and necessary conditions for detailed balance in classical-quantum systems and checked it agrees with the known conditions for classical and quantum detailed balance, we finally turn to show that the dynamics presented in Section \ref{sec: construct} satisfies classical-quantum detailed balance.

Considering first the overdamped dynamics, we begin by finding the form of the symmetric and antisymmetric parts of the generator of the unitary dynamics, as well as the reversible and irreversible parts of the back-reaction operators. Comparing the form of \eqref{eq: overdamped} to \eqref{eq: L_alternate} to read off $G$ and $K_x$, and using the definitions \eqref{eq: symmetric_G_def} to \eqref{eq: irreversible_K_def}, we find the dynamics to be entirely characterised by the following operators
\begin{equation}
\begin{split}
    G^S&=H + \frac{\mu \beta}{8}M_{xx}\quad \quad G^A=0\\
    K_x^{rev}=0&\quad \quad K_x^{irr}=-\frac{\mu}{2}L_x \quad  \quad D_2=\frac{2\mu}{\beta}
\end{split}
\end{equation} Here we see that both $H$ and $M_{xx}$ are time-reversal invariant, while the anti-symmetric part of the unitary generator $G^A$ vanishes. As expected for an overdamped particle dynamics, the reversible part of the dynamics also vanishes. 

To check detailed balance in the overdamped dynamics, we use the sufficient and necessary conditions derived in Sections \ref{subsec: pure_DB_conditions} and \ref{subsec: general_DB_conditions}. Since this dynamics saturates the trade-off, in this case we may use the simpler conditions given in \eqref{eq: diffusion_constraint} to \eqref{eq: antisymmetric_effective_H} to check whether detailed balance is satisfied. Since the diffusion coefficient is constant and $x$ is an even variable under time-reversal, it is immediate to see that the diffusion constraint \eqref{eq: diffusion_constraint} is satisfied. Substituting in the definition of $L_x$ given in \eqref{eq: L_op_def} it is straightforward to verify the the irreversible back-reaction constraint \eqref{eq: K_irr_constraint} holds, while $K^{rev}=0$ implies that the reversible back-reaction constraint also holds with $a(x)=0$. To see that the symmetric unitary generator constraint \eqref{eq: symmetric_effective_H} holds, we first compute $O^S$. Since here $D_2$ is full rank, $a$ is zero, and $L_x$ satisfies \eqref{eq: L_op_property}, we see that $O^S$ simplifies to simply $(\mu \beta/16)[L_x^\dag L_x,\pi^{1/2}]$. Comparing the resulting symmetric unitary generator constraint with the definition of $M_{xx}$ provided in \eqref{eq: M_op_def2}, we see that Equation \eqref{eq: symmetric_effective_H} holds, with the time-reversal invariant Hermitian operator $X$ being identified with the classical-quantum Hamiltonian $H$ divided by $\hbar$. Finally, one may use the same relations to see that $O^A=0$, which since the antisymmetric part of the unitary generator $G^S$ is zero implies that the antisymmetric unitary generator constraint \eqref{eq: antisymmetric_effective_H} is also satisfied. The overdamped dynamics thus satisfies classical-quantum detailed balance with respect to $\pi$ as claimed.

Turning now to the underdamped dynamics, we again first find the explicit forms of the symmetric and anti-symmetric parts of $G$ and the reversible and irreversible parts of $K$. Comparing \eqref{eq: underdamped} to \eqref{eq: L_alternate} and using the defintions as before we find
\begin{equation}
\begin{split}
    G^S&=H + \frac{\beta}{8\gamma} M_{qq}\quad \quad \ \ G^A=\frac{i\hbar}{16}\beta\frac{p}{m}(L_q^\dag - L_q)\\
    K^{rev}&=\dfrac{1}{2}\begin{pmatrix}
        p/m\mathds{1}\\
        -L_q
    \end{pmatrix}\quad \quad K^{irr}=\dfrac{1}{2}\begin{pmatrix}
        0\\
        -\gamma p/m\mathds{1}
    \end{pmatrix}\\
    \ & \quad \quad \quad\quad\quad D_2=\begin{pmatrix}
        0 & 0\\
        0 &  2\gamma/\beta
        \end{pmatrix}
\end{split}
\end{equation} Here we see that the reversible parts of the dynamics are associated to terms that previously made up the Alexandrov bracket in the high temperature limit (c.f. Section \ref{subsec: back_reaction_limit}), while the irreversible part of the dynamics relates to the friction. The appearance of an anti-symmetric component of the unitary generator $G^A$ arises due the inclusion of the classical drift into the the back-reaction operators, which leads to an additional term in the unitary part of the dynamics that acts to cancel the new cross terms in the decoherence part of the dynamics.

To show that the underdamped dynamics also satisfies detailed balance, we check the dynamics against the conditions in \eqref{eq: diffusion_constraint} to \eqref{eq: antisymmetric_effective_H}. As before, the diffusion constraint is straightforward to check that it is satisfied, given that the only non-zero component is $D_{2,pp}$, which does not change under time-reversal. To check the reversible back-reaction constraint, we note again that the property of the $L_z$ operators \eqref{eq: L_op_property} means that the constraint is trivially satisfied with $a(q,p)=(0,0)^T$, while the irreversible back-reaction constraint is straightforward to check is satisfied using the fact that the thermal state in this dynamics is assumed to contain a classical kinetic term $p^2/(2m)\mathds{1}$. To check the symmetric unitary generator constraint, we first compute $O^S$, which in this case simplifies to $\beta/(16\gamma)[L_q^\dag L_q,\pi^{1/2}]$. As in the overdamped case, comparing this to the definition of $M_{xx}$ we see that this is indeed satisfied, with $X$ identified with $H/\hbar$. Finally, we check the anti-symmetric unitary generator constraint. To do so, we first compute $O^A$. In this case, $D_2$ is not full-rank, and thus we must retain a number of terms that vanished in the overdamped case. Computing this explicitly, we find that $O^A$ takes the form $(\beta p)/(16 m)[L_q^\dag -L_q,\pi^{1/2}]$. This final constraint is therefore exactly satisfied, given the value of $G^A$, thus demonstrating that the underdamped dynamics we provide also satisfies classical-quantum detailed balance.

\section{Discussion}

\

In this work, we introduced classical-quantum dynamics compatible with the second law of thermodynamics. There were three main technical contributions to achieve this: (i) the proof in Section \ref{sec: entropyetc.} that thermal-state preserving, completely-positive and linear dynamics necessarily obeys the second law of thermodynamics; (ii) the identification in Section \ref{sec: construct} of the $L_z$ and $M_{xy}$ classes of operators, given in Eqs. \eqref{eq: L_op_def} and \eqref{eq: M_op_def1}, which we showed could be used to construct such dynamics; and (iii) the definition and characterisation of detailed-balance in classical-quantum systems in Section \ref{sec: DB}, which we showed also applied to our $L_z$, $M_{xy}$ constructed dynamics. While we presented basic forms of such dynamics for the case of overdamped and underdamped classical systems in Eqs. \eqref{eq: overdamped} and \eqref{eq: underdamped}, the $L_z$ and $M_{xy}$ operators may be used to construct dynamics to treat a wide variety of systems, with some straightforward generalisations presented in Appendix \ref{app: generalised_dynamics}.

The fact that our dynamics may always be implemented using continuous measurement and feedback  \cite{layton2022healthier,tilloy2024general} provides a direct method of studying experimental realisations of our framework, with thermal states and the second law arising at level of the combined system of quantum system plus classical measurement signals. A simple example is provided in the overdamped model described in Sec. \ref{sec: analytic_model}, whose analytic solution \eqref{eq: analytic_sol_varrho00} to \eqref{eq: analytic_sol_varrho11} may be used to directly check experimental results obtained using the measurement and feedback procedure introduced in \cite{annby2022quantum}, as discussed in Appendix \ref{app: measfeedbackmodel1}. More generally, we expect our results to be valuable to experimentally bound possible stochastic theories of gravity \cite{kafri2014classical,tilloy2016sourcing,oppenheim2023postquantum,layton2022healthier,weller2024classical,oppenheim2023gravitationally,layton2023weak,Galley2023anyconsistent,oppenheim2023covariant,oppenheim2024diffeomorphism,weller2024quantum,oppenheim2024emergence,angeli2025positivity,angeli2025probing,kryhin2025distinguishable,piccione2025newtonian}, or to study experimental realisations of models designed to probe classical-quantum coupling \cite{eglinton2024stochastic}, extended using our results to non-perturbative regimes. An attractive feature of the present theory is that it provides an unambiguous notion of thermodynamic variables such as heat and entropy at the stochastic level, since the unravelled trajectories are objective \cite{layton2022healthier}. In principle this same approach can be extended to incorporate time-dependent driving, which would  facilitate a natural way of measuring fluctuating work and fluctuation theorems \cite{campisi2011colloquium,funo2018quantum} through monitoring of the conditional classical trajectories.

It is worth emphasising that the dynamics in this work have not been attempted to be derived from a full quantum theory. Indeed, a valid question to ask is whether consistent classical-quantum dynamics can even arise from quantum theory, in some appropriate limit. This question was taken up recently in \cite{layton2024classical}, and it was found that for sufficiently strong decoherence, completely-positive and linear dynamics in the form of Eq. \eqref{eq: L_general} was found. Interestingly, in this context, the form of operators determining the decoherence and back-reaction were found to take the form 
\begin{equation}
    L^H_z=i E_f\frac{\partial}{\partial z}\big( e^{-\frac{i}{E_f}H}\big) e^{\frac{i}{E_f}H}
\end{equation} where here $H$ is the classical-quantum Hamiltonian corresponding to the classical limit of one subsystem, and $E_f$ is an energy characterising the ratio of $\hbar$ and the decoherence timescale $\tau$. It is immediate from the above definition that this operator coincides with the operator $L_z$ introduced in Eq. \eqref{eq: L_op_def}, up to the Wick rotation
\begin{equation}
    \frac{i}{E_f}\mapsto \frac{\beta}{2}.
\end{equation} The entirely independent appearance of the $L_z$ operator in the context of dynamics arising from a fully quantum theory suggests some deeper connection between the current thermodynamic theory and those derived from classical-quantum limits. Understanding this, and the differences between these dynamics outside of the regime where the two operators coincide, is an interesting question to address in future work.

Finally, given that thermodynamics fundamentally is a theory that puts limitations on the allowable transformations of a system \cite{callen1991thermodynamics,brandao2013resource}, it is interesting to consider what general bounds may be derivable from a consistent thermodynamic theory of classical and quantum systems. For the class of dynamics introduced in Section \ref{sec: construct}, it was remarked that since both the Einstein relation and the decoherence-diffusion trade-off relate back-reaction in the classical-quantum system to the diffusion, they could be used to derive bounds on decoherence rates, given in Eqs. \eqref{eq: overdamped_D0_bound} and \eqref{eq: underdamped_D0_bound} for the overdamped and underdamped cases. Beyond these basic model-dependent bounds, we expect the theory we present here to provide a number of general thermodynamic bounds on systems described by a classical-quantum framework, and thus bounds on the achievability of a number of transformations, from electronic transitions in molecules to allowable state preparation in measurement-based feedback. 

\textit{\textbf{Acknowledgements:}} The authors would like to thank Maite Arcos, Kay Brandner, Basile Curchod, Andreas Dechant, Ken Funo, Masato Itami, Jonathan Oppenheim,    Emanuele Panella, Andrea Russo, Takahiro Sagawa, Keiji Saito, Andy Svesko, Cesare Tronci, Zach Weller-Davies and Toshihiro Yada for many invaluable discussions. I.L. acknowledges financial support from EPSRC and the Japan Society for the Promotion of Science (JSPS) as an International Research Fellow. H.M. acknowledges funding from a Royal Society Research Fellowship (URF/R1/231394).

\bibliography{thebib.bib}

\appendix

\widetext

\section{A broader class of thermal state preserving CQ dynamics} \label{app: generalised_dynamics}

In this appendix we provide a explicit forms of dynamics satisfying both $\mathcal{L}(\pi)=0$ and detailed balance that generalise the models provided in Section \ref{sec: construct}.

\subsection{Overdamped dynamics with correlated noise}

Consider a thermal fixed point $\pi=e^{-\beta H(z)}/\mathcal{Z}$ and consider a general classical-quantum dynamics that both saturates the trade-off and satisfies detailed balance i.e. takes the form \eqref{eq: L_alternate} and satisfies the equations \eqref{eq: diffusion_constraint} to \eqref{eq: antisymmetric_effective_H}. If we additionally assume that $D_2$ is full rank, $K^{rev}=0$ and $K_i^{irr}\pi^{1/2} = \pi^{1/2} {K_i^{irr}}^{\dag}$, then the dynamics significantly simplifies, and may be written in a closed form. 

To begin with, one may use $K^{rev}=0$ and $K_i^{irr}\pi^{1/2} = \pi^{1/2} {K_i^{irr}}^{\dag}$ to show that the Lindblad operators $K_i$ take the form
\begin{equation}
    K_i=\frac{1}{4}\frac{\partial D_{2,ij}}{\partial z_j} +\frac{1}{2}D_{2,ij}\frac{\partial \pi^\frac{1}{2}}{\partial z_j} \pi^{-1/2}=\frac{1}{4}\frac{\partial D_{2,ij}}{\partial z_j} -\frac{\beta}{4}D_{2,ij}L_j,
\end{equation} where here we use $L_j$ to denote the $L_z$ operator for $z=z_j$. On the other hand, since $D_2$ is full rank, $\mathbb{I}-D_2 D_2^{-1}=0$, which along with $K^{rev}=0$ and $K_i^{irr}\pi^{1/2} = \pi^{1/2} {K_i^{irr}}^{\dag}$ implies that $O^A=0$. This means that there are no additional constraints on $K_i$ in terms of $D_{2,ij}$, and additionally that $G^A$ is 0 up to an arbitrary component that we set equal to zero. This means that the unitary part of the dynamics is described by a unitary generator $G$ that is purely symmetric under time-reversal, which takes the form
\begin{equation}
    G=H -\frac{i\hbar\beta}{32}D_{2,ij}\frac{\partial D_{2,il}}{\partial z_l}D_{2,jk}(L_k^\dag-L_k)+\frac{\beta^2}{16}D_{2,ij}M_{ij},
\end{equation} where $M_{ij}$ is $M_{xy}$ for $x=z_i$ and $y=z_j$. When we take the case of uncorrelated noise that satisfies the Einstein relation $D_{2,ij}=\frac{2\mu}{\beta}\delta_{ij}$, and take the number of degrees of freedom $n=1$, we recover the model given in \eqref{eq: overdamped}. 

\subsection{Underdamped dynamics for arbitrary $H(q,p)$}

Consider a thermal fixed point $\pi=e^{-\beta H(q,p)}/\mathcal{Z}$, where $H(q,p)$ may have any functional dependence on $q$ and $p$ provided it is invariant under time-reversal. In this case, there may be backreaction on both the classical momentum $p$ and classical position $q$, and thus for positivity of the dynamics, there must be corresponding diffusion in both momentum and position. Although phase-dependent and correlated noise may be studied using the methods discussed in the previous section, a simple choice is
\begin{equation}
    D_2=\frac{2}{\beta}\begin{pmatrix}
        \gamma_q & 0 \\
        0& \gamma_p
    \end{pmatrix},
\end{equation} where $\gamma_q$ and $\gamma_p$ are constants.  Taking the following choices for $K=K^{rev}+K^{irr}$ and $S$
\begin{equation}
    K^{rev}=\frac{1}{2}\begin{pmatrix}
        L_p \\
        -L_q
    \end{pmatrix} \quad \quad
    K^{irr}=\frac{1}{2}\begin{pmatrix}
        -\gamma_q L_q \\
        -\gamma_p L_p
    \end{pmatrix} \quad \quad
    S=\frac{\beta}{8}( \gamma_q + \gamma_p^{-1})M_{pp} + \frac{\beta}{8}( \gamma_p + \gamma_q^{-1})M_{qq}+ H,
\end{equation} it is straightforward to check that these satisfy the detailed balance constraints, and thus that the dynamics satisfies detailed balance, and hence $\mathcal{L}(\pi)=0$.

\section{Invariance of interaction under changes to the classical Hamiltonian} \label{app: invariance}

For the interaction between the classical and quantum subsystems to be meaningfully defined, it is reasonable to expect that it is independent of the particular choice of classical Hamiltonian. In other words, changing the classical potential should only affect the drift of the classical system, and not affect the structure of the dynamics related to the quantum system.

To see that this is indeed the case, consider the affect of modifying the classical-quantum Hamiltonian by 
\begin{equation} \label{eq: transform_H_by_classical}
    H\mapsto H+ V(z)\mathbb{I}.
\end{equation} The corresponding change to $L_z$ is easily computed to be
\begin{equation}
    L_z\mapsto L_z + \frac{\partial V}{\partial z}\mathds{1},
\end{equation} and the change in $M_{zz}$ may also be solved, using \eqref{eq: M_op_def2} and \eqref{eq: L_op_property}, to give
\begin{equation}
    M_{zz}\mapsto M_{zz} -\frac{i \hbar}{2} \frac{\partial V}{\partial z}(L_z - L_z^\dag).
\end{equation} As expected, the change to $L_z$ leads to an additional drift term in the classical part of the dynamics. However, at first glance the change in $M_{zz}$ makes it appear that the unitary part of the quantum dynamics is affected by the choice of classical potential $V$. However, the change in $L_z$ also affects the decoherence part of the dynamics, and it turns out that the additional term that arises exactly cancels the change in $M_{zz}$ described above. The overall effect of the transformation \eqref{eq: transform_H_by_classical} is thus simply captured by
\begin{equation}
    \mathcal{L}(\varrho)\mapsto \mathcal{L}(\varrho) + \mu \partial_x ( (\partial_x V))
\end{equation} or
\begin{equation}
    \mathcal{L}(\varrho)\mapsto \mathcal{L}(\varrho) + \{V,\varrho\}.
\end{equation} Aside from being physically reasonable, this property means that purely classical parts of the dynamics may be included by including additional classical drift terms, independent of $L_z$ and $M_{zz}$, rather than needing to include such effects directly via these operators.

\section{Monotonicity of the CQ relative entropy}\label{app: CQmonotonicity}

\

In this section, we sketch a proof of the monotonicity of the classical-quantum relative entropy. We first model our hybrid system with a set of \textit{discrete} classical variables $\{x\}$. This is sufficient for our proof since one may always obtain a diffusive Markovian equation as a limit of a discrete one \cite{diosi2023hybrid}, so any statement about monotonicity is maintained in such a limit. In the discrete case a general evolution map can be expressed in its Kraus form \cite{diosi2014hybrid}
\begin{align}\label{eq:map}
\rho'(x)=\sum_{\alpha,y}   K_\alpha(x,y)\rho(y) K_\alpha^\dagger(x,y),
\end{align}
where
\begin{align}\label{eq:normalise}
    \sum_{\alpha,x}   K_\alpha^\dagger(x,y) K_\alpha(x,y)=\hat{\mathbb{I}}
\end{align}
We introduce the CQ relative entropy between two states,
\begin{align}
    S[\rho(x)|| \sigma(x)]:=\sum_x \tr{\rho(x) [\text{log} \ \rho(x)-\text{log} \  \sigma(x)]}
\end{align}
This generalises the standard relative entropy between two density operators,
\begin{align}
    S[\rho|| \sigma]:=\tr{\rho [\text{log} \ \rho-\text{log} \  \sigma]},
\end{align}
which has the important property that it is contractive under CPTP maps such as $T(.)$ \cite{sagawa2013second}, ie.
\begin{align}\label{eq:mono}
    S[T(\rho)||T( \sigma)]\leq S[\rho|| \sigma].
\end{align}
We can prove that the same monotonicity property applies to the CQ relative entropy.  First note that it's possible to define a density matrix and Kraus operators on an extended separable Hilbert space using a basis $\{x\}\mapsto \{\ket{x}\}$
\begin{align}
    &\hat{\varrho}:=\sum_y \rho(y)\otimes \ket{y}\bra{y}. \\
    &\hat{\mathcal{K}}_\alpha=\sum_{x,y}   K_\alpha(x,y)\otimes\ket{x}\bra{y}
\end{align}
We now expand the following trace functional using the Taylor series for $\text{log}(x)$ around $x=1$,
\begin{align}
    \nonumber\tr{T(\rho) \ \text{log} \ T( \sigma)}&=\tr{\sum_\alpha  K_\alpha\rho K^\dagger_\alpha \ \text{log} \ \sum_\beta K_\beta \sigma K^\dagger_\beta} \\
    \nonumber&=\sum_{n=1}^\infty\frac{(-1)^{n+1}}{n}\tr{\sum_\alpha  K_\alpha\rho K^\dagger_\alpha \ \bigg( \sum_\beta K_\beta \sigma K^\dagger_\beta-\hat{\mathbb{I}}\bigg)^n} \\
    \nonumber&=\sum_{n=1}^\infty\frac{(-1)^{n+1}}{n}\text{Tr}\bigg(\sum_\alpha\sum_{y,x',y',x'',y''}  K_\alpha(x'',y'')\otimes \ket{x''}\bra{y''}\rho(y)\otimes \ket{y}\bra{y} \\
    \nonumber& \ \ \ \ \ \ \  \times K^\dagger_\alpha(x',y')\otimes \ket{y'}\bra{x'} \ \bigg( \sum_\beta\sum_{w,z',w',z'',w''} K_\beta(w'',z'')\otimes\ket{w''}\bra{z''} \\
    \nonumber& \ \ \ \ \ \ \  \ \ \ \ \ \  \times \sigma(w)\otimes\ket{w}\bra{w} K^\dagger_\beta(w',z')\otimes\ket{z'}\bra{w'}-\hat{\mathbb{I}}\bigg)^n \bigg) \\
    \nonumber&=\sum_{x'}\sum_{n=1}^\infty\frac{(-1)^{n+1}}{n}\text{Tr}\bigg(\sum_{\alpha,y} K_\alpha(x',y)\rho(y) K^
    \dagger_\alpha(x',y) \\
    \nonumber& \ \ \ \ \ \ \ \ \ \ \ \ \ \ \ \ \ \   \times\bigg(\sum_{\beta,w} K_\alpha(x',w) \sigma(w) K^
    \dagger_\beta(x',w)-\hat{\mathbb{I}}\bigg)^n\bigg) \\
    &=\sum_{x'}\tr{\rho'(x)\text{log} \  \sigma'(x)}
\end{align}
This implies
\begin{align}\label{eq:ineq}
S[\rho(x)|| \sigma(x)]=S[\rho|| \sigma]\geq S[T(\rho)||T( \sigma)]=S[\rho'(x)|| \sigma'(x)],
\end{align}
where we used the monotonicity~\eqref{eq:mono}.

\section{Measurement and feedback representation of Model I} \label{app: measfeedbackmodel1}

In this appendix, we show explicitly how the model described in Section \ref{sec: analytic_model} coincides with a specific case of a measurement and feedback system involving a detector with finite bandwidth, that was introduced in \cite{annby2022quantum}. In doing so, we prove that this measurement and feedback system satsifies detailed balance and the second law on the level of both quantum system and classical controller. We also discuss how the analytic solution in \eqref{eq: analytic_sol_varrho00} to \eqref{eq: analytic_sol_varrho11} is modified in the case of inefficient detection.

First, from \cite{annby2022quantum}, we note that the dynamics of a continuously monitored quantum system and its corresponding measurement signal $D$ can be characterised by a classical-quantum master equation, refered to as the ``quantum Fokker-Planck master equation", which takes the form
\begin{equation}
\frac{\partial \hat{\varrho}}{\partial t}=\mathcal{L}^{FB}\hat{\varrho} + k (\hat{A}\hat{\varrho} \hat{A} - \frac{1}{2}\{\hat{A}^2,\hat{\varrho}\}_+)+g\frac{\partial}{\partial D}  (D \hat{\varrho})-\frac{g}{2}\frac{\partial}{\partial D}\{\hat{A},\hat{\varrho}\}_+ + \frac{g^2}{8 k}\frac{\partial^2 \hat{\varrho}}{\partial D^2}
\end{equation} where here, $\hat{A}$ is the monitored observable, $k$ denotes the measurement strength, $g$ denotes the bandwidth of the detector, and $\mathcal{L}^{FB}$ denotes a superoperator characterising the feedback in the system which can depend on the measurement signal $D$. Let us additionally assume that the feedback takes the form  $\mathcal{L}^{FB}(\cdot) =-\frac{i}{\hbar}[H(D),\cdot] $ with $H(D)=- s D \hat{A}$, where $s$ characterises the strength of the feedback \cite{prech2025quantum}. Turning to the dynamics we introduce in Sec. \ref{sec: construct}, we may explicitly write our master equation \eqref{eq: overdamped} with the Hamiltonian given in Eq. \eqref{eq: H_overdamped} to show that the dynamics takes the form
\begin{equation}
    \frac{\partial \hat{\varrho}}{\partial t}=-\frac{i}{\hbar}[\lambda (x\hat{\mathds{1}}-l\hat{\sigma}_z)^2,\hat{\varrho}] + \frac{\mu \beta \lambda^2 l^2}{2}(\hat{\sigma}_z \hat{\varrho} \hat{\sigma}_z -\hat{\varrho})+ 2 \lambda \mu \frac{\partial}{\partial x}( x \hat{\varrho}) - \mu l \lambda \frac{\partial}{\partial x}\{\hat{\sigma}_z,\hat{\varrho}\}_+ + \frac{\mu}{\beta}\frac{\partial^2 \hat{\varrho}}{\partial x^2}
\end{equation}
Comparing the two equations, it is straightforward to see that for
\begin{equation}
D=x,\quad 
s=2\lambda, \quad 
    A=l\sigma_z, \quad g=2\mu\lambda,\quad k=\frac{\mu\beta \lambda^2}{2},
\end{equation} the two dynamics coincide. Rearranging for $\beta$ and $\mu$ we find that
\begin{equation}
    \beta=\frac{8 k}{s g},\quad \quad 
    \mu=\frac{g}{s},
\end{equation} i.e. this particular case of quantum Fokker-Planck equation dynamics describe an overdamped classical-quantum system with effective temperature and mobility determined by the measurement strength $k$, detector bandwidth $g$ and feedback strength $s$. The classical-quantum Hamiltonian corresponding to this system is given by
\begin{equation}
    \hat{H}(D)=\frac{s}{2}(D\hat{\mathds{1}}-\hat{A})^2,
\end{equation} for which the corresponding thermal state is a fixed point of the quantum Fokker-Planck master equation. By the results of Secs. \ref{sec: entropyetc.} and \ref{sec: DB}, and by showing the equivalence of this measurement and feedback with our dynamics, we thus prove that this type of measurement and feedback set-up satisfies the second law and detailed balance at the level of the entire classical-quantum system.

Finally, we note that any experimental implementations of this measurement and feedback set-up will necessarily lead to additional decoherence due to some inefficiencies in the detection. Denoting the efficiency $\eta$, when $\eta<1$ the results are modified by $k\mapsto \eta k$ and the addition of a decoherence term $D_0(\hat{\sigma}_z \hat{\varrho} \hat{\sigma}_z -\hat{\varrho})$ where $\tilde{D}_0=k(1-\eta)$. This leads to a change in the effective temperature $\beta \mapsto \eta \beta$, while the analytic solution in \eqref{eq: analytic_sol_varrho00} to \eqref{eq: analytic_sol_varrho11} must be modified by an additional term $-2\tilde{D}_0 t$ in the exponential of the off-diagonal components.

\section{Deriving the M operator for a harmonic oscillator}\label{app:Mop}

\

Here we will derive the explicit form of the operator
\begin{equation}\label{eq:Mapp}
    \hat{M}_{qq}=\frac{i\hbar}{2}\int_0^\infty  e^{-s\pi^\frac{1}{2}} [\hat{L}_q^{\dag}\hat{L}_q,\hat{\pi}^\frac{1}{2}] e^{-s\hat{\pi}^\frac{1}{2}} ds,
\end{equation}
for the oscillator model of Section~\ref{sec: oscillator_model}, without relying on any initial assumption regarding its form. To start we will use the following series expansion \cite{liu2016quantum}
\begin{align}
    \int^s_0 ds' \ e^{-s \hat{B}}\hat{A} e^{-s \hat{B}}=\sum^\infty_{k=0}\sum^k_{l=0}\frac{(-s)^{k+1}}{(k+1)!} C^l_k \hat{B}^l \hat{A} \hat{B}^{k-l}
\end{align}
which upon substitution into~\eqref{eq:Mapp} gives a formula
\begin{align}\label{eq:Mapp2}
    \hat{M}_{qq} 
    =\lim_{s\to\infty} \frac{i\hbar}{2}\sum^\infty_{k=0}\sum^k_{l=0}\frac{(-s)^{k+1}}{(k+1)!} C^l_k \bigg((\sqrt{\hat{\pi}})^l \hat{L}_q^\dagger \hat{L}_q(\sqrt{\hat{\pi
    }})^{k-l+1}-(\sqrt{\hat{\pi}})^{l+1} \hat{L}_q^\dagger \hat{L}_q(\sqrt{\hat{\pi}})^{k-l}\bigg).
\end{align} Here $C^{l}_k$ are the binomial coefficients $C^{l}_k=k!/(l!(k-l)!)$. The operator $\hat{L}_q$ from the main text~\eqref{eq: L_oscillator} is of the form
\begin{align}
\hat{L}_q= \frac{2 m_q \omega}{\hbar \beta } \sinh{\frac{\hbar \omega \beta}{2}} (q\hat{\mathds{1}}- \hat{q}) + \frac{2i}{\hbar \beta  }(1 -\cosh{\frac{\hbar \omega \beta}{2}})\hat{p}+ m_c\Omega^2 q \hat{\mathds{1}}
\end{align}
The commutation relations~\eqref{eq: sqrtpi_q_identity} and~\eqref{eq: sqrtpi_p_identity} can be straightforwardly extended to show
\begin{equation} \label{eq: sqrtpi_q_identity2}
    (q\hat{\mathds{1}}- \hat{q})(\sqrt{\hat{\pi}})^n =(\sqrt{\hat{\pi}})^n \big[\cosh{\frac{n\hbar\omega \beta}{2}} (q\hat{\mathds{1}}- \hat{q}) +\frac{i}{m_q \omega} \sinh{\frac{n\hbar\omega \beta}{2}}\hat{p}\big] ,
\end{equation} and
\begin{equation}  \label{eq: sqrtpi_p_identity2}
    \hat{p}(\sqrt{\hat{\pi}})^n = (\sqrt{\hat{\pi}})^n\big[\cosh{\frac{n\hbar\omega \beta}{2}}  \hat{p}-i m_q \omega \sinh{\frac{n\hbar\omega \beta}{2}}(q\hat{\mathds{1}}- \hat{q})\big] .
\end{equation}
Putting this together we have and simplifying we get
\begin{align}
    \nonumber \hat{L}_q(\sqrt{\hat{\pi}})^n&=(\sqrt{\hat{\pi}})^n\bigg(\frac{4 }{\hbar \beta } \sinh{\frac{\hbar \omega \beta}{4}}\bigg)\bigg(m_q \omega\cosh{\frac{(2n-1)\hbar\omega \beta}{4}} (q\hat{\mathds{1}}- \hat{q})  \\
    & \ \ \ \ \ \ \ \ \ \ \ \ \ \ \ \ \ \ \ \ \ \ \ \ \ \ \ \ \ \ \ \ \ \ \ \ \ \ \ \ \ \ \    - i \sinh{\frac{(2n-1)\hbar\omega \beta}{4}}\hat{p}\bigg)+ m_c\Omega^2 q (\sqrt{\hat{\pi}})^n
\end{align}
Multiplying this by its hermitian conjugate we get a quadratic expression
\begin{align}
    (\sqrt{\hat{\pi}})^n \hat{L}_q^\dagger \hat{L}_q(\sqrt{\hat{\pi}})^{-n}=m_c^2\Omega^4q^2\hat{\mathds{1}}+A_n(q \hat{\mathds{1}}- \hat{q})^2+B_n\hat{p}^2+iC_n\{q \hat{\mathds{1}}- \hat{q},\hat{p}\}+D_n (q \hat{\mathds{1}}- \hat{q})+i E_n \hat{p}
\end{align}
with coefficients
\begin{align}
    \nonumber&A_n=\frac{8 m_q^2\omega^2}{\hbar^2\beta^2}\sinh^2{\frac{\hbar \omega \beta}{4}}\bigg(\cosh{\frac{\hbar \omega \beta}{2}}+\cosh{n\hbar \omega \beta}\bigg), \\
    \nonumber&B_n=\frac{8 }{\hbar^2\beta^2}\sinh^2{\frac{\hbar \omega \beta}{4}}\bigg(\cosh{\frac{\hbar \omega \beta}{2}}-\cosh{n\hbar \omega \beta}\bigg), \\
    \nonumber&C_n=\frac{16 m_q\omega}{\hbar^2\beta^2}\sinh^2{\frac{\hbar \omega \beta}{4}}\cosh{\frac{(2n-1)\hbar \omega \beta}{4}}\sinh{\frac{(2n+1)\hbar \omega \beta}{4}}, \\
    \nonumber&D_n=\frac{8 m_q \omega m_c \Omega^2 q}{\hbar \beta } \sinh{\frac{\hbar \omega \beta}{2}} \cosh{\frac{n\hbar\omega \beta}{2}}, \\
    &E_n=\frac{4  m_c \Omega^2 q}{\hbar \beta } \sinh{\frac{\hbar \omega \beta}{2}}\sinh{\frac{n\hbar\omega \beta}{2}}.
\end{align}
Now we will need the following exponential series, setting $y=e^{\hbar\omega \beta/2}>1$ ,
\begin{align}
\nonumber\lim_{s\to\infty}\sum^\infty_{k=0}\sum^k_{l=0}\frac{(-s)^{k+1}}{(k+1)!} C^l_k \big[\sinh{\frac{l\hbar\omega \beta}{2}}-\sinh{\frac{(l+1))\hbar\omega \beta}{2}}\big] x^{k+1}&=\lim_{s\to\infty}\frac{(y-1)}{2(y+1)}\big(2-e^{-sx (1+y)-e^{-sx(1+y)/y}}\big), \\
&=\frac{(y-1)}{(y+1)},
\end{align}
and
\begin{align}
\nonumber\lim_{s\to\infty}&\sum^\infty_{k=0}\sum^k_{l=0}\frac{(-s)^{k+1}}{(k+1)!} C^l_k \big[\cosh{\frac{(2l-1)\hbar \omega \beta}{4}}\sinh{\frac{(2l+1)\hbar \omega \beta}{4}} \\
\nonumber& \ \ \ \ \ \ \ \ \ \ \ \ \ \ \ \ \ \ \ \ \ \ \ \ \ \ \ \ \ \ \ \ \ \ \ \ \ \  -\cosh{\frac{(2l+1)\hbar \omega \beta}{4}}\sinh{\frac{(2l+3)\hbar \omega \beta}{4}}\big] x^{k+1} \\
&=\frac{1}{2}\frac{y^2-1}{y^2+1}.
\end{align}
Similar calculations show that the other series for $A_n$, $B_n$ and $D_n$ vanish, so we are left with an expression that includes only $\{q\hat{\mathds{1}}- \hat{q},\hat{p}\}$ and $\hat{p}$. Plugging this in to the series formula~\eqref{eq:Mapp2} we arrive at
\begin{align}
    \hat{M}_{qq}=\frac{2 m_q\omega}{\hbar\beta^2}\big(\tanh{\frac{\hbar \omega \beta}{2}}-\sinh{\frac{\hbar \omega \beta}{2}}\big)\{q\hat{\mathds{1}}- \hat{q},\hat{p}\}+\frac{2  m_c \Omega^2 q}{ \beta }\big(1-\cosh{\frac{\hbar \omega \beta}{2}}\big)\hat{p} ,
\end{align} as claimed.

\section{Thermalisation additional material} \label{app: thermalisation_additional}

In this appendix, we include additional material showing evidence that the coherences in the adiabatic basis, correlations between position and momentum, and the correlations between the classical and quantum systems, all vanish over time in the relative position representation. Along with the thermalisation properties shown in Section \ref{subsec: thermalisation}, this provides evidence that the classical-quantum state for this model does indeed thermalise for typical initial conditions and model parameters.

\begin{figure}[h]
    \centering
    
    {\includegraphics[width=0.64\linewidth]{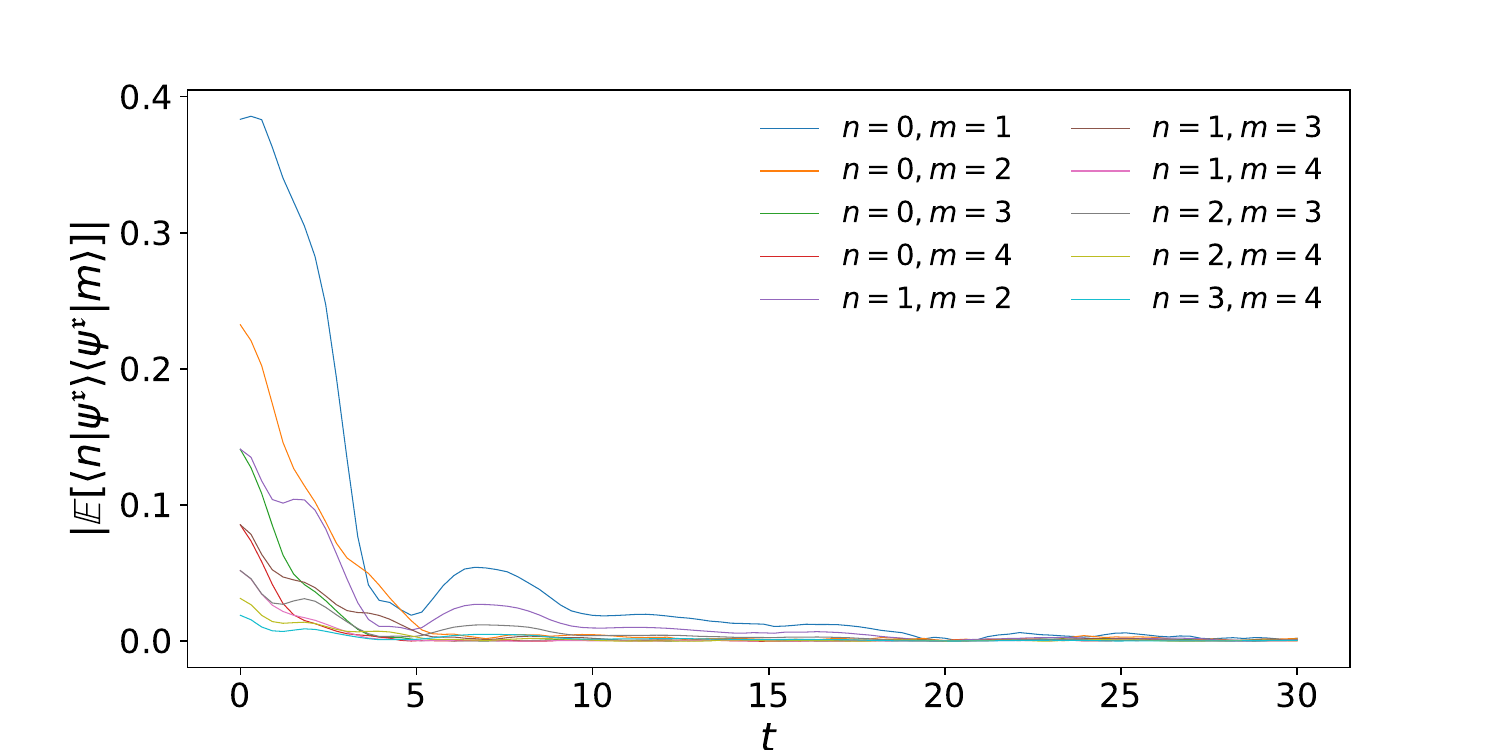}}
    \caption{Absolute value of the coherences between the first five energy levels plotted from $t=0$ to $t=30$. Here we take parameters $\omega=m_C=m_Q=\hbar=\beta=\gamma=\Omega=1$, while $N_{max}=20$ and $N_{steps}=10^4$, and compute the average over $10^4$ trajectories.}
    \label{fig: coherence_thermalisation}
\end{figure}

We first turn to look at the coherences in the adiabatic basis in the relative position representation i.e. the coherences in the number state basis when expressed in terms of $|\psi^\mathfrak{r}\rangle$. To study this, we consider a pure state whose probabilities of occupying the adiabatic basis states is given by the thermal state, namely that $|\psi^\mathfrak{r}\rangle=\sum_n\sqrt{p_n}|n\rangle$ where $p_n\propto \exp(-\beta \epsilon_n)$. Plotting the coherences between the first 5 adiabatic energy levels in Fig. \ref{fig: coherence_thermalisation} we see that their each of their values appear to tend to zero. The same behaviour may be found for other initial conditions with coherence in this basis.

\begin{figure}[h]
    \centering
    \includegraphics[width=0.6\linewidth]{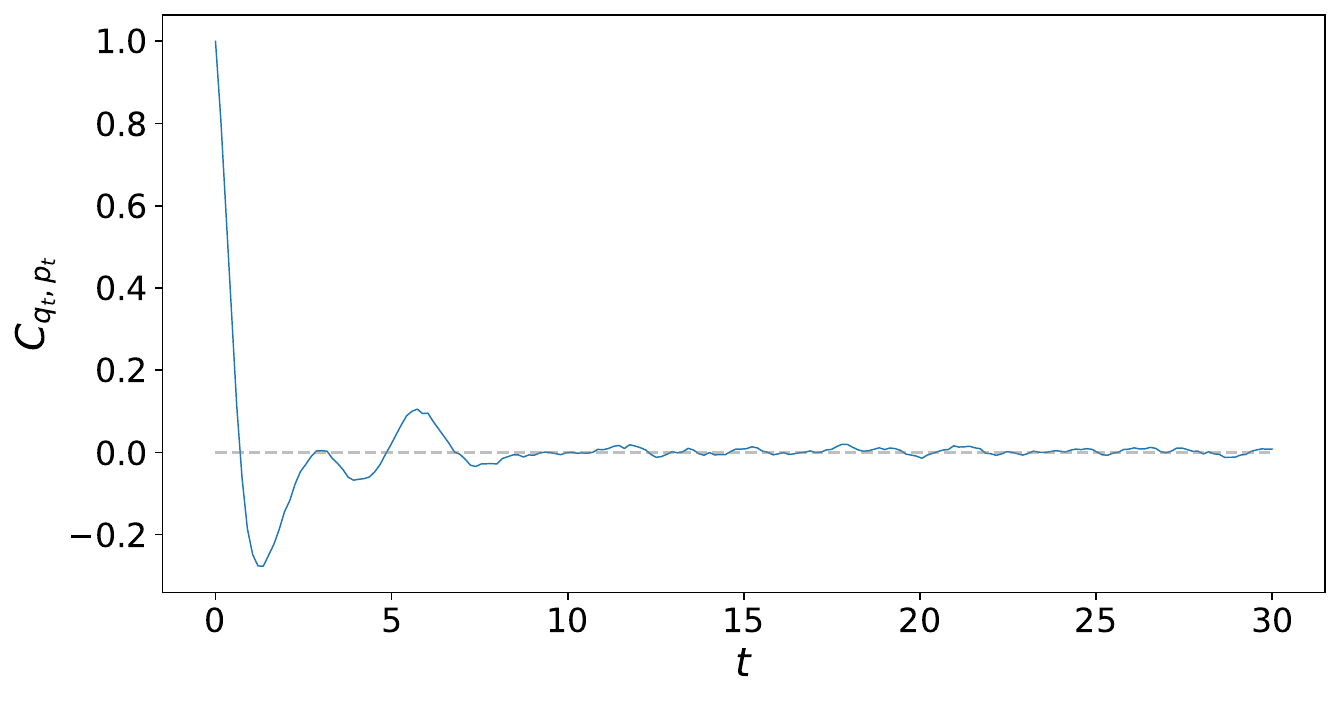}
    \caption{The covariance between $q_t$ and $p_t$ from $t=0$ to $t=30$ for an initial state given in Eq. \eqref{eq: correlated_CQ_state_IC} with $\omega=m_C=m_Q=\hbar=\beta=\gamma=\Omega=1$, while $N_{max}=10$ and $N_{steps}=5000$, and computed over $2\times 10^4$ trajectories. }
    \label{fig: correlations_qp}
\end{figure}

\begin{figure*}[!tbp]
    \centering
    \includegraphics[width=0.48\linewidth]{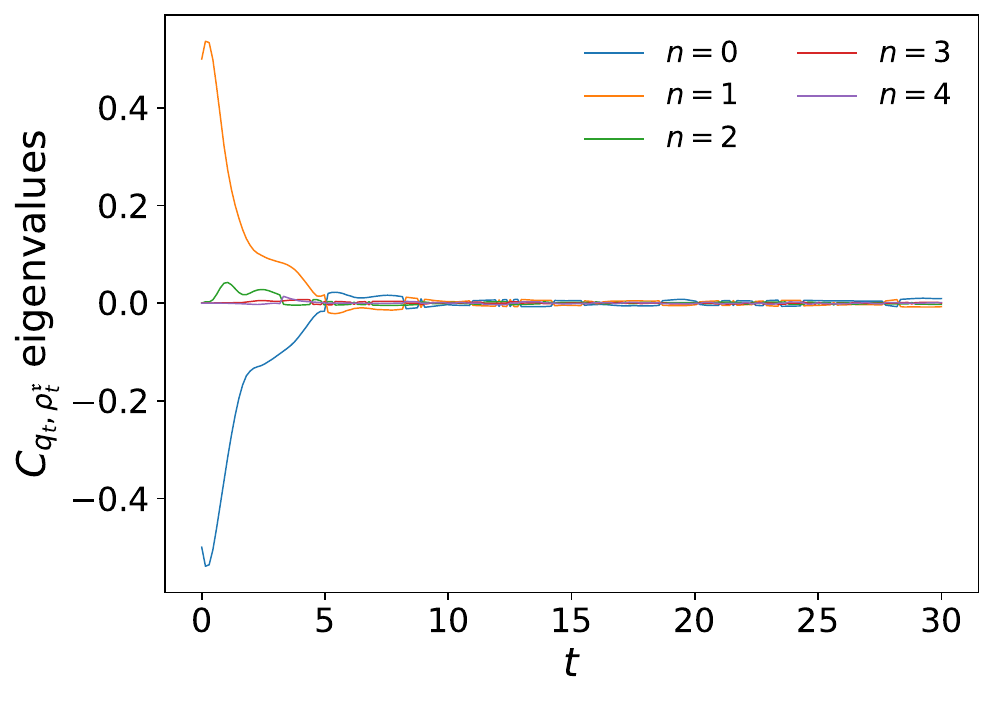}
    \includegraphics[width=0.48\linewidth]{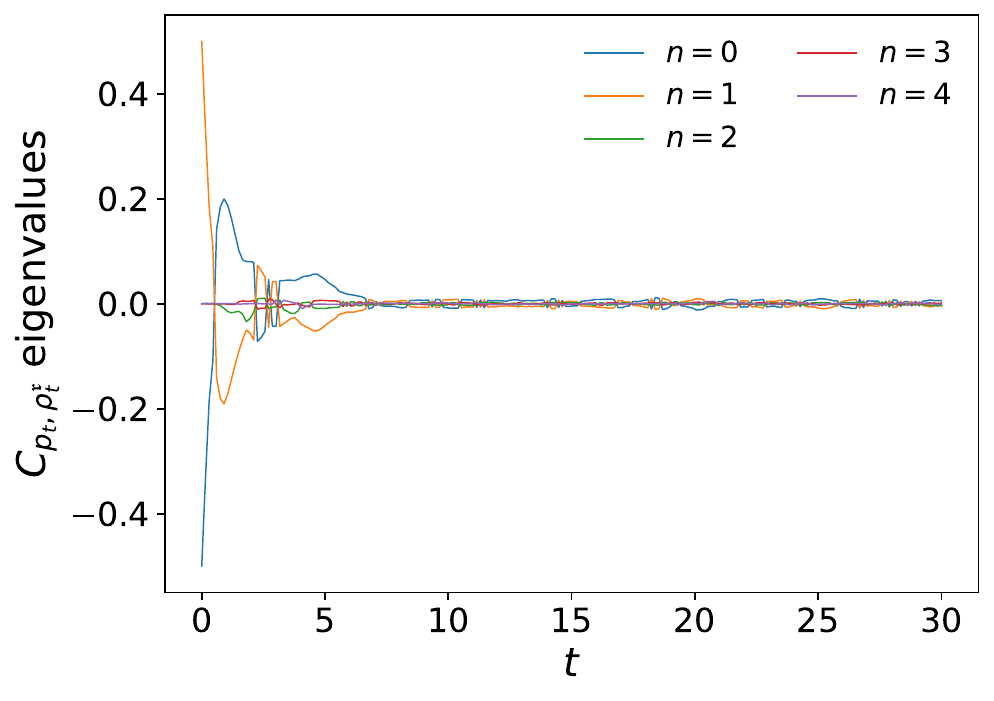}
    \caption{Eigenvalues of the covariance between $q_t$ (left) and $q_t$ (right) with the quantum state in the relative position basis $\hat{\rho}^\mathfrak{r}_t$, with the same initial conditions and parameters as in Figure \ref{fig: correlations_qp}.  }
    \label{fig: covariance_Q_C}
\end{figure*}
We must also check that the classical variables $q_t$ and $p_t$ are indeed uncorrelated at long times. To check this, we plot the covariance function between the two variables in Figure \ref{fig: correlations_qp} for an initial condition in which the system either starts at $q_0=p_0=1$ with $|\psi^\mathfrak{r}\rangle=|1\rangle$ or  $q_0=p_0=-1$ with $|\psi^\mathfrak{r}\rangle=|0\rangle$ with equal probability i.e. with the initial condition
\begin{equation} \label{eq: correlated_CQ_state_IC}
{\hat{\varrho}}^{\mathfrak{r}}(q,p)=\frac{1}{2}\delta(q-1)\delta(p-1)|1\rangle\langle 1| +\frac{1}{2}\delta(q+1)\delta(p+1)|0\rangle\langle 0|. 
\end{equation} We see here that the initially non-zero covariance between the two classical variables vanishes at long times, up to remaining fluctuations limited by the achievable number of trajectories in the simulation.

Finally, we study the correlations between the classical and quantum degrees of freedom in the relative position representation, using the same correlated distribution as described by Eq. \eqref{eq: correlated_CQ_state_IC}. As before, we may compute the covariance, here between the random variable $q_t$ or $p_t$ and the density matrix $\hat{\rho}^\mathfrak{r}_t=|\psi^\mathfrak{r}\rangle_t\langle\psi^\mathfrak{r}|_t$. Since this gives in both cases a Hermitian matrix, plotting the eigenvalues of this matrix provides a measure of the covariance between the classical and quantum degrees of freedom – if all of the eigenvalues tend to zero then this is sufficient to conclude that the covariance between the classical and quantum degrees vanishes. Plotting the first five eigenvalues of these matrices in Figure \ref{fig: covariance_Q_C} we see that the initial correlations due to the initial conditions rapidly die off, again up to fluctuations due to limitations in the number of trajectories that are averaged over.

Strictly speaking, the vanishing of the covariance function in the above two cases does not guarantee that the two variables in each case are not correlated, since we have not also demonstrated that they also are not jointly Gaussian distributed. While this appears likely, given that the drift and back-reaction operators are linear in both $q,p$ and in $ \hat{q},\hat{p}$, we leave this, and a more rigorous theoretical characterisation of thermalisation, to future work.

\section{An equivalent general form of classical-quantum dynamics} \label{app: L_equivalent}
For the study of general detailed balance, it will be necessary to have a form of classical-quantum dynamics for which we know both the sufficient and necessary conditions for positivity. Otherwise, the general form of dynamics satisfying detailed balance will only have known sufficient conditions for positivity, but not necessary ones. However, using the form of dynamics provided in Eq. \eqref{eq: L_general}, written in terms of a fixed basis of traceless and orthonormal operators $L_\alpha$, is not convenient for computing the form of dynamics satisfying detailed balance. This is because any operator $L_\alpha$ that includes terms proportional to the identity will necessarily not be included in this form, even if this is a natural choice, such as for the $L_z$ operators.

Fortunately, one is able to characterise the completely positive sufficient and necessary conditions by an alternative representation of the general form of dynamics, which is given
\begin{equation} \label{eq: L_alternate_2}
\begin{split}
\frac{\partial \varrho}{\partial t}=& -i[S,  \varrho ]- \frac{\partial }{\partial z_{i}} \left( K_i  \varrho +  \varrho K_i^\dag \right) \\
&+\frac{1}{2} \frac{\partial^2 }{\partial z_i \partial z_j} ( D_{2, i j}  \varrho )+ {D_{2,ij}^{-1}} \big( K_i  \varrho K_j^{\dag} - \frac{1}{2}  \{ K_j^{\dag} K_i,  \varrho \}_+ \big) \\
& +\tilde{D}_0^{\alpha \beta}\big( L_{\alpha}  \varrho L_{\beta}^{\dag} - \frac{1}{2}  \{ L_{\beta}^{\dag} L_{\alpha},  \varrho \}_+ \big) \\
\end{split}
\end{equation} and is completely positive \textit{if and only if} 
\begin{equation}
    \tilde{D}_0\succeq 0, \quad D_2\succeq 0, \quad S=S^\dag, 
\end{equation} and there exists a phase space dependent vector $v$ of length $n$ such that if $K$ denotes the operator valued vector $K=(K_1,\ldots, K_n)^T$ then
\begin{equation}
    (\mathbb{I}-D_2^{-1}D_2)K=v\mathds{1}.
\end{equation} Here the $L_\alpha$ are an arbitrary set of traceless and orthogonal operators that may have phase space dependence.

To see that this dynamics is equivalent to the general form of dynamics above, we note that defining
\begin{equation}
    K_i=\frac{1}{2}D^C_{1,i}\mathds{1}+(D_{1,i}^\alpha)^*L_\alpha
\end{equation} brings the generator of Eq. \eqref{eq: L_general} into the form of Eq. \eqref{eq: L_alternate_2} when
\begin{equation}
    S=\bar{H}-\frac{i}{4} {D^C_{1,i}} D_{2,ij}^{-1}({D_{1,j}^\alpha}^* L_\alpha-{D_{1,j}^\alpha} L_\alpha^\dag) \quad \quad \tilde{D}_0=D_0-D_1^\dag D_2^{-1}D_1.
\end{equation} To see that the above conditions are necessary and sufficient for complete positivity, we note that $S$ is Hermitian iff $\bar{H}$ is Hermitian, and that $\tilde{D}_0$ by definition is positive semi definite iff $D_0\succeq D_1^\dag D_2^{-1} D_1$. Finally, when $(\mathbb{I}- D_2 D_2^{-1})D_1 =0$ then
\begin{equation}
    (\mathbb{I}- D_2 D_2^{-1})K=\frac{1}{2}(\mathbb{I}- D_2 D_2^{-1})D_1^C \mathds{1} + (\mathbb{I}- D_2 D_2^{-1})D_1 L =  \frac{1}{2}(\mathbb{I}- D_2 D_2^{-1})D_1^C \mathds{1}
\end{equation} i.e. $v=\frac{1}{2}(\mathbb{I}- D_2 D_2^{-1})D_1^C $. Conversely, if $(\mathbb{I}- D_2 D_2^{-1})K=v\mathds{1}$, then we see that
\begin{equation}
    \frac{1}{2}(\mathbb{I}- D_2 D_2^{-1})D_1^C \mathds{1} + (\mathbb{I}- D_2 D_2^{-1})D_1 L = v \mathds{1}
\end{equation} which since the $L_\alpha$ are linearly independent and traceless, implies that $(\mathbb{I}- D_2 D_2^{-1})D_1=0$.

Finally, to find the form of the additional decoherence as given in Eq. \eqref{eq: L_alternate}, we note that when $\tilde{D}_0$ 
is positive semi-definite, one may diagonalise the dissipator to write it in terms of a set of traceless (but not necessarily orthonormal) operators $\tilde{L}_\alpha$. These two forms are equivalent, since when written in diagonal form, one may always rewrite in a non-diagonal form, with positive semi-definite $\tilde{D}_0$, by decomposing each $\tilde{L}_\alpha$ in terms of an orthonormal basis.

\section{Derivation of the detailed balance conditions} \label{app: DB_derivation}

In this section we demonstrate how the detailed balance conditions \eqref{eq: diffusion_constraint} to \eqref{eq: antisymmetric_effective_H} may be derived from the definition of detailed balance \eqref{eq: classical-quantum_DB}, using the positivity conditions of \eqref{eq: alt_pos_conditions_1} and \eqref{eq: alt_pos_conditions_2}. Note that here we assume the trade-off is saturated i.e. $\tilde{L}_\alpha=0$.

We start by explicitly writing out the forms of both the generator $\pi^{-\frac{1}{2}}\mathcal{L}(\pi^{\frac{1}{2}}A\pi^{\frac{1}{2}})\pi^{-\frac{1}{2}}$ and the adjoint generator under time-reversal $\mathcal{L}^\dag_{\epsilon}(A)$, in terms of the operators $K_i^{rev}$, $K_i^{irr}$, $G^S$, and $G^A$. It is straightforward to check that these take the form
\begin{equation}
\begin{split}
\pi^{-\frac{1}{2}}\mathcal{L}(\pi^{\frac{1}{2}}A\pi^{\frac{1}{2}})\pi^{-\frac{1}{2}}=& -\frac{i}{\hbar}\pi^{-\frac{1}{2}}[G^S + G^A,  \pi^{\frac{1}{2}}A\pi^{\frac{1}{2}} ]\pi^{-\frac{1}{2}} - \pi^{-\frac{1}{2}}\frac{\partial }{\partial z_{i}} \left( K_i^{rev}  \pi^{\frac{1}{2}}A\pi^{\frac{1}{2}} +  \pi^{\frac{1}{2}}A\pi^{\frac{1}{2}} K_i^{rev\dag} \right)\pi^{-\frac{1}{2}} \\
&- \pi^{-\frac{1}{2}}\frac{\partial }{\partial z_{i}} \left( K_i^{irr}  \pi^{\frac{1}{2}}A\pi^{\frac{1}{2}} +  \pi^{\frac{1}{2}}A\pi^{\frac{1}{2}} K_i^{irr \dag} \right)\pi^{-\frac{1}{2}} +\frac{1}{2} \pi^{-\frac{1}{2}}\frac{\partial^2 }{\partial z_i \partial z_j} ( D_{2, i j}  \pi^{\frac{1}{2}}A\pi^{\frac{1}{2}} )\pi^{-\frac{1}{2}}\\
&+ {D_{2,ij}^{-1}} \pi^{-\frac{1}{2}}\big[ (K_i^{rev} +K_i^{irr}) \pi^{\frac{1}{2}}A\pi^{\frac{1}{2}} (K_j^{rev} +K_j^{irr})^{\dag} - \frac{1}{2}  \{ (K_j^{rev} +K_j^{irr})^{\dag} (K_i^{rev} +K_i^{irr}),  \pi^{\frac{1}{2}}A\pi^{\frac{1}{2}} \}_+ \big]\pi^{-\frac{1}{2}} 
\end{split}
\end{equation}
\begin{equation} 
\begin{split}
\mathcal{L}_\epsilon^\dag(A)-2i[X,A]=& \ -2i[X,A] + \frac{i}{\hbar}[G^S - G^A,  A ] + (K_i^{irr}- K_i^{rev})^\dag\frac{\partial A}{\partial z_{i}} + \frac{\partial A}{\partial z_{i}} (K_i^{irr}- K_i^{rev}) + \frac{1}{2}\epsilon_i\epsilon_j  D_{2, i j}(\epsilon z)  \frac{\partial^2 A}{\partial z_i \partial z_j}\\
&+ \epsilon_i \epsilon_j {D_{2,ij}^{-1}}(\epsilon z) \big[ (K_i^{irr}-K_i^{rev})^\dag  A (K_j^{irr}-K_j^{rev}) - \frac{1}{2}  \{ (K_j^{irr}-K_j^{rev})^\dag (K_i^{irr}-K_i^{rev}),  A \}_+ \big],
\end{split}
\end{equation} where here all quantities are dependent on $z$ rather than $\epsilon z$ unless indicated otherwise. 

Since the detailed balance condition \eqref{eq: classical-quantum_DB} must hold for all $A$, the terms containing each order of derivative of $A$ must cancel separately. Starting with the second order derivative terms, we see that all the dependence on $\pi$ cancels to give
\begin{equation}
    \frac{1}{2} D_{2, i j} \frac{\partial^2 A}{\partial z_i \partial z_j} =\frac{1}{2}\epsilon_i\epsilon_j  D_{2, i j}(\epsilon z)  \frac{\partial^2 A}{\partial z_i \partial z_j}.
\end{equation} This holds for all $A$ if and only if
\begin{equation} \label{eq: app_diffusion_constraint}
     D_{2, i j}( z)=\epsilon_i\epsilon_j  D_{2, i j}(\epsilon z)
\end{equation} which is exactly equivalent to the diffusion constraint of \eqref{eq: diffusion_constraint}. 

Turning now to the first derivatives of $A$, collecting terms and defining the operators $B_i$ as
\begin{equation}
    B_i=-\pi^{-\frac{1}{2}}K_i^{rev}\pi^{\frac{1}{2}} -\pi^{-\frac{1}{2}}K_i^{irr}\pi^{\frac{1}{2}} + \frac{1}{2}\frac{\partial D_{2,ij}}{\partial z_j} + D_{2,ij} \pi^{-\frac{1}{2}} \frac{\partial \pi^{\frac{1}{2}}}{\partial z_j} -K_i^{irr \dag}+K_i^{rev \dag},
\end{equation} we see that
\begin{equation}
    B_i \frac{\partial A}{\partial z_i} + \frac{\partial A}{\partial z_i} B_i^\dag = 0
\end{equation} must hold for every choice of operator $A$, which is true if and only if $B_i C +  C B_i^\dag=0$
holds for every choice of operator $C$. Taking the trace with $C=\mathds{1}$, we see that each $B_i$ must be anti-Hermitian. Decomposing this as $B=iJ$ where $J=J^\dag$, we see that the above relation implies that $J$ commutes with every operator i.e. is proportional to the identity operator. We thus see that we can write this condition in the form
\begin{equation} \label{eq: app_herm_and_antiherm_backreaction_constraint}
   -\pi^{-\frac{1}{2}}(K_i^{rev}+K_i^{irr})\pi^{\frac{1}{2}} + \frac{1}{2}\frac{\partial D_{2,ij}}{\partial z_j} + D_{2,ij} \pi^{-\frac{1}{2}} \frac{\partial \pi^{\frac{1}{2}}}{\partial z_j} -(K_i^{irr }-K_i^{rev })^\dag= -ia_i\mathds{1},
\end{equation} where here $a_i$ is some real number that may depend on phase space. Acting on the left with the operator $\pi^{\frac{1}{2}}$ and setting the Hermitian and anti-Hermitian parts equal separately, we see that we recover the reversible and irreversible backreaction constraints given in \eqref{eq: K_rev_constraint} and \eqref{eq: K_irr_constraint}. For these constraints to be self-consistent, we see that $a_i$ must transform as the reversible back-reaction operators do i.e. $a_i(\epsilon z)=-\epsilon_i a_i(z)$. 

The final conditions that follow from the terms proportional to $A$ are a little more complex to compute. To start with, we consider the expression $\mathcal{E}_1$, made up of moving all the terms which have operators acting on both sides of $A$ to the left hand side of the equation, which takes the form
\begin{equation}
\begin{split}
    \mathcal{E}_1=&-\pi^{-\frac{1}{2}}(K_i^{rev} + K_i^{irr})\pi^{\frac{1}{2}}A \frac{\partial \pi^{\frac{1}{2}}}{\partial z_i} \pi^{-\frac{1}{2}}-\pi^{-\frac{1}{2}} \frac{\partial \pi^{\frac{1}{2}}}{\partial z_i} A \pi^{\frac{1}{2}}(K_i^{rev} + K_i^{irr})^\dag \pi^{-\frac{1}{2}} +D_{2,ij} \pi^{-\frac{1}{2}} \frac{\partial \pi^{\frac{1}{2}}}{\partial z_i} A \frac{\partial \pi^{\frac{1}{2}}}{\partial z_i} \pi^{-\frac{1}{2}}\\
    &+ D_{2,ij}^{-1}\big[\pi^{-\frac{1}{2}}(K_i^{rev} + K_i^{irr})\pi^{\frac{1}{2}}]A\big[\pi^{\frac{1}{2}}(K_i^{rev} + K_i^{irr})^\dag \pi^{-\frac{1}{2}} \big] -D_{2,ij}^{-1}(K_i^{irr}-K_i^{rev})^\dag A (K_i^{irr}-K_i^{rev}).
\end{split}
\end{equation} Here we have here used the diffusion constraint \eqref{eq: app_diffusion_constraint} to rewrite $\epsilon_i \epsilon_j {D_{2,ij}^{-1}}(\epsilon z)$ as simply $D_{2,ij}^{-1}$. From the above expression, it is apparent that some of the terms will cancel upon replacing appearances of $\pi^{-\frac{1}{2}}(K_i^{rev} + K_i^{irr})\pi^{\frac{1}{2}}$ and its conjugate with $(K_i^{irr }-K_i^{rev })^\dag$ and its conjugate using Eq. \eqref{eq: app_herm_and_antiherm_backreaction_constraint}. Substituting these in, and using the properties of $D_2$ and its pseudoinverse that follow from the positivity condition $D_2\succeq0$,  such as $D_2 D_2^{-1}=D_2^{-1}D_2$ and $D_2 D_2 D_2^{-1}= D_2$, we find that this expression takes the form
\begin{equation}
\begin{split} \label{eq: app_almost_all_left_or_right}
    \mathcal{E}_1=&-(\mathbb{I}-D_2 D_2^{-1})_{ij}\big[(K_i^{rev}-K_i^{irr})^\dag + \frac{1}{2}\frac{\partial D_{2,ik}}{\partial z_k} + i a_i\big] A \frac{\partial \pi^{\frac{1}{2}}}{\partial z_j} \pi^{-\frac{1}{2}} + h.c. \\
    &+D_{2,ij}^{-1}(K_i^{rev}-K_i^{irr})^\dag (\frac{1}{2}\frac{\partial D_{2,jk}}{\partial z_k} - i a_j) A  +\frac{1}{8}D_{2,ij}^{-1}\frac{\partial D_{2,ik}}{\partial z_k}\frac{\partial D_{2,jl}}{\partial z_l} A + \frac{1}{2}D_{2,ij}^{-1} a_i a_j A  + h.c. \\
\end{split}
\end{equation} where here $h.c.$ denotes the Hermitian conjugate (treating $A$ as Hermitian) of all of the terms explicitly written on a given line. We see from this expression that the only terms that are not of the form of only a single operator to the left or right of $A$ are on the top line. However, let us recall that in addition to the positivity condition $D_2\succeq 0$, we also have the condition of \eqref{eq: alt_pos_conditions_2}. Writing this out explicitly, we see that this implies the existence of a phase space dependent vector $v$ such that
\begin{equation} \label{eq: app_positivity_condition}
    (\mathbb{I}-D_2 D_2^{-1})(K^{rev}+K^{irr})=v\mathds{1}.
\end{equation} Acting on the left with $\pi^{-\frac{1}{2}}$ and the right with $\pi^{\frac{1}{2}}$, we may use the expression in \eqref{eq: app_herm_and_antiherm_backreaction_constraint} to rewrite this as
\begin{equation} 
    (\mathbb{I}-D_2 D_2^{-1})\big[(K^{rev}-K^{irr})^\dag + \frac{1}{2}\frac{\partial D_{2}}{\partial z} + ia \big] = v \mathds{1},
\end{equation} where here we have dropped a term using $D_2 D_2^{-1} D_2=D_2$. We thus see that the main expression of the top line of \eqref{eq: app_almost_all_left_or_right} is proportional to the identity, and thus may be commuted with $A$ to give an expression of $\mathcal{E}_1$ entirely in the form of operators on the left or right hand side of $A$. Returning now to include the terms already of this form, we see that the constraint setting the entirety of the terms proportional to $A$ to zero takes the form
\begin{equation}
    F A + A F^\dag = 0
\end{equation} for all operators $A$, where $F$ is defined as the operator
\begin{equation}
\begin{split}
    F=&-\frac{i}{\hbar}(\pi^{-\frac{1}{2}}G^S\pi^{\frac{1}{2}}+G^S) -\frac{i}{\hbar}(\pi^{-\frac{1}{2}}G^A\pi^{\frac{1}{2}}-G^A) +2iX\\
    &-\pi^{-\frac{1}{2}} \frac{\partial}{\partial z_i}\bigg((K_i^{rev}+K_i^{irr})\pi^{\frac{1}{2}}\bigg)
    +\frac{1}{4}\frac{\partial^2 D_{2,ij}}{\partial z_i \partial z_j} + \pi^{-\frac{1}{2}}\frac{\partial D_{2,ij}}{\partial z_i}\frac{\partial \pi^{\frac{1}{2}}}{\partial z_j} + \frac{1}{2}\pi^{-\frac{1}{2}}D_{2,ij}\frac{\partial^2 \pi^{\frac{1}{2}}}{\partial z_i \partial z_j}\\
    &-\frac{1}{2}D_{2,ij}^{-1}\bigg[\pi^{-\frac{1}{2}}(K_i^{rev}+K_i^{irr})^\dag(K_j^{rev}+K_j^{irr})\pi^{\frac{1}{2}}+(K_i^{irr}-K_i^{rev})^\dag(K_j^{irr}-K_j^{rev})\bigg] \\
    &-(\mathbb{I}-D_2 D_2^{-1})_{ij} \big[(K_i^{rev}-K_i^{irr}) + \frac{1}{2}\frac{\partial D_{2,ik}}{\partial z_k} - i a_i\big] \pi^{-\frac{1}{2}} \frac{\partial \pi^{\frac{1}{2}}}{\partial z_j} \\
    &+D_{2,ij}^{-1}(K_i^{rev}-K_i^{irr})^\dag (\frac{1}{2}\frac{\partial D_{2,jk}}{\partial z_k} - i a_j)  +\frac{1}{8}D_{2,ij}^{-1}\frac{\partial D_{2,ik}}{\partial z_k}\frac{\partial D_{2,jl}}{\partial z_l} + \frac{1}{2}D_{2,ij}^{-1} a_i a_j.
\end{split}
\end{equation} Noting as before that this implies that $F=ib\mathds{1}$, where $b$ is a real number dependent on phase-space, we may act with $\pi^{\frac{1}{2}}$ on the left to find the analogue of \eqref{eq: app_herm_and_antiherm_backreaction_constraint} for the zeroth order derivative terms
\begin{equation} \label{eq: app_unitary_generator_constraints_both}
\begin{split}
    &-\frac{i}{\hbar}(G^S\pi^{\frac{1}{2}}+\pi^{\frac{1}{2}}G^S) -\frac{i}{\hbar}(G^A\pi^{\frac{1}{2}}-\pi^{\frac{1}{2}}G^A) +i(X\pi^{\frac{1}{2}}+\pi^{\frac{1}{2}}X)\\
    &-\frac{\partial}{\partial z_i}\bigg((K_i^{rev}+K_i^{irr})\pi^{\frac{1}{2}}\bigg)
    +\frac{1}{4}\pi^{\frac{1}{2}}\frac{\partial^2 D_{2,ij}}{\partial z_i \partial z_j} + \frac{\partial D_{2,ij}}{\partial z_i}\frac{\partial \pi^{\frac{1}{2}}}{\partial z_j} + \frac{1}{2}D_{2,ij}\frac{\partial^2 \pi^{\frac{1}{2}}}{\partial z_i \partial z_j}\\
    &-\frac{1}{2}D_{2,ij}^{-1}[K_i^{rev\dag}K_j^{rev} +K_i^{irr\dag}K_j^{irr},\pi^{\frac{1}{2}}]-\frac{1}{2}D_{2,ij}^{-1}\{K_i^{rev\dag}K_j^{irr} + K_i^{irr\dag}K_j^{rev},\pi^{\frac{1}{2}}\}_+\\
    &-(\mathbb{I}-D_2 D_2^{-1})_{ij} \big[(K_i^{rev}-K_i^{irr}) + \frac{1}{2}\frac{\partial D_{2,ik}}{\partial z_k} - i a_i\big]  \frac{\partial \pi^{\frac{1}{2}}}{\partial z_j} \\
    &+D_{2,ij}^{-1}(\frac{1}{2}\frac{\partial D_{2,ik}}{\partial z_k} - i a_i)\pi^{\frac{1}{2}}(K_j^{rev}-K_j^{irr})^\dag   +\frac{1}{8}D_{2,ij}^{-1}\frac{\partial D_{2,ik}}{\partial z_k}\frac{\partial D_{2,jl}}{\partial z_l} \pi^{\frac{1}{2}} + \frac{1}{2}D_{2,ij}^{-1} a_i a_j \pi^{\frac{1}{2}} = i b \pi^{\frac{1}{2}}.
\end{split}
\end{equation} In the above, we have used the property that $X$ commutes with $\pi$ to rewrite the term containing $X$ as an anti-Hermitian term, as well as rewriting the $D_{2,ij}^{-1}$ term into commutator and anti-commutator parts. To derive the symmetric and anti-symmetric unitary generator constraints from this expression, we consider the anti-Hermitian and Hermitian parts of the above expression separately. The anti-Hermitian part of this equation takes the form
\begin{equation}
\begin{split}
    &-\frac{i}{\hbar}(G^S\pi^{\frac{1}{2}}+\pi^{\frac{1}{2}}G^S) + i(X\pi^{\frac{1}{2}}+  \pi^{\frac{1}{2}}X) \\
    &- \frac{1}{2} \frac{\partial}{\partial z_i}(K_i^{rev}\pi^{\frac{1}{2}}-\pi^{\frac{1}{2}} K_i^{rev\dag} )- \frac{1}{2} \frac{\partial}{\partial z_i}(K_i^{irr}\pi^{\frac{1}{2}}-\pi^{\frac{1}{2}} K_i^{irr\dag })
    \\
    &-\frac{1}{2}D_{2,ij}^{-1}[K_i^{rev\dag}K_j^{rev} +K_i^{irr\dag}K_j^{irr},\pi^{\frac{1}{2}}] \\
    &-\frac{1}{2}(\mathbb{I}-D_2 D_2^{-1})_{ij}\big[(K_i^{rev}-K_i^{irr}) \frac{\partial \pi^{\frac{1}{2}}}{\partial z_j} - \frac{\partial \pi^{\frac{1}{2}}}{\partial z_j}(K_i^{rev}-K_i^{irr})^\dag\big]
    +ia_i(\mathbb{I}-D_2 D_2^{-1})_{ij} \frac{\partial \pi^{\frac{1}{2}}}{\partial z_j} \\
    &-\frac{i}{2} a_i D_{2,ij}^{-1}(K_j^{rev}\pi^{\frac{1}{2}} +\pi^{\frac{1}{2}}K_j^{rev\dag} )+\frac{i}{2} a_i D_{2,ij}^{-1}(K_j^{irr}\pi^{\frac{1}{2}} +\pi^{\frac{1}{2}}K_j^{irr\dag} )\\
    &-\frac{1}{4}D_{2,ij}^{-1}\frac{\partial D_{2,ij}}{\partial  z_k}(K_j^{rev}\pi^{\frac{1}{2}} -\pi^{\frac{1}{2}}K_j^{rev\dag} ) +\frac{1}{4}D_{2,ij}^{-1}\frac{\partial D_{2,ij}}{\partial  z_k}(K_j^{irr}\pi^{\frac{1}{2}} -\pi^{\frac{1}{2}}K_j^{irr\dag} )= i b \pi^{\frac{1}{2}}.
\end{split}
\end{equation} Substituting in the reversible and irreversible backreaction constraints \eqref{eq: K_rev_constraint} and \eqref{eq: K_irr_constraint} wherever $K_i^{rev}\pi^{1/2}-\pi^{1/2}K_i^{rev\dag}$ or $K_i^{irr}\pi^{1/2}+\pi^{1/2}K_i^{irr\dag}$ explicitly appear, one finds that the equation simplifies to
\begin{equation}
\begin{split}
    -\frac{i}{\hbar}(G^S\pi^{\frac{1}{2}}+\pi^{\frac{1}{2}}G^S)&= - i(X\pi^{\frac{1}{2}}+  \pi^{\frac{1}{2}}X) +\ \frac{1}{2}\frac{\partial}{\partial z_i}(K_i^{irr}\pi^{1/2}-\pi^{1/2}{K_i^{irr}}^\dag)\\
    &-\frac{1}{2}(\mathbb{I}-D_2 D_2^{-1})_{ij}(K_i^{irr}\frac{\partial \pi^{\frac{1}{2}}}{\partial z_j}-\frac{\partial \pi^{\frac{1}{2}}}{\partial z_j}{K_i^{irr}}^\dag) -\frac{1}{4}D_{2,ij}^{-1}\frac{\partial D_{2,ik}}{\partial z_k}(K_j^{irr} \pi^{1/2}- \pi^{1/2}{K_j^{irr}}^\dag) \\
    &+\frac{i}{2}D_{2,ij}^{-1}a_i (K_i^{rev} \pi^{1/2} + \pi^{1/2} {K_i^{rev}}^\dag) +\frac{1}{2}D_{2,ij}^{-1}[{K_i^{rev}}^\dag K_j^{rev}+{K_i^{irr}}^\dag K_j^{irr},\pi^{1/2}]\\
    &-\frac{i}{2}a_i(\mathbb{I}-D_2 D_2^{-1})_{ij} \frac{\partial \pi^{\frac{1}{2}}}{\partial z_j} + \frac{1}{2}(\mathbb{I}-D_2 D_2^{-1})_{ij}(K_i^{rev}\frac{\partial \pi^{\frac{1}{2}}}{\partial z_j}-\frac{\partial \pi^{\frac{1}{2}}}{\partial z_j}{K_i^{rev}}^\dag) +i(b+\frac{1}{2}\frac{\partial a_i}{\partial z_i})\pi^{\frac{1}{2}}
\end{split}
\end{equation} Since the left-hand side of this equation is symmetric under time-reversal, the right-hand side must also be. Performing the time-reversal transformation, noting the previously found form of time-reversal for $a_i$, we see that all of the terms are symmetric, other than the three terms on the final line. Returning to the positivity condition \eqref{eq: app_positivity_condition}, we note that since this holds for both the reversible and irreversible components separately i.e.
\begin{equation} \label{eq: app_positivity_condition_split}
    (\mathbb{I}-D_2 D_2^{-1})K^{rev}=v^{rev}\mathds{1},\quad \quad (\mathbb{I}-D_2 D_2^{-1})K^{irr}=v^{irr}\mathds{1},
\end{equation} one may rearrange the two terms containing $\mathbb{I}-D_2 D_2^{-1}$ on the final line into the form
\begin{equation}
    \frac{1}{2}(\mathbb{I}-D_2 D_2^{-1})_{ij} \big[K_i^{rev} \pi^{\frac{1}{2}} -  \pi^{\frac{1}{2}} K_i^{rev} - i a_i \pi^{\frac{1}{2}} \big]\pi^{-\frac{1}{2}}\frac{\partial \pi^{\frac{1}{2}}}{\partial z_j},
\end{equation} which clearly vanishes due to the reversible back-reaction constraint. For the final term containing $b$ to vanish, we see that $b$ must have an anti-symmetric component equal to $-\frac{1}{2}\partial_i a_i$ such that this final term cancels. Since the remaining component is purely symmetric and real, this term can be absorbed into the definition of $X$ by including a term $\frac{b}{2}\mathds{1}$; since the above condition required the existence of a real-number $b$, and detailed balance required the existence of the operator $X$, we see that we may do this without loss of generality. Doing so, and defining the operator $O^S$ as in \eqref{eq: OS_operator_def}, we find the form of the symmetric unitary generator constraint quoted in \eqref{eq: symmetric_effective_H}. 

Finally, turning to the Hermitian part of \eqref{eq: app_unitary_generator_constraints_both}, we find this takes the form
\begin{equation}
\begin{split}
\begin{split}
    & -\frac{i}{\hbar}(G^A\pi^{\frac{1}{2}}-\pi^{\frac{1}{2}}G^A) \\
    &-\frac{1}{2}\frac{\partial}{\partial z_i}(K_i^{rev}\pi^{\frac{1}{2}}+\pi^{\frac{1}{2}}K_i^{rev\dag})-\frac{1}{2}\frac{\partial}{\partial z_i}(K_i^{irr}\pi^{\frac{1}{2}}+\pi^{\frac{1}{2}}K_i^{irr\dag})\\
    &+\frac{1}{4}\pi^{\frac{1}{2}}\frac{\partial^2 D_{2,ij}}{\partial z_i \partial z_j} + \frac{\partial D_{2,ij}}{\partial z_i}\frac{\partial \pi^{\frac{1}{2}}}{\partial z_j} + \frac{1}{2}D_{2,ij}\frac{\partial^2 \pi^{\frac{1}{2}}}{\partial z_i \partial z_j}\\
    &-\frac{1}{2}D_{2,ij}^{-1}\{K_i^{rev\dag}K_j^{irr} + K_i^{irr\dag}K_j^{rev},\pi^{\frac{1}{2}}\}_+\\
    &-\frac{1}{2}(\mathbb{I}-D_2 D_2^{-1})_{ij}\big[(K_i^{rev}-K_i^{irr}) \frac{\partial \pi^{\frac{1}{2}}}{\partial z_j} + \frac{\partial \pi^{\frac{1}{2}}}{\partial z_j}(K_i^{rev}-K_i^{irr})^\dag\big] - \frac{1}{2}(\mathbb{I}-D_2 D_2^{-1})_{ij}  \frac{\partial D_{2,ik}}{\partial z_k}   \frac{\partial \pi^{\frac{1}{2}}}{\partial z_j}\\
    &+\frac{i}{2} a_i D_{2,ij}^{-1}(K_j^{rev}\pi^{\frac{1}{2}} -\pi^{\frac{1}{2}}K_j^{rev\dag} )-\frac{i}{2} a_i D_{2,ij}^{-1}(K_j^{irr}\pi^{\frac{1}{2}} -\pi^{\frac{1}{2}}K_j^{irr\dag} )\\
    &+\frac{1}{4}D_{2,ij}^{-1}\frac{\partial D_{2,ij}}{\partial  z_k}(K_j^{rev}\pi^{\frac{1}{2}} +\pi^{\frac{1}{2}}K_j^{rev\dag} ) -\frac{1}{4}D_{2,ij}^{-1}\frac{\partial D_{2,ij}}{\partial  z_k}(K_j^{irr}\pi^{\frac{1}{2}} +\pi^{\frac{1}{2}}K_j^{irr\dag} )\\
    & +\frac{1}{8}D_{2,ij}^{-1}\frac{\partial D_{2,ik}}{\partial z_k}\frac{\partial D_{2,jl}}{\partial z_l} \pi^{\frac{1}{2}} + \frac{1}{2}D_{2,ij}^{-1} a_i a_j \pi^{\frac{1}{2}} = 0.
\end{split}
\end{split}
\end{equation} As before, substituting in the expressions for $K_i^{rev}\pi^{1/2}-\pi^{1/2}K_i^{rev\dag}$ or $K_i^{irr}\pi^{1/2}+\pi^{1/2}K_i^{irr\dag}$ from \eqref{eq: K_rev_constraint} and \eqref{eq: K_irr_constraint}, we find that this equation simplifies to take the form
\begin{equation}
\begin{split}
    -\frac{i}{\hbar}(G^A\pi^{\frac{1}{2}}-\pi^{\frac{1}{2}}G^A)=
    &\ \frac{1}{2}\frac{\partial}{\partial z_i}(K_i^{rev}\pi^{1/2}+\pi^{1/2}{K_i^{rev}}^\dag)\\
    &+\frac{1}{2}(\mathbb{I}-D_2 D_2^{-1})_{ij}(K_i^{rev}\frac{\partial \pi^{\frac{1}{2}}}{\partial z_j}+\frac{\partial \pi^{\frac{1}{2}}}{\partial z_j}{K_i^{rev}}^\dag) \\
    &-\frac{1}{4}D_{2,ij}^{-1}\frac{\partial D_{2,ik}}{\partial z_k}(K_j^{rev} \pi^{1/2}+ \pi^{1/2}{K_j^{rev}}^\dag) \\
    &+\frac{i}{2}D_{2,ij}^{-1}a_i (K_i^{irr} \pi^{1/2} - \pi^{1/2} {K_i^{irr}}^\dag)\\
    &+\frac{1}{2}D_{2,ij}^{-1}\{{K_i^{rev}}^\dag K_j^{irr}+{K_i^{irr}}^\dag K_j^{rev},\pi^{1/2}\}_+\\
    &+\frac{1}{4}(\mathbb{I}-D_2 D_2^{-1})_{ij} \frac{\partial D_{2,ik}}{\partial z_k}\frac{\partial \pi^{\frac{1}{2}}}{\partial z_j} -\frac{1}{2}(\mathbb{I}-D_2 D_2^{-1})_{ij}(K_i^{irr}\frac{\partial \pi^{\frac{1}{2}}}{\partial z_j}+\frac{\partial \pi^{\frac{1}{2}}}{\partial z_j}{K_i^{irr}}^\dag). \\
\end{split}
\end{equation} As before, we perform the time-reversal operation on both sides to see that all but the final line as written are antisymmetric under time-reversal. Using now the relation for $K^{irr}$ in \eqref{eq: app_positivity_condition_split} we see that this final time-reversal symmetric line vanishes in an analogous way to before, this time by using the irreversible back-reaction constraint. Defining the operator $O^A$ in \eqref{eq: OA_operator_def}, we thus recover the antisymmetric unitary generator constraint \eqref{eq: antisymmetric_effective_H}.

\end{document}